\documentclass[useAMS,usegraphicx,usenatbib,times,hyperref]{mn2e}

\usepackage{graphicx}
\usepackage{subfigure}
\usepackage{amsmath,amssymb}
\usepackage{algorithm}
\usepackage{algorithmicx}
\usepackage{algpseudocode}

\usepackage[usenames]{color}
\definecolor{grey}{rgb}{0.35,0.35,0.35}
\definecolor{dblue}{rgb}{0.05,0.05,0.35}

\usepackage[ps2pdf,colorlinks=true,breaklinks=true,bookmarks=true,hypertexnames=false,bookmarksnumbered=true,backref=page,pagebackref=true,citecolor= dblue,linkcolor= grey,bookmarksopen=false,bookmarks=true,bookmarksnumbered=true]{hyperref} 

\newtheorem{mydef}{Definition}[section]
\newtheorem{myprop}{Property}[mydef]

\def\cofaceof#1{\left<#1\right>}

\def\progname{DisPerSE\,}

\def\papertwo{Sousbie, Pichon, Kawahara (2010)\,}

\newcommand{\kpn}[2][k]{( #1 \!+\! #2 )}
\newcommand{\kmn}[2][k]{( #1 \!-\! #2 )}
\def\kp1{$\kpn{1}$}
\def\km1{$\kmn{1}$}

\def\Mpc {{\,h^{-1}\,{\rm Mpc}}}

\def\and  {\it {et al.} \rm}

\def\nsig#1{{\rm #1-}\sigma}

\def\ppair#1#2{\left[#1,#2\right]}

\def\eg {{\it e.g. }}
\def\ie {{\it i.e. }}
\def\ni{\noindent}

\def\vx{{\mathbf x}}
\def\vp{{\mathbf p}}
\def\vL{{\mathbf L}}
\def\gradx{ \nabla_\vx }
\def\vH{{\cal H}}
\def\reals{{\mathbb{R}}}
\def\torus{{\mathbb{T}}}

\def\eqvspace{\nonumber\\\noalign{\vskip 1mm}}

\usepackage[usenames]{color}
\definecolor{Blue}{rgb}{0,0.08,0.65}
\definecolor{Red}{rgb}{0.65,0.08,0.05}
\definecolor{Green}{rgb}{0.15,0.65,0.25}

\begin{document}
\title[Persistent cosmic web I: 
Theory and implementation]{The persistent cosmic web and its filamentary structure \\ I: 
Theory and implementation}
\author[T. Sousbie]
{\ni T. Sousbie$^{1,2}$\\
$^{1}$Department of Physics, The University of Tokyo, Tokyo 113-0033, Japan,\\
 $^{2}$Institut d'astrophysique
de Paris \& UPMC (UMR 7095), 98, bis boulevard Arago, 75 014, Paris.\\
{tsousbie@gmail.com, sousbie@utap.phys.s.u-tokyo.ac.jp}
}
\maketitle
\begin{abstract}
We present \progname, a novel approach to the coherent multi-scale identification of all types of astrophysical structures, and in particular the filaments, in the large scale distribution of matter in the Universe. This method and corresponding piece of software allows a genuinely scale free and parameter free identification of the voids, walls, filaments, clusters and their configuration within the cosmic web, directly from the discrete distribution of particles in N-body simulations or galaxies in sparse observational catalogues. To achieve that goal, the method works directly over the Delaunay tessellation of the discrete sample and uses the DTFE density computed at each tracer particle; no further sampling, smoothing or processing of the density field is required. 
  
  The idea is based on recent advances in distinct sub-domains of computational topology, namely the {\em discrete} Morse theory which allows a rigorous application of topological principles to astrophysical data sets, and the theory of persistence, which allows us to consistently account for the intrinsic uncertainty and Poisson noise within data sets. Practically, the user can define a given persistence level in terms of robustness with respect to noise (defined as a ``number of sigmas'') and the algorithm returns the structures with the corresponding significance as sets of critical points, lines, surfaces and volumes corresponding to the clusters, filaments, walls and voids; filaments, connected at cluster nodes, crawling along the edges of walls bounding the voids. From a geometrical point of view, the method is also interesting as it allows for a robust quantification of the topological properties  of a discrete distribution in terms of Betti numbers or Euler characteristics, without having to resort to smoothing or having to define a particular scale.
  
In this paper, we  introduce the necessary mathematical background and describe the method and implementation, while we address the application to 3D simulated and observed data sets to the companion paper, \papertwo\!.

\end{abstract}

\section{Introduction}

\label{sec:intro}
The existence of an intricate network of filaments in the large scale distribution of matter is now considered an established fact. Its was first observed by \citet{lapparent86} \citep[see also \eg][]{2Df} and latter theorized \citep[see \eg][]{pogo96,bond96}: under-dense void regions bounded by sheet-like walls embedded in a web like filamentary network branching on high density dark matter haloes and galaxy clusters form the so called cosmic web \citet{bond96}, that spans over a wide range of scales larger than the Megaparsec. Dark matter halos and galaxy clusters have arguably been the most studied component, and there exist a wide range of methods to identify them in simulations or observational catalogues such as the classical friend-of-friend (FOF) \citep{FOF}, HFOF and 6D minimal spanning tree\citep{gottlober98}, SUBFIND \citep{springel01}, VOBOZ \citep{neyrinck05} or  ADAPTAHOP \citep{ADHOP,ADHOP2} (the list is not exhaustive).  Cosmological voids were first observed by \citet{kirshner81} and theoretical models were latter developed \citep[see \eg][]{hoffman82,icke84,bertschinger85}. Although they have been the subject of less attention, there still exist a large number of references describing their features and introducing numerical void finders such as for instance \citet{neyrinck}, \citet{platen} or \citet{calvo_voids} (see also the references therein). Because of the intrinsic difficulty of even defining the concepts of wall and filament, not to mention designing consistent identification algorithms (especially in the case of observational data), their generic properties still remain relatively uncertain. One can for instance refer to \citet{calvo10} for a nice review of the different identification techniques and a study of the filaments properties in dark matter N-body simulations \citep[see also \eg][]{gay10}, and \citet{stoica10} or  \citet{SDSSskel} for recent attempts at identifying filaments properties in the SDSS and 2dFGRS galaxy catalogues, using the CANDY model \citep{stoica05} and skeleton formalism \citep{skel} respectively. In this paper, we  present a general framework within which the physically meaningful objects that are the voids, walls, filaments and haloes are rigorously and consistently defined and we also detail the corresponding numerical method that allows for their direct identification in simulated as well as observational data sets. We focus in particular on what is probably the most striking feature of matter distribution on large scales in the Universe, its filamentary structure.\\

\begin{figure*}
\centering\includegraphics[width=0.48\textwidth]{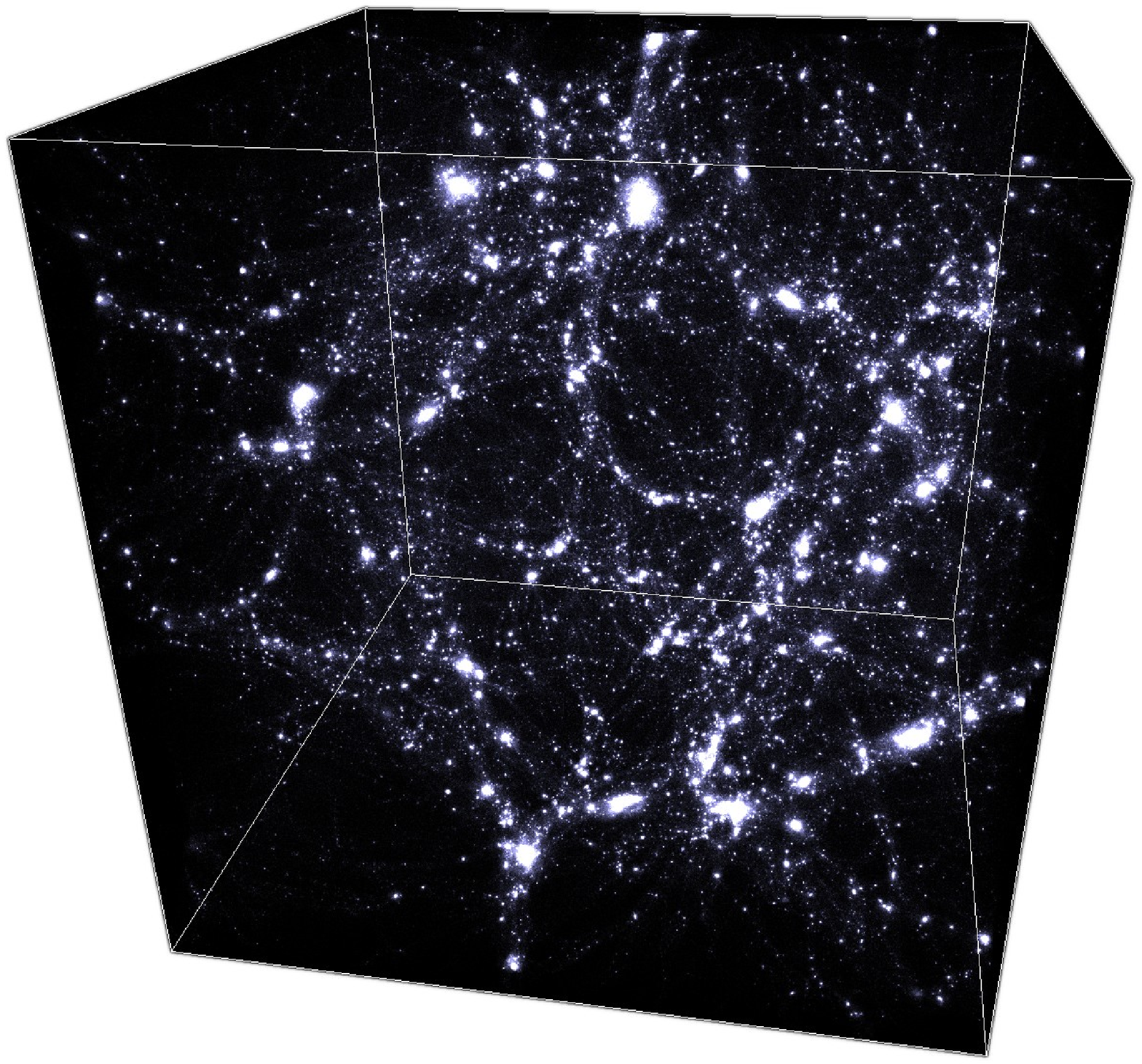}
\includegraphics[width=0.48\textwidth]{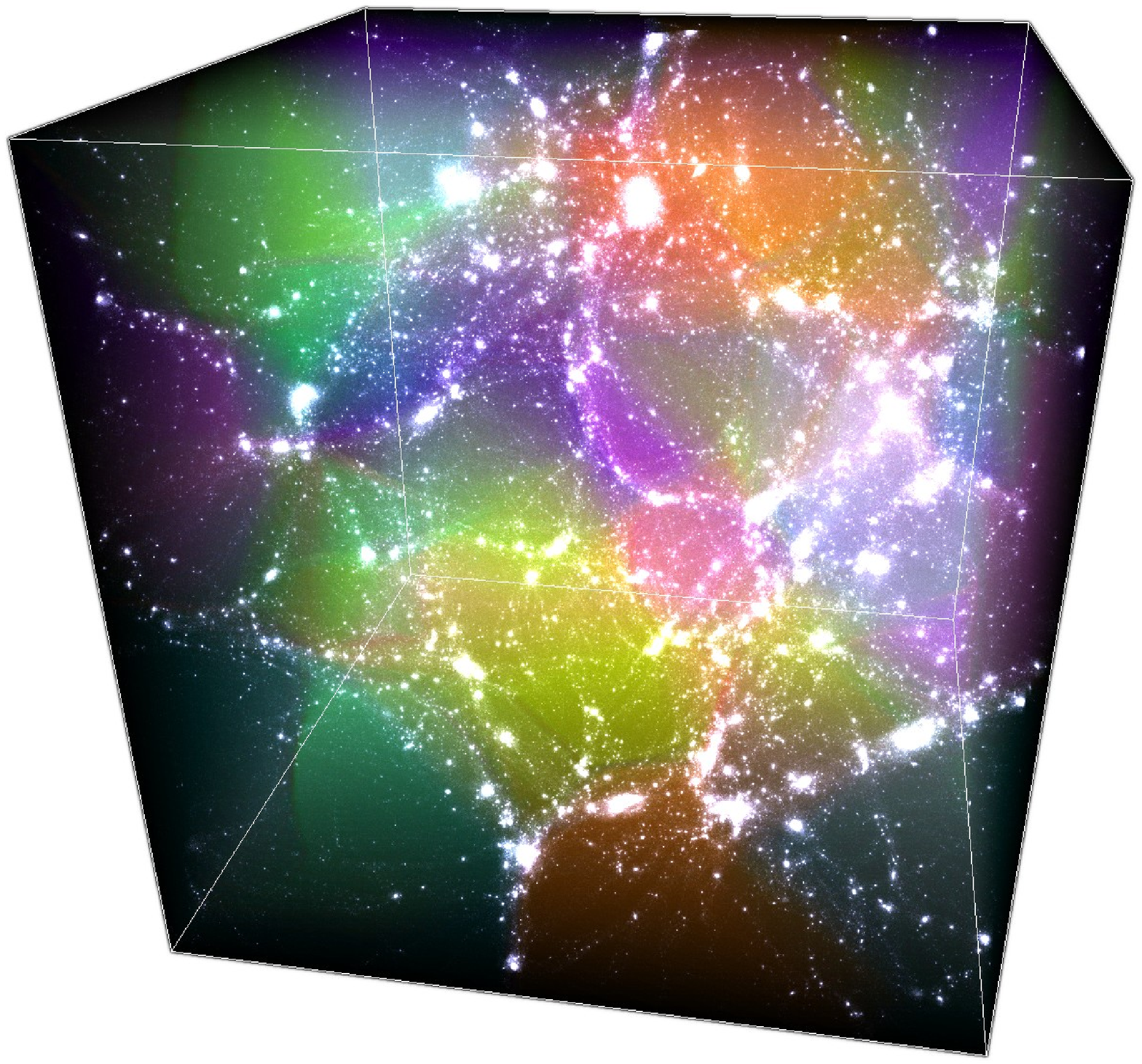}\\
\centering\includegraphics[width=0.48\textwidth]{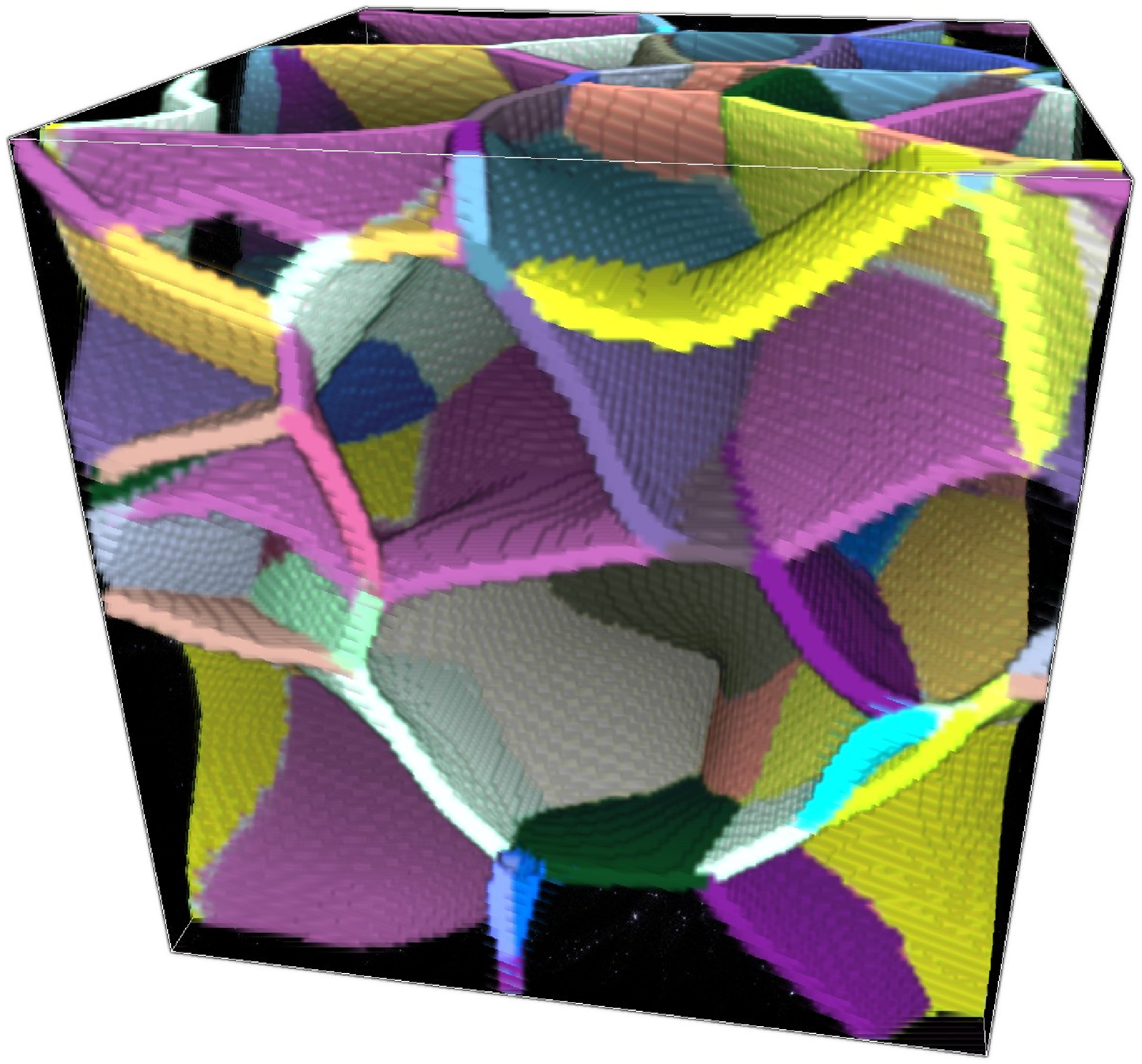}
\includegraphics[width=0.48\textwidth]{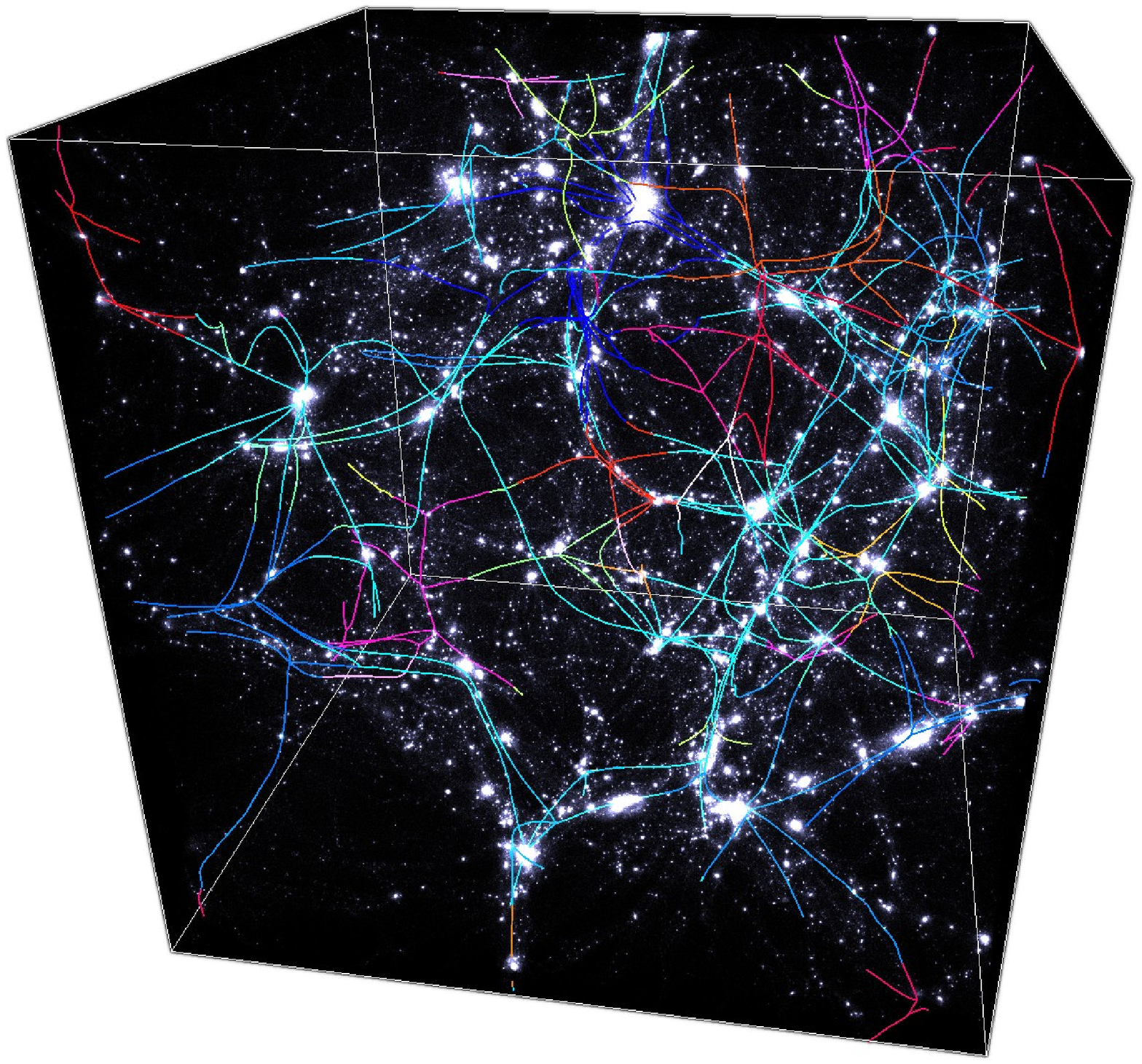}\\
\caption{The dark matter density distribution in a $50\Mpc$ large cosmological simulation (top left frame), with its ascending $3$-\hyperref[defmanifold]{manifolds} (\ie the voids, top right frame), ascending $2$-\hyperref[defmanifold]{manifold}s (\ie the walls, bottom left frame) and ascending $1$-\hyperref[defmanifold]{manifold}s (\ie the filaments, bottom right frame). The \hyperref[defmanifold]{manifold}s were computed using the method introduced in \citet{rsex}.\label{fig_example_morse_simu}}
\end{figure*}

During the last few years, Morse theory  (\eg ~\citet{milnor,jost}) has been recognized as a very promising approach to the global identification of all types of astrophysically significant features of the large scale galaxy distribution in the universe \citep[see \eg][]{skel2D,hahn,SDSSskel,rsex, skel,spine,jaime}. The main reason for this strong interest comes from the fact that all the salient features of the web-like pattern of galaxies have a direct, mathematically well defined equivalent in Morse theory. In fact, Morse theory mainly relies on the definition of so-called ascending and descending $k$-\hyperref[defmanifold]{manifold}s, which {\sl partition} space into series of $k$-dimensional domains defined by the gradient of a function (in the present case, the density field), and the network whose branches are formed by their intersections and whose nodes are the critical points, the so-called Morse-complex (see section \ref{sec_morse}). As illustrated on figure \ref{fig_example_morse_simu}, each of those can be directly associated to an astrophysical objects of interest: an ascending $3$-\hyperref[defmanifold]{manifold} defines a void, an ascending $2$-\hyperref[defmanifold]{manifold} defines a wall, and ascending $1$-\hyperref[defmanifold]{manifold} defines a filaments, a descending $3$-\hyperref[defmanifold]{manifold} defines a peak-patch of peak theory \citep{BBKS}, ... and the Morse complex defines some sort of hierarchy and a notion of neighbourhood between them (see section \ref{sec_morse} for more details).\\

Nevertheless, and as promising as it may seem, all the efforts toward applying Morse theory to astrophysical data sets such as galaxy catalogues have so far been plagued by major difficulties. Those difficulties are a direct consequence of the fact that Morse theory, although very attractive, is fundamentally a mathematical theory defined for idealized, well defined and properly behaved smooth functions, which of course is not generally the case of any physical data set resulting from actual measurements. At least two critical issues can be identified in the case of the large scale structure identification problem. The first results from the presence of Poisson noise and large observational biases in galaxy catalogues, which should be dealt with from the start, especially when the data set is relatively sparse as it becomes even more difficult in that case to distinguish between noise features and the actual features of the sampled data set. The second issue arises from the fact that Morse theory applies to so called \hyperref[defMF]{Morse functions} (see definition \ref{def_morse_function}), which are sufficiently smooth twice differentiable {\em continuous} functions (whose critical points are non-degenerate) whereas the galaxy distribution is discrete by nature. This incompatibility is fundamental, as it means that the theoretical notions of Morse theory may actually not apply to any practical data set. A more detailed discussion of this problem is presented in appendix \ref{sec_appwatershed} as well as an example of the consequences of neglecting this inconsistency in the case of watershed based methods such as \cite{rsex,spine}.\\

In this paper, we focus on presenting \progname, a formalism and corresponding software specifically designed for analyzing  the cosmic web and its filamentary network. This formalism is based on Morse theory,  while the aforementioned incompatibilities with astrophysical data sets are overcome by relying on relatively recent advances in distinct sub-domains of computational topology. These domains are discrete Morse theory  (a distinct though related theory developed by Forman see ~\citet{forman98,forman} and references therein) and persistent homology, first introduced in \citet{edel00,edel}. We therefore start by introducing the corresponding necessary notions of computational topology in sections \ref{sec_morse}, \ref{sec_DMtheory} and \ref{sec_persistence} respectively. Note that no previous knowledge in the field of computational topology is assumed here, the goal of those sections being mainly to introduce the required mathematical vocabulary that we use extensively in the following sections, and  give a glimpse at how those theories can help deepen our understanding of the structure of the cosmic web. The reader interested in pursuing this investigation further should refer to the aforementioned references for a more detailed and involved introduction. In particular, we strongly recommend the reading of ~\citet{phd} and especially ~\citet{zomo} for a very didactic presentation of these concepts. Indeed, the particular method and implementation presented in this paper are inspired by the work presented in those two references.\\ 

We then proceed by showing in section \ref{sec_implementation} how it is possible, relying on the previously mentioned theories, to design an algorithm that rigorously computes the {\em discrete} Morse complex of a discrete density field, obtained using DTFE technique \citep{DTFE} from the delaunay tesselation of a given discretely sampled data set, such as the distribution of galaxies in the universe. Within our approach, the Morse complex is directly computed from the delaunay tessellation which means it is scale adaptive and parameter free. The problem of dealing with Poisson noise and measurement errors is addressed in section \ref{sec_toposimp}, where we make use of persistence theory to remove spurious topological features from the Morse complex. Practically, the filamentary network (and associated voids, walls, ...) computed from the initial distribution is simplified by canceling pairs of critical points according to a persistence criterion, that can be restated in terms of significance relative to shot noise. Finally, in section \ref{sec_det_imp}, we address technical questions such as dealing with boundary conditions, smoothing the identified voids, walls and filaments and important implementation problems before concluding in section \ref{sec_conclusion}.

 Importantly, let us  emphasize that within this framework, the mathematical theories that we use are fundamentally discrete and readily apply to the measured raw data; the unique supplementary but critical step consists in defining heuristically  a consistent labeling of the segments, triangle and tetrahedron of the delaunay tesselation with regards to the DTFE densities computed at the sampling points (see section \ref{sec_disc_grad}). This warrants  that all the well known and extensively studied mathematical properties of the Morse complex are ensured {\em by construction} at the mesh level, and that the corresponding cosmological structures therefore correspond to well defined mathematical objects with known mathematical properties. It also provides a consistent way of reconnecting  the corresponding network after the removal of  insignificant (non-persistent) pairs of critical points.  

Note that a reference is given on the last two page, in which most mathematical terminology introduced in sections \ref{sec_morse}, \ref{sec_DMtheory} and \ref{sec_persistence} is defined in relatively simple terms. As we only aim here to introduce the necessary mathematical notions and giving a detailed description of the computation pipeline, extensively illustrating each step, the application to actual data sets is presented in a less technical companion paper, \papertwo\!. In that paper, we show the potential of this approach by applying it to typical cosmological data-sets: a large scale dark matter cosmological N-body simulation and the $7^{\rm th}$ data release (DR7) of the SDSS galaxy catalogue \citep{Abazajian09}. 

\section{Morse theory for smooth manifolds}
\label{sec_morse}

\begin{figure*}
\centering\includegraphics[width=0.45\textwidth]{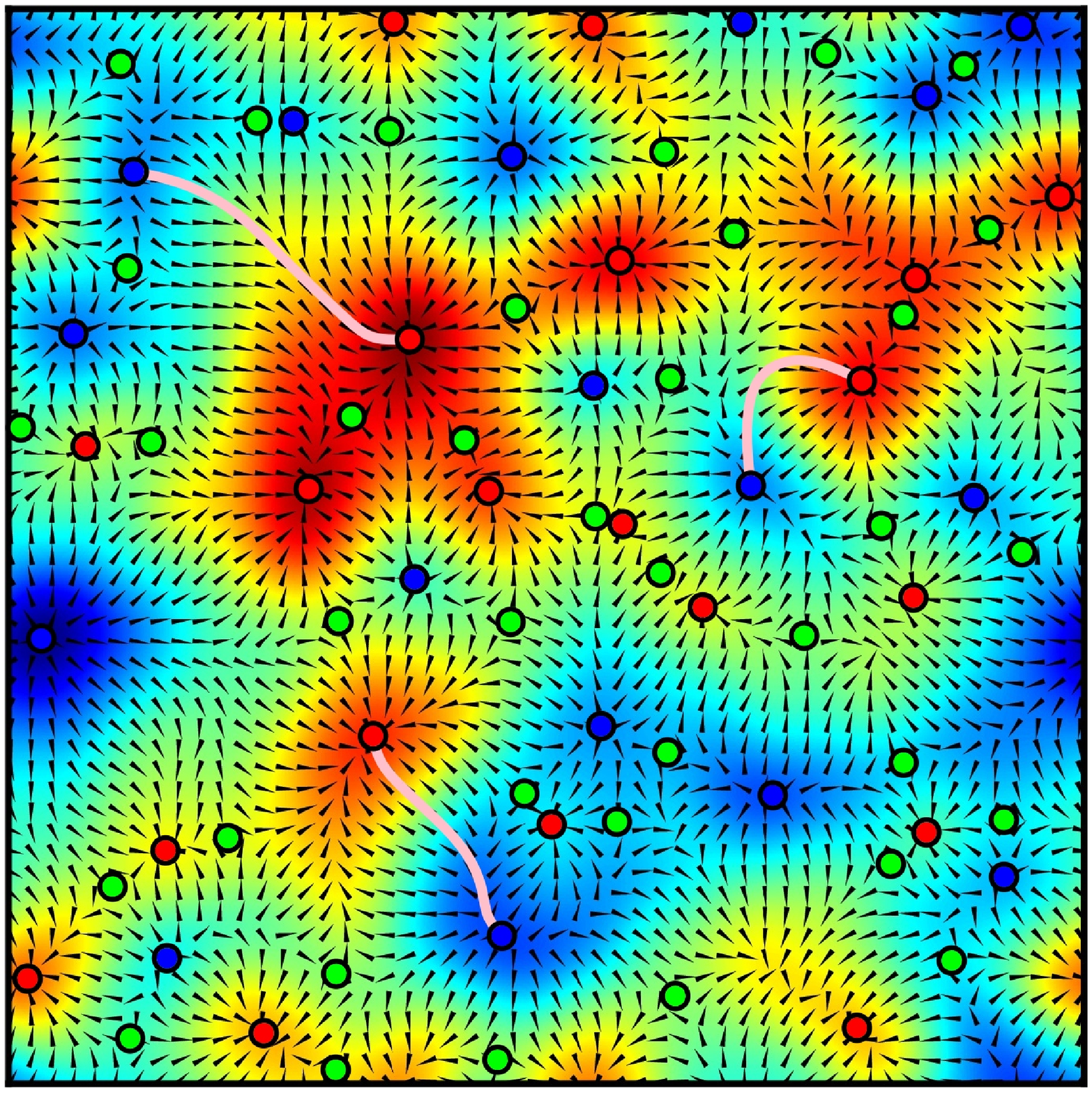}
\includegraphics[width=0.45\textwidth]{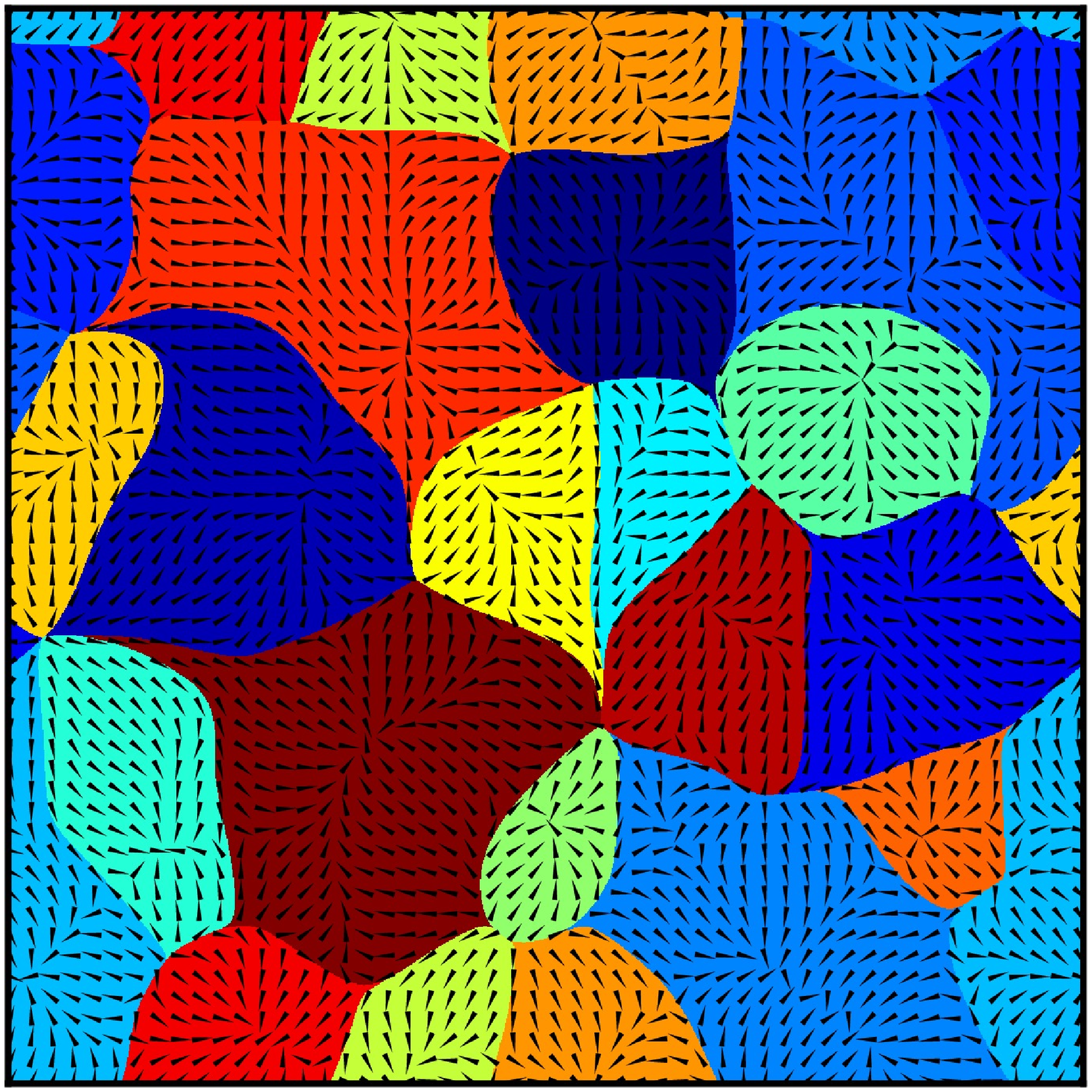}\\
\centering\includegraphics[width=0.45\textwidth]{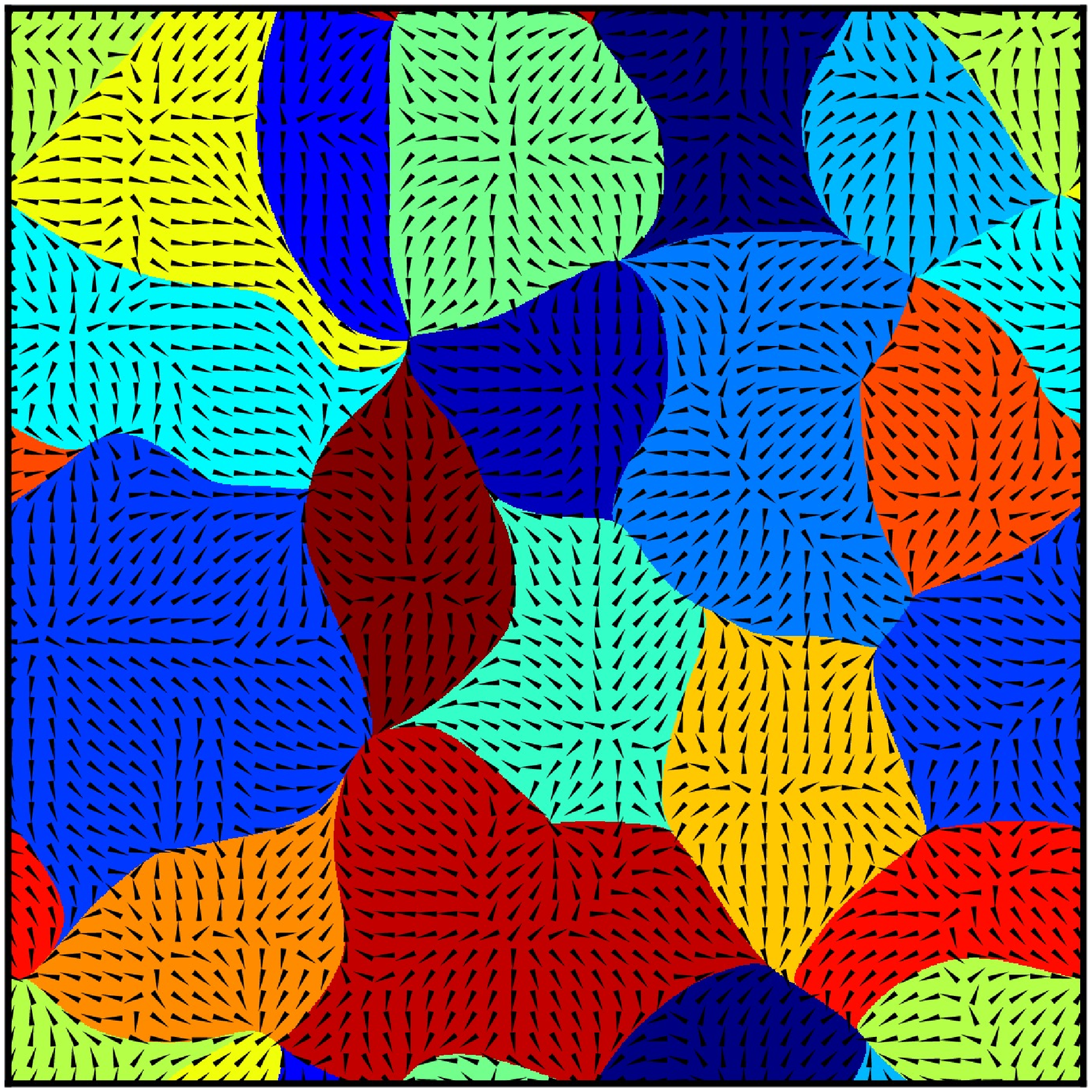}
\includegraphics[width=0.45\textwidth]{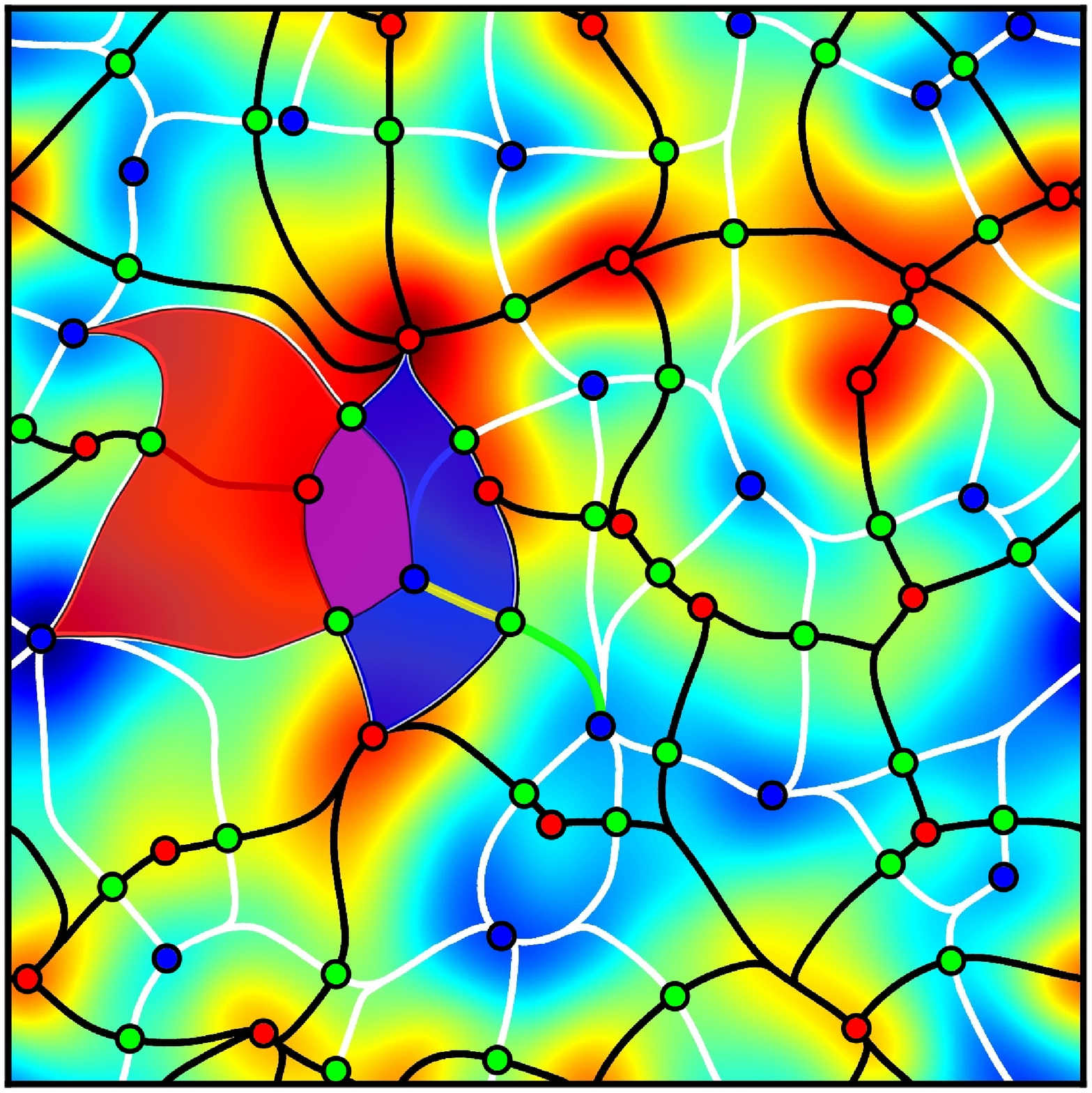}\\
\caption{A 2D density field with its gradient (top left), its descending $2$-\hyperref[defmanifold]{manifold}s (top right), its ascending $2$-\hyperref[defmanifold]{manifold}s (bottom left), and its Morse-Smale complex (bottom right, see the black and white network). The maxima/saddle points/minima are represented as  red/green/blue circled disks respectively and three integral lines are drawn in pink on the top left frame. On the central left part of the bottom right frame, an arc (\ie a $1$-cell) is represented in yellow (intersection of a green ascending $1$-\hyperref[defmanifold]{manifold} and a blue descending $2$-\hyperref[defmanifold]{manifold}) and a quad (\ie a $2$-cell) in purple (intersection of a red descending $2$-\hyperref[defmanifold]{manifold} and a blue ascending $2$-\hyperref[defmanifold]{manifold}).
 \label{fig_example_morse_2D}}
\end{figure*}

Mathematically speaking, Morse theory is concerned with smooth scalar functions (say height of a mountain, or the temperature in a room) defined over generic \hyperref[defmanifold]{manifold}s. In the present case we are mainly interested in density fields: real valued functions defined over $d$ dimensional Euclidian spaces\footnote{This is actually not generally true. Numerical simulations for instance often use periodic boundary conditions, which amounts to defining density on a torus $\torus^d\subset\reals^d$.} $\reals^d$. We will therefore restrict the present discussion to such geometries for the sake of simplicity. Morse theory provides a way to capture the intricate relation between the geometrical and topological properties of a function. What one means by geometrical property is basically any property unaffected by rigid motions such as translations or rotations. If $h$ is the altitude function of a mountain landscape for instance, the altitude of the highest peak or its total surface are geometrical properties. Topology on the other hand captures how points are connected to each other with notions such as that of neighborhood. Topological properties are invariant under smooth continuous transformations.  Sometimes topology is coined to be rubber geometry. Sticking to the landscape analogy and defining a mountain as the set of points that can be reached from its summit by going down the slope (\ie following the gradient of $h$), then the mountain itself is in some sense a topological property of the altitude function. Indeed, in winter, when covered with snow, or during summer, after the snow melted, the altitude map slightly changes, but the underlying mountain can still be easily identified as the same mountain. For the same reasons, a crest linking two mountains or a valley for instance are also topological properties of the landscape. When it comes to characterizing a function such as the matter density $\rho$ on large scales in the universe, both topological and geometrical properties are interesting. While topological properties such as the number of galaxy clusters or dark matter haloes in a given volume are robust with respect to changes in the precise measured value of $\rho$, geometrical properties such as the density profile and precise location of a halo or a filament are more specific and characterize better the properties of $\rho$.\\ 

The relation between geometry and topology is intricate, and while modifying topology certainly requires a modification of geometry, the reverse is not generally true. For instance, the shape of a mountain may only slightly change with season, but more drastic events such as the explosion of a volcano (\ie a drastic change in geometry) could actually erase it. Morse theory captures this relation for a generic function $f$ by relying on the gradient $\gradx f\left(\vx\right) = {d f}/{d\vx}\left(\vx\right)$ and its flow. The gradient defines a preferential direction at every point (the direction of steepest ascent) except where it vanishes (\ie where $\gradx f=0$). Those particular points are called critical points and can be classified according to the sign of the Hessian matrix, the $d\times d$ matrix of the second derivatives $\vH_f\left(\vx\right) = {d^2 f}/{dx_idx_j}\left(\vx\right)$: 

\begin{mydef}[critical point of order $k$]
\label{def_crit}
Let $f$ be a function defined over $\reals^d$ and $P$ a point with coordinate $\vp\in\reals^d$. Then $P$ is  a critical point of $f$ if $\gradx f\left(\vp\right)= 0$. It is said to be of order $k$ if the Hessian matrix $\vH_f\left(\vp\right)$ has exactly $k$ negative eigenvalues.\vspace{1mm} 
\end{mydef}
Intuitively, in $2$D, the top of a mountain is a maximum (order $2$), a pass is a saddle-points (order $1$) and the bottom of a valley a minimum (order $0$). The top left frame of figure \ref{fig_example_morse_2D} shows the gradient and critical points of a function defined over $\reals^2$. On this picture, the blue, green and red circles stand for the critical points of order $0$ (minima), $1$ (saddle points) and $2$ (maxima) respectively. Note that according to definition \ref{def_crit}, the order of a critical point is defined by the sign of the eigenvalues of the Hessian, which must therefore be non null. This condition is essential to Morse theory: a function $f$ which obeys Morse theory must necessarily satisfy this constraint.
 Conversely, such functions are called \hyperref[defMF]{Morse functions}:
\begin{mydef}[Morse function]
\label{def_morse_function}
A Morse function is a smooth function whose critical points are non-degenerate. This means that for any $P$ such that $\gradx f\left(\vp\right)= 0$, $det\,\vH_f\left(\vp\right) \neq 0$.\vspace{1mm} 
\end{mydef}

We will assume from now on that $f$ is a \hyperref[defMF]{Morse function}.\footnote{this is a strong requirement in practice, as shown in appendix \ref{sec_appwatershed} .}
 At the location of any non critical point, the gradient indicates a prefered direction, and one can therefore define specific lines, the integral lines, by following the gradient flow:
\begin{mydef}[Integral line or field line]
\label{def_int_line}
An integral line (also called field line) is a curve $\vL\left(t\right)\in\reals^d$ such that
\begin{equation}
\frac{d \vL\left( t \right)}{dt}=\gradx f \,.\label{eq_fieldline}
\end{equation}
Its origin and destination are defined as $\lim_{t\rightarrow -\infty} \vL\left(t\right)$ and $\lim_{t\rightarrow +\infty} \vL\left(t\right)$ respectively.\vspace{1mm} 
\end{mydef}
The pink curves on top left frame of figure \ref{fig_example_morse_2D} show examples of integral lines: the lower order critical point at their extremity is their origin and the higher one their destination. The integral lines of a \hyperref[defMF]{Morse function} actually always have critical points as origin and destination. Let us consider the case of an integral line passing through a base point $P$. One can show that such integral line obeys certain properties:
\begin{myprop}[Integral lines of a {Morse function}]
\label{prop_int_line}
The integral lines of a \hyperref[defMF]{Morse function} $f$ defined over $\reals^d$ and passing through a given point $P$ is obtained by folowing the gradient and minus the gradient from $P$. It obeys certain properties:
\begin{itemize}
\item The origin and destination of an integral line is a critical point.
\item Two integral lines passing through points $P$ and $P^\prime$ respectively may only be identical or fully distinct : two integral lines cannot intersect (they can share their origin and/or destination though).
\item The set of all the integral lines cover all of $\reals^d$ and each point $P$ of space belong to exactly one integral line. It may be the origin/destination of several integral lines if it is a critical point though. 
\item An integral line with base point a critical point $P$ is reduced to that point $P$. 
\end{itemize}
\end{myprop}
The combination of the first and second properties is particularly interesting, as it allows classifying each points of space according to the origin {\em or} destination of its (unique) integral line. Such classification defines distinct regions of space called ascending and descending manifolds:
\begin{mydef}[Ascending/Descending $n$-manifold]
\label{def_manifolds}
Let $P$ be a critical point of order $k$ of the \hyperref[defMF]{Morse function} $f$ defined over $\reals^d$. The ascending $\kmn[d]{k}$-manifold defines a region of space with dimension $\kmn[d]{k}$: the set of points reached by integral lines with origin $P$. The descending $k$-manifold defines a region of space with dimension $k$, the set of points reached by integral lines with destination $P$.\vspace{1mm} 
\end{mydef}
There exist exactly $d$ different classes of ascending and descending \hyperref[defmanifold]{manifold}s, classified according to the order of the critical point at their origin or destination. Note that an ascending or descending $d$-\hyperref[defmanifold]{manifold} of a \hyperref[defMF]{Morse function} always spans a domain of dimension $d$ (\ie a $0$-\hyperref[defmanifold]{manifold} is a (critical) point, a $1$-\hyperref[defmanifold]{manifold} a line, a $2$-\hyperref[defmanifold]{manifold} a surface, a $3$-\hyperref[defmanifold]{manifold} a volume, ...). The central frames of figure \ref{fig_example_morse_2D} show the ascending and descending $2$-\hyperref[defmanifold]{manifold}s of the 2D function on the upper frame. The notions of ascending and descending \hyperref[defmanifold]{manifold}s are actually at the core of Morse theory and the set of the descending or ascending \hyperref[defmanifold]{manifold}s is usually called the Morse complex\footnote{weather one chooses to use the ascending or the descending \hyperref[defmanifold]{manifold}s is only a matter of convention, as the descending $n$-\hyperref[defmanifold]{manifold}s of $f$ are the ascending $n$-\hyperref[defmanifold]{manifold}s of $-f$.}:
\begin{mydef}[Morse complex]
\label{def_morse_complex}
the Morse complex of a \hyperref[defMF]{Morse function} $f$ is the set of its ascending (or descending) \hyperref[defmanifold]{manifold}s.\vspace{1mm} 
\end{mydef}
The notion of Morse complex can actually be extended by following Smale and adding one more condition to a \hyperref[defMF]{Morse function}:
\begin{mydef}[Morse-Smale function]
\label{def_morseSmale_function}
A Morse-Smale function is a \hyperref[defMF]{Morse function} whose ascending and descending \hyperref[defmanifold]{manifold}s intersect only transversely,
\end{mydef}
where the word ``transverse'' can be understood as the opposite of ``tangent'', in the sense that there exist no point where two transverse \hyperref[defmanifold]{manifold}s are tangent. In other words, two ascending an descending \hyperref[defmanifold]{manifold}s should not be tangent and they should always penetrate into each other where they cross (\ie they should ``distinctly'' intersect where they do). This additional condition ensures that the intersection of the ascending and descending \hyperref[defmanifold]{manifold}s is properly defined everywhere, so that the intersection of a $p$-ascending \hyperref[defmanifold]{manifold} and a  $q$-ascending \hyperref[defmanifold]{manifold} may only have dimension $n=min\left(p,q\right)$ or be void. Such a non-null intersection is called a Morse-Smale \hyperref[defcell]{$n$-cell}:
\begin{mydef}[Morse-Smale $n$-cell]
\label{def_morse_ncell}
A Morse-Smale $n$-cell is the non void intersection of a $p$-ascending and a $q$-ascending \hyperref[defmanifold]{manifold} of a Morse-Smale function such that $n=\min\left(p,q\right)$. A $1$-cell is generally called \hyperref[defarc]{arc}, a $2$-cell is a \hyperref[defarc]{quad} and a $3$-cell a \hyperref[defarc]{crystal}.\vspace{1mm} 
\end{mydef}
A $n$-cell is a refinement of the concept of an ascending/descending \hyperref[defmanifold]{manifold}. Whereas the descending and ascending \hyperref[defmanifold]{manifold}s are defined by the sets of integral lines having common origin {\em or} common destination respectively, a $n$-cell is defined by the sets of integral lines with common origin {\em and} destination. The bottom right frame of figure \ref{fig_example_morse_2D} displays examples of $n$-cells in 2D. The purple region for instance is the quad defined by the intersection of the red descending $2$-\hyperref[defmanifold]{manifold} and the blue ascending $2$-\hyperref[defmanifold]{manifold}: all integral lines within this region have the minimum on its boundary as origin and the maximum as destination (see also the upper right and lower left frames). Similarly, the yellow curve defines an arc at the intersection of the blue ascending $2$-\hyperref[defmanifold]{manifold} and the green descending $1$-\hyperref[defmanifold]{manifold}, as only one integral line has the minimum and saddle point at its extremities as origin and destination. The set of all $n$-cells defines the  \hyperref[defMSC]{Morse-Smale complex}:
\begin{mydef}[Morse-Smale complex]
\label{def_mscomplex}
The Morse-Smale complex of a Morse-Smale function $f$ is the set of all the $n$-cells of $f$.\vspace{1mm} 
\end{mydef}
On the same picture, the \hyperref[defMSC]{Morse-Smale complex} is described by the critical points and the black and white curves. Basically, the critical points are its $0$-cells, the set of black or white curves linking two critical points are its arcs ($1$-cells) and the regions bounded by a black and a white border are its quads ($2$-cells). In the 3D case we will consider in the next sections, the Morse Smale complex is also composed of $3$-cells (the so called cristals). Note that the notion of $n$-cell is very interesting as it defines a natural partition of space induced by the flow of the gradient, literally dividing it into a so-called cell complex (a generalization of the concept of simplicial complex presented in section \ref{sec_DMtheory}). We do not give further details here though as only the concept of arc is really needed for our purpose, the arcs really defining how critical points are connected to each other by integral lines. Actually, and although this is not formally correct, the reader may find it simpler to only consider the nodes (critical points) and arcs of the \hyperref[defMSC]{Morse-Smale complex} complex, each arc connecting the critical points at their extremities, two critical points being potentially connected only if their order differ by $1$ (\ie a minimum and a $1$-saddle, a $1$-saddle and a $2$-saddle, or a $2$-saddle and a maximum). For instance, the arcs connecting maxima to saddle points are sub-sets of the ascending $1$-\hyperref[defmanifold]{manifold}s and they enclose the information on how each filament (represented by its saddle point) connects exactly two maxima. Note that the geometry of an arc is determined by the integral lines whose origin and destination are the two critical points the arc connects. The \hyperref[defMSC]{Morse-Smale complex} obey the following ``combinatorial"\footnote{in this context, the term {\em combinatorial} is used to signify the discrete properties of the network formed by the Morse-complex: its number of nodes, their types, the number of branches and cycles, see below.} properties:
\begin{myprop}[Morse-Smale complex arcs]
\label{prop_mscomplex}the arcs (\ie $1$-cells) in the \hyperref[defMSC]{Morse-Smale complex} connect critical points in such a way that:
\begin{itemize}
\item two arcs may only intersect at a critical point,
\item an arc in the \hyperref[defMSC]{Morse-Smale complex} links two critical points with index difference $1$,
\item there are exactly two descending arcs reaching a given critical point of order $1$ (each departing from not necessarily distinct minima),
\item there are exactly two ascending arcs departing from a given critical point of order $d-1$ (each reaching not necessarily distinct maxima).\\
\end{itemize}
\end{myprop}
Figure \ref{fig_example_morse_simu} illustrates how the theoretical concepts of Morse theory apply to cosmology. On this figure, the dark matter density distribution in a cosmological simulation is displayed on the top frame, together with its $3$, $2$ and $1$ ascending \hyperref[defmanifold]{manifold}s on the second, third and fourth frame from top respectively. The ascending $3$-\hyperref[defmanifold]{manifold}s associated to minima clearly trace the under dense regions usually denominated voids. The type $1$ critical points trace the geometry of the walls through their ascending $2$-\hyperref[defmanifold]{manifold}s and the filaments are traced by the ascending $1$ \hyperref[defmanifold]{manifold}s, associated to critical points of type $1$. As stated as the beginning of this section, Morse complex actually establishes the link between the geometrical (where are the critical points? what path does each arc follows?) and topological (how are critical-points connected? how many of each type are there?) properties of the cosmic web. Supposing the large scale matter density distribution $\rho$ were a \hyperref[defMF]{Morse function}, each critical point of $\rho$ could in fact be associated to a topological feature of the cosmic web whose geometry would then described by an ascending or descending \hyperref[defmanifold]{manifold}, the arcs defining a hierarchical neighborhood relation between them (the so called combinatorial property). The purpose of this paper is to construct, from the particles, a discrete Morse function which closely resemble\footnote{Conversely, this construction would bias the reconstructed Morse-Smale complex if the underlying density was far from being a Morse function.} the sampled density (which in fact match it at the vertex of the tessellation) and which will therefore warrant all the corresponding discrete topological features.

\section{Discrete Morse theory}
\label{sec_DMtheory}
Even though the idea of applying Morse theory directly to the analysis of the cosmic web is quite appealing a priori, the task is actually not straightforward in practice. Indeed, Morse theory is defined for a \hyperref[defMF]{Morse function}, which is basically a smooth and at least twice differentiable real valued function satisfying the Morse criterion (definition \ref{def_morse_function}). Whether it is because they result from fundamentally discrete processes, as in the case of galaxy distribution, or obtained through sampling, as for numerical simulation or observational data, typical astrophysical data sets  typically do not comply to those criteria in general.
In contrast, Discrete Morse theory, first introduced by \citet{forman98,forman}, is a very powerful theory which manages to capture the essence of  the smooth Morse theory while still being readily applicable to discrete or sampled data commonly available to scientists. It is basically a combinatorial  adaptation of Morse theory that applies to intrinsically discrete functions defined over simplicial complexes\footnote{actually, discrete Morse theory applies to the broader class of topological spaces called CW-complexes, which also include functions sampled over a regular cubic grid for instance.}.\\

Let us start by defining the basic building block of such spaces, the simplex. A \hyperref[defksimplex]{$k$-simplex} is the simplest possible geometrical figure of dimension $k$, or simply speaking the $k$ dimensional analog of a triangle: A $0$-simplex for instance is a point, a $1$-simplex a segment, a $2$-simplex a triangle, a $3$-simplex a tetrahedron, ... More formally:
\begin{mydef}[$k$-simplex]
\label{def_simplex}
A $k$-simplex $\sigma_k$ is the convex hull of $k+1$ affinely independent points $S=\lbrace p_0, ..., p_k \rbrace$. In other words, it is the set of points within the smallest possible solid with summits the $k+1$ points in $S$. It may be noted $\sigma_k=\lbrace p_0, ..., p_k \rbrace$.\vspace{1mm} 
\end{mydef}
A simplex may have  \hyperref[defface]{faces} and  \hyperref[defcoface]{cofaces}:
\begin{mydef}[face/coface of a $k$-simplex]
\label{def_face}
A face of a $k$-simplex $\sigma_k$ with vertice $S=\lbrace p_0, ..., p_k \rbrace$ is any $l$-simplex $\gamma_l$ with $l\leq k$, such that its vertice $P=\lbrace p_0, ..., p_l \rbrace\subset S$. If $\gamma_l$ is a face of $\sigma_k$, then $\sigma_k$ is a coface of $\gamma_l$. In general, when $k$ and $l$ only differ by $1$, a face is called a facet and a coface is called a cofacet.\vspace{1mm} 
\end{mydef}
Simply speaking, considering a tetrahedron in 3D (\ie a $3$-simplex) with $4$ vertices as summits, its $2$-faces are four triangles (\ie its facets, any possible combination of three vertice), its $1$-faces are $6$ segments (\ie any possible combination of two vertice) and its $0$-faces are fours points (\ie any possible combination of one vertice). Reciprocally, the tetrahedron is a \hyperref[defcoface]{coface} of any of those triangle, segments or points, and in particular it is a cofacet of any of the triangles. In general, a $k$-simplex has $C_{l+1}^{k+1}$ faces of dimension $l$. Finally, a simplicial complex is a set of $k$-simplexes that comply to specific criteria:
\begin{mydef}[simplicial complex]
\label{def_simp_complex}
A simplicial complex $K$ is a finite union of simplexes such that
\begin{itemize}
\item Any face of a simplex in $K$ also belongs to $K$.
\item The intersection of two simplexes in $K$ is empty or a simplex of dimension lower or equal, to the highest dimensional simplex they share.
\end{itemize}
\end{mydef}

\begin{figure}
\includegraphics[width=\linewidth]{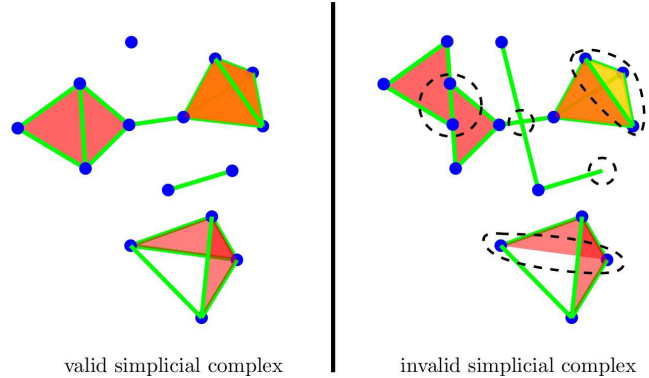}
\caption{Illustartion of two sets of 3D simplexes, $K$ and $K^\prime$, forming a valid (left) and an invalid (right) simplicial complex. It is invalid because, from left to right and top to bottom, the interesection of the two $2$-simplexes is not a simplex in $K^\prime$, two $1$-simplexes intersect, a $3$-simplex (light yellow mostly hidden tetrahedron),  a $1$-simplex and a $2$-simplex each lack one of their facets.\label{fig_simp_complex}} 
\end{figure}

Figure \ref{fig_simp_complex} shows an example of a combination of simplexes that form a simplicial complex (left frame) and a different combination that do not (right frame). A common example of a simplicial complex in astrophysics is the delaunay tessellation \citep[see \eg][]{okabe00,DTFE} of a set of discretely sampled points.\\

As stated previously, discrete Morse theory directly applies to functions defined over a simplicial complexes. Those particular functions are called discrete functions, and for discrete Morse theory to apply, they also need to comply certain criteria:
\begin{mydef}[Discrete Morse function]
\label{def_DMfunction}
A discrete function $f$ defined over a simplicial complex $K$ associates a real value $f\left(\sigma_k\right)$ to each simplex $\sigma_k\in K$. The discrete function $f$ is a \hyperref[defDMF]{discrete Morse function} if and only if, for each $\sigma_k\in K$, 
\begin{enumerate}
\item there exist {\em at most} one facet $\alpha_{k-1}$ of $\sigma_k$ such that $f\left(\sigma_k\right)\leq f\left(\alpha_{k-1}\right)$,\\
\item there exist {\em at most} one cofacet $\beta_{k+1}$ of $\sigma_k$ such that $f\left(\sigma_k\right)\geq f\left(\beta_{k+1}\right)$.
\end{enumerate}
\end{mydef}
In other words, the Hessian non-degeneracy condition of smooth Morse theory (definition \ref{def_morse_function}), becomes a condition on the value of the functions in the discrete theory: locally, a simplex has a higher value than its facets and a lower value than its cofacets, and there can only be one exception at most in each case. The reason for such a condition is not obvious at first sight but it is actually essential to the existence of a discrete gradient, the counterpart of the gradient in the smooth theory. In fact, if condition (i) and (ii) of the definition \ref{def_morse_function} of a \hyperref[defDMF]{discrete Morse function} are satisfied then, locally, the discrete gradient of $f$ (see below) can only define at most one preferential direction, as does the gradient of the corresponding smooth theory. Following this line of thought, the analog of a critical point of order $k$ (see definition \ref{def_crit}), a critical $k$-simplex of $f$, is a simplex for which $f$ does {\sl not} have any preferential relationship with one of its direct neighborhood (\ie its facets and cofacets):
\begin{mydef}[Critical $k$-simplex]
\label{def_critical_simplex}
A \hyperref[defksimplex]{$k$-simplex} $\sigma_k$ is critical for the \hyperref[defDMF]{discrete Morse function} $f$ if
\begin{enumerate}
\item there exist {\em no} facet $\alpha_{k-1}$ of $\sigma_k$ such that $f\left(\sigma_k\right)\leq f\left(\alpha_{k-1}\right)$,\\
\item there exist {\em no} cofacet $\beta_{k+1}$ of $\sigma_k$ such that $f\left(\sigma_k\right)\geq f\left(\beta_{k+1}\right)$.
\end{enumerate}
It is important here to realize that the equivalent in discrete Morse theory of a critical point of order $k$ is a critical \hyperref[defksimplex]{$k$-simplex} : in $2$D, a minimum is a critical vertex ($0$-simplex), a saddle-point is a critical segment ($1$-simplex) and a maximum is a critical triangle ($2$-simplex). 
\end{mydef}
Moreover, one can show that if definition \ref{def_DMfunction} is satisfied, then at least one of the two conditions of definition \ref{def_critical_simplex} is verified, which leaves only two possible configurations for a simplex $\sigma_k$: exactly one of its \hyperref[defcofacet]{cofacets} and all its facets have a lower value or exactly one of its faces and all its \hyperref[defcofacet]{cofacets} have a higher value. In both cases, a preferential relation is established between $\sigma_k$ and one of its facets or cofacets, which also defines a preferential direction, and leads to the following definition:
\begin{mydef}[Discrete gradient vector field]
\label{def_DGradient}
A discrete gradient vector field can be defined for a \hyperref[defDMF]{discrete Morse function} $f$ over $K$ by coupling simplexes in gradient arrows (also called gradient pairs):
\begin{itemize}
\item if a simplex $\sigma_k$ has exactly one lower valued \hyperref[defcofacet]{cofacet} $\alpha_{k+1}$, then $\left[\sigma_k,\alpha_{k+1}\right]$ form a gradient arrow, 
\item if a simplex $\sigma_k$ has exactly one higher valued facet $\beta_{k-1}$, then $\left[\sigma_k,\beta_{k-1}\right]$ form a gradient arrow, 
\item if a simplex $\sigma_k$ satisfies definition \ref{def_critical_simplex}, it is critical, and does not belong to a gradient arrow.
\end{itemize} 
Note that other configurations are impossible precisely because $f$ is a \hyperref[defDMF]{discrete Morse function}. Also, within a gradient arrow, the lowest valued simplex is the tail and the highest valued one the head (\ie the discrete gradient actually points in the opposite direction of its smooth counterpart).\vspace{1mm} 
\end{mydef}

\begin{figure*}
\begin{minipage}[c][\textheight]{\linewidth}
\begin{centering}
\subfigure[A discrete Morse function over a 2D simplicial complex]{\centering \includegraphics[width=0.49\linewidth]{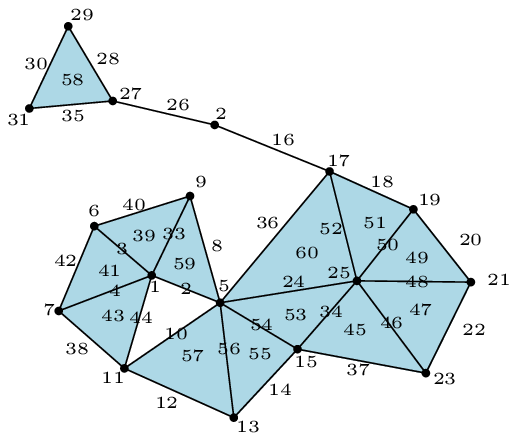}\label{fig_discrete_grad_introA}}
\subfigure[The corresponding discrete gradient and critical simplexes (red triangles, green segments, and blue vertices)]{\centering \includegraphics[width=0.49\linewidth]{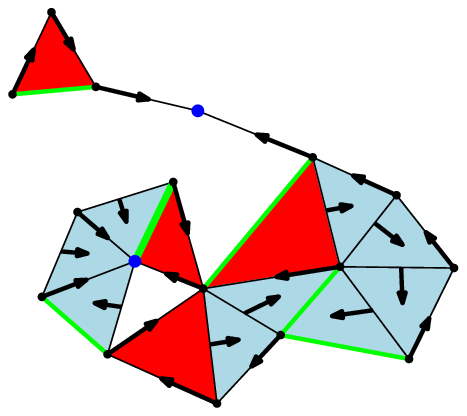}\label{fig_discrete_grad_introB}}\\
\subfigure[Example of two V-pathes (purple shade)]{\centering \includegraphics[width=0.49\linewidth]{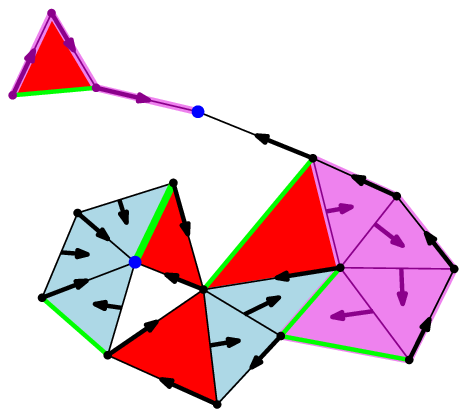}\label{fig_discrete_grad_introC}}
\subfigure[Example of an ascending $2$-manifold (union of the blue vertices, green segments and red triangles)]{\centering \includegraphics[width=0.49\linewidth]{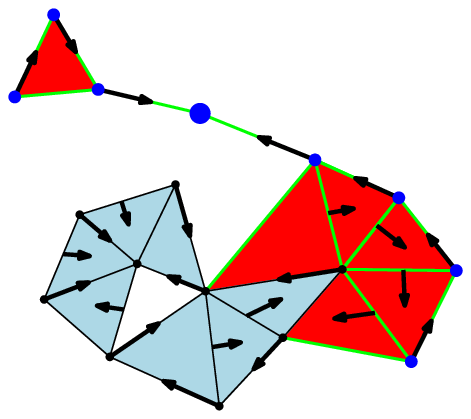}\label{fig_discrete_grad_introD}}
\end{centering}
\caption{\label{fig_discrete_grad_intro} Illustration of the notions introduced by discrete Morse-theory. On the upper left panel (figure \ref{fig_discrete_grad_introA}), the numbers associated to each \hyperref[defksimplex]{$k$-simplex} (\ie vertexes, segments and triangles) of the underlying simplicial complex define a discrete Morse-function. Note that a discrete Morse function must comply to definition \ref{def_DMfunction}, which is relatively restrictive, and in the present case, the function has been designed to illustrate notions of discrete Morse theory on a relatively small complex. We show in section \ref{sec_disc_grad} how a \hyperref[defDMF]{discrete Morse function} can be defined to mimic the properties of a smooth function (such as the density or an altitude field for instance). The corresponding \hyperref[defDG]{discrete gradient} (see definition \ref{def_DGradient}) is represented by the arrows on the upper right frame (figure \ref{fig_discrete_grad_introB}), each arrow associating a $k-1$-simplex (the tail) to a \hyperref[defksimplex]{$k$-simplex} (the head). On the same frame (see also figure \ref{fig_discrete_grad_introC}), the red, green and blue shaded simplexes are the critical $2$, $1$ and $0$-simplexes of the discrete function respectively (\ie the equivalent of the maxima, saddle-points and minima of smooth theory). On the lower left frame (figure \ref{fig_discrete_grad_introC}), the two purple shaded sets of simplexes correspond to two \hyperref[defVP]{V-pathes} of the \hyperref[defDMF]{discrete Morse function} (the discrete analog of an integral line, see definition \ref{def_vpath}). Intuitively, a \hyperref[defVP]{V-path} is a set of simplexes linked by \hyperref[defDG]{discrete gradient} arrows, similarly to the integral lines of the smooth theory. Finally, the extended ascending \hyperref[defmanifold]{manifold} (see definition \ref{def_extdmanifolds}) of the minimum with value $2$ (the large blue disk) is shown on the lower right frame (figure \ref{fig_discrete_grad_introD}). Similarly to the smooth theory, the corresponding ascending $0$-\hyperref[defmanifold]{manifold} (definition \ref{def_dmanifolds}) is defined by the set of simplexes that one can reach by following the gradient arrows from the minimum (\ie all the blue vertexes and green segments that belong to a gradient pair - \ie an arrow - ). For the sake of consistency, one needs to define discrete extended \hyperref[defmanifold]{manifold}s (definition \ref{def_extdmanifolds}), which also include recursively the \hyperref[defcoface]{cofaces} of any simplex in the discrete \hyperref[defmanifold]{manifold}, as well as the ascending \hyperref[defmanifold]{manifold}s of those \hyperref[defcoface]{cofaces} that are critical. The resulting discrete extended ascending $0$-\hyperref[defmanifold]{manifold}s is the set of blue vertices, green segments and red triangles on the bottom right frame.} 
\end{minipage}
\end{figure*}

Figure \ref{fig_discrete_grad_intro} shows a \hyperref[defDMF]{discrete Morse function} defined over a 2D simplicial complex (upper left frame), and its corresponding \hyperref[defDG]{discrete gradient} vector field and critical simplexes (upper right frame). One can note the similarity in the relation between the \hyperref[defDG]{discrete gradient} flow and the critical simplexes and that between the gradient and critical points on top frame of figure \ref{fig_example_morse_2D}. Finally, one last important definition is that of the discrete integral line. In the terminology of \cite{formanvfield}, it is called a \hyperref[defVP]{V-path}:
\begin{mydef}[V-path]
\label{def_vpath} A V-path is a strictly decreasing alternating sequence of \hyperref[defksimplex]{$k$-simplexes} $\alpha^i_k$ and {\kp1}-simplexes $\beta^j_{k+1}$
\begin{displaymath}
\alpha^0_k,\beta^0_{k+1},\alpha^1_k,\beta^1_{k+1},... ,\alpha^n_k,\beta^n_{k+1},
\end{displaymath}  
where each pair $\lbrace\alpha^i_k,\beta^i_{k+1}\rbrace$ forms a gradient pair and $\alpha^{i+1}_k$ is a facet of $\beta^{i}_{k+1}$.
\end{mydef}
Tracing a \hyperref[defVP]{V-path} basically consists in intuitively following the direction of the gradient pairs, as one can see on the lower left frame of figure \ref{fig_discrete_grad_intro} where two \hyperref[defVP]{V-pathes} are highlighted in purple.\\

Using the previously introduced concepts, it becomes relatively straightforward to define a discrete \hyperref[defMSC]{Morse-Smale complex}, and contrary to the smooth case, no \hyperref[defmanifold]{manifold} transversality condition (definition \ref{def_morseSmale_function}) needs to be enforced, as this is naturally achieved by the tesselation itself. In fact, following definition \ref{def_manifolds}:
\begin{mydef}[Discrete A./D. $n$-manifold]
\label{def_dmanifolds}
Let $\sigma_k$ be a critical simplex of order $k$ of the \hyperref[defDMF]{discrete Morse function} $f$ defined over a simplicial complex $K$. The discrete ascending $\kmn[d]{k}$-\hyperref[defmanifold]{manifold} is the set of \hyperref[defksimplex]{$k$-simplexes} that belong to at least one $V$-path with origin $\sigma_k$. The discrete descending $k$-\hyperref[defmanifold]{manifold} is the set of \hyperref[defksimplex]{$k$-simplexes} reached by field lines with destination $\sigma_k$.\vspace{1mm} 
\end{mydef}
Note that according to that definition, a discrete $k$-\hyperref[defmanifold]{manifold} only contains \hyperref[defksimplex]{$k$-simplexes} (those in the \hyperref[defVP]{V-pathes} of $\sigma_k$). This makes it difficult to define discrete $n$-cells (see definition \ref{def_morse_ncell}) by intersecting \hyperref[defmanifold]{manifold}s, as they are made of simplexes with different dimensions. Following \citet{phd}, this definition is therefore extended to:  
\begin{mydef}[Extended Discrete A./D. $n$-manifold]
\label{def_extdmanifolds}
An extended discrete ascending (resp. descending) $n$-\hyperref[defmanifold]{manifold}s is a discrete ascending (resp. descending) $n$-\hyperref[defmanifold]{manifold}, together with its \hyperref[defcoface]{cofaces} (resp. faces) and their extended discrete ascending (resp. descending) $n$-\hyperref[defmanifold]{manifold}s.
\vspace{1mm} 
\end{mydef}
This literally fills lower dimensional ``holes'' in the \hyperref[defmanifold]{manifold}, making the intersection of two extended \hyperref[defmanifold]{manifold}s a very simple operation. On the lower right frame of figure \ref{fig_discrete_grad_intro} for instance, the discrete ascending $2$-\hyperref[defmanifold]{manifold} is represented by the blue dots only. Their \hyperref[defcoface]{cofaces}, the green segments, are included in the extended \hyperref[defmanifold]{manifold}, as well as their extended ascending \hyperref[defmanifold]{manifold}s (red triangles). The definition of the discrete Morse complex is therefore similar to the one in the smooth case:
\begin{mydef}[Discrete morse complex]
\label{def_Dmorse_complex}
the discrete Morse complex of a \hyperref[defMF]{Morse function} $f$ is the set of its extended ascending (or descending) \hyperref[defmanifold]{manifold}s.\vspace{1mm} 
\end{mydef}

Similarly, a discrete $n$-cell is the intersection of two extended ascending and descending discrete \hyperref[defmanifold]{manifold}s (definition \ref{def_morse_ncell}), and the discrete \hyperref[defMSC]{Morse-Smale complex} remains the set of the discrete $n$-cells (definition \ref{def_mscomplex}). As in the smooth case, the discrete \hyperref[defMSC]{Morse-Smale complex} is really a combinatorial object as it describes a particular way of grouping critical simplexes in pairs, quads, cristals ..., associating to each of those combination the geometry spanned by intersections of ascending and descending \hyperref[defmanifold]{manifold}s. We conclude by noting that, neglecting the effect of boundary conditions, the arcs of the discrete \hyperref[defMSC]{Morse-Smale complex} (\ie the \hyperref[defVP]{V-pathes} linking critical simplexes) obey the same properties as those of the \hyperref[defMSC]{Morse-Smale complex} (definition \ref{prop_mscomplex}).

\section{Topological persistence }
\label{sec_persistence}
The concept of \hyperref[defpers]{persistence} was first formalized in \citet{edel} (see also \citet{robins}). It is basically a method to quantify the importance of the topological features of a space, and was initially developed as a way to robustly measure topological properties when noise is present, and to enable topological simplification (\ie the modification of a function or a space so that its less significant {\it topological} features are removed). The theory was originally described in the context of simplicial homology (for functions defined over a simplicial complex, see appendix \ref{sec_simp_homo}) and was very nicely exposed in \citet{edel}. Let us stress here that the concept of persistence itself is largely independent of the fact that a function is smooth or not, as it only quantifies the robustness of its topological properties given one can measure them, whatever the nature of the function itself. Let us illustrate here the idea behind the concept of \hyperref[defpers]{persistence} using simple examples.\\

 For  smooth functions, \hyperref[defpers]{persistence} theory is based on the evolving properties of the so-called sub-level sets of a function $\rho$, as they change with the value of the level $\rho_0$. A sub-level set is the set of points where $\rho\left({\mathbf x}=\left(x_1,..x_n\right)\right)$ is higher than or equal to a certain value $\rho_0$:
\begin{mydef}[Level set and Sub-level set]
\label{def_sublevel}
A level set (also called isocontour) of a function $\rho\left({\mathbf x}\right)$ of $n$ variables $x_i$ at level $\rho_0$ is defined as 
\begin{displaymath}
\left(x_1,..x_n\right)|\rho\left(x_1,..x_n\right) =\rho_0.
\end{displaymath}
A sub-level set (also called excursion set) is defined as 
\begin{displaymath}
\left(x_1,..x_n\right)|\rho\left(x_1,..x_n\right)\geq \rho_0.
\end{displaymath}
\end{mydef}
Using this definition, \hyperref[defpers]{persistence} can be interpreted as a measure of the ``life-time'' of topological features, the so called $k$-cycles, in the sub-level sets. When the value of $\rho_0$ skims through the image of $\rho$ (\ie the set of values $\rho\left({\mathbf x}\right)$ may take), the corresponding sub-level set grows, and the way it is connected evolves. In 3D, isolated islands (also called components or $0$-cycles) first appear around the maxima. Those islands later merge into each other at saddle points of type $1$ to finally form rings bordering holes (the $1$-cycles). For lower values of $\rho_0$, those holes get filled at saddle points of type $2$, destroying the corresponding $1$-cycles, to later form spherical shells around minima (the $2$-cycles), when a sufficient number of holes have been filled and those spherical shell also end up being filled at minima, therefore destroying the corresponding $2$-cycles. \hyperref[defpers]{Persistence} then relates the importance of a given $k$-cycle to the length of the interval of values $\rho_0$ can take and for which a given $k$-cycle exists within the growing sub-level sets.\\ 

\begin{figure*}
\centering
\includegraphics[width=\linewidth]{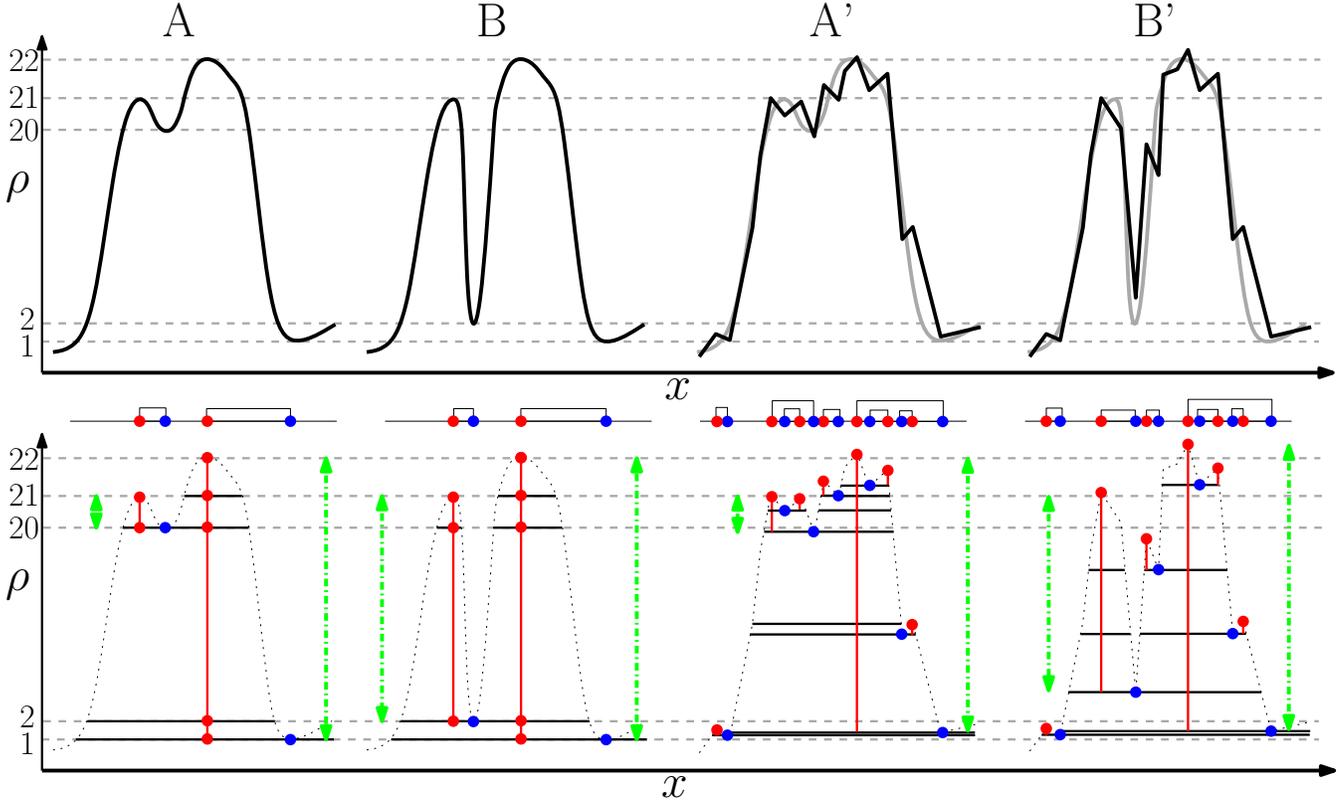}
\caption{\label{fig_per1D} Illustration of the concept of persistence over a 1D functions. The upper panel shows two functions (left) and their discretely sampled counterparts, with noise added (right). The lower panel displays the evolution of sublevel-sets of these functions at the level of different critical points as one spans densities from high to low. The green dash-dotted vertical arrows emphasize the lifetime of components in the sublevel-sets, the persistence pairs are displayed in the central part over the function's Morse-Smale complex.}
\end{figure*}

Figure \ref{fig_per1D} illustrates how \hyperref[defpers]{persistence} works in 1D. On this figure, the upper part displays four different functions, where the two on the right (labeled $A^\prime$ and $B^\prime$) were obtained by discretely sampling the two on the left (labeled $A$ and $B$), adding random noise, and linearly interpolating between the sample points. The lower part of the figure shows the different sub-level sets of these functions for values corresponding to their critical points. On the bottom left frame  for instance, the sub-level sets of $\rho\left(x\right)$ are empty for levels $\rho_0>22$. At level $\rho_0=22$ though, a new component (\ie a $0$-cycle) appears, which corresponds to the highest maximum of the function. This component grows for levels $22>\rho_0>21$ and a new independent components appears at the level of the second highest maximum, $\rho_0=21$. Those two components remain independent while $\rho_0>20$ but merge when reaching $\rho_0=20$, the value of the first minimum. Basically, the minimum {\em destroyed} a component that was {\em created} by a maximum. By convention, we say that it destroys the most recently created one (i.e the maximum with lowest density), and that the minimum and left maximum therefore form a \hyperref[defperspair]{persistence pair} (as illustrated on the central sketch) with \hyperref[defpers]{persistence} $21-20=1$. The four sketches on the bottom part illustrate this pairing process for the four different function. One should note that a given critical point may not always be paired in the process, and that because the 1D case is very simple, a given type of critical point always create or always destroy, but this is not the case in general, for critical points that are not extrema.\\

A very common task when studying galaxy distributions or cosmological N-body simulations involves identifying galaxy clusters or dark matter haloes. This is often achieved using relatively simple but robust methods, such as the friend-of-friend algorithm \citep{FOF}, that basically involve carefully selecting a global level $\rho_c$ and considering each independent component in the sub-level set $\rho_c$ of the density field $\rho\left({\mathbf x}\right)$ as an independent cosmological structure. Applied to the functions $A$ of figure \ref{fig_per1D} for example, such a method may detect one or two different peaks with $\rho_c=19$ or $\rho_c=20.5$ respectively, but it will not yield any information on weather those peaks are comparable or if one of them is more meaningful than the other. \hyperref[defpers]{Persistence} on the other hand can make such distinction, because it is built using information present in {\em all} the sub-level sets: while function B contains two comparably persistent peaks, function $A$ really only contains one (the peaks \hyperref[defpers]{persistence} is symbolized by the length of the green arrows on the figure). The remarkable fact is that this stays true even if the sampling is poor and noise is present, as illustrated by function $A^\prime$ and $B^\prime$. Because of noise, many spurious peaks exist in this two functions, which may potentially lead to numerous fake identifications, but even in that case, whereas a density selection method would clearly fail to count the peaks correctly, \hyperref[defpers]{persistence} easily identifies the presence of only one persistent peaks in $A^\prime$, and two in $B^\prime$, as in the case of functions $A$ and $B$.\\

Although a very simple 1D case was illustrated here, the general idea remains the same in higher dimensional spaces.
 In general, one studies how components of sub-level sets are created or destroyed, but in higher dimensions, one also has to keep track of more complex structures than independent components, such as the formation of 2D holes or 3D shells in the structure (\ie the $1$-cycles and $2$-cycles).\\ 

As was mentioned earlier, \hyperref[defpers]{persistence} can equally be computed directly for discrete functions defined over simplicial complexes, given that one can define a concept similar to that of growing sub-level sets in such context. \hyperref[deffiltration]{Filtration} is such a concept. A discrete function $v$ associates a value to each simplex in a complex, and one can for instance define a \hyperref[deffiltration]{filtration} $F$ of a simplicial complex $K$ as the sets of sub simplicial complexes $K_i$ such that only the simplexes $\sigma$ with value $v(\sigma)<v_i$ belong to each $K_i$. More generally, 
\begin{mydef}[Filtration]
\label{def_filtration}
A filtration of a finite simplicial complex $K$ is a sequence of $N+1$ sub-complexes $K^i$ of $K$ such that:
\begin{displaymath}
\begin{tabular}{rl}
$\left(1\right)$&$\;\emptyset = K^0 \subseteq K^1 \subseteq ... \subseteq K^{N-1}\subseteq K^{N}=K,$\\
$\left(2\right)$&$\; K^{i+1}=K^ {i} \cup \delta^i,$
\end{tabular}
\end{displaymath}
where $\delta^i$ is a subset of the simplexes in $K$, and $A\subseteq B$ means that $A$ is included in or equal to $B$. One can in particular define a special filtration induced by a function $v$ that associates a value to each $\sigma\in K$, such that each $K_i$ is the set of simplexes in $K$ with value $v\left(\sigma \right)$ less or equal to a given threshold $v_i$. \vspace{1mm} 
\end{mydef}
In that case, each $K_i$ is the discrete equivalent of the growing sub-level sets of definition \ref{def_sublevel} (see also figure~\ref{fig_persistence} for an example of a filtration) for the discrete function $v$ that defines the order of entrance of simplexes within the filtration. As the \hyperref[deffiltration]{filtration} grows with increasing value of $v_i$, new components, loops, shells, ... appear. As for the  smooth function counterpart, those topological features are generally called \hyperref[defkcycle]{$k$-cycles}, and we define them formally in the context of discrete theory (this definition would conceptually be very close in the context of smooth functions though):
\begin{mydef}[$k$-cycle]
\label{def_kcycle}
a $k$-cycle in a simplicial complex $K$ is a $k$ dimensional topological feature with $0\leq k < D$, where D is the number of dimensions. When $D=3$ for instance, a $0$-cycle is an independent component (\ie a set of simplexes non-linked to the rest of the complex), a $1$-cycle is a loop (a set of simplexes that form a ring with a hole in the middle) and a $2$-cycle is a shell (a set of simplexes bounding a 3D empty region).
\end{mydef}
As for a smooth function, one can therefore track the creation and destruction of $k$-cycles in $F$ as simplexes enter the \hyperref[deffiltration]{filtration}, pairing critical simplexes into persistence pairs:
\begin{mydef}[Persistence]
\label{def_persistence}
persistence measures the ``life-time'' of topological features (\ie $k$-cycles) in a \hyperref[deffiltration]{filtration} of a finite simplicial complex $K$ induced by a discrete function $v$ or equivalently in the growing sub-level sets of a smooth function $\rho$. The arrival of each critical simplex in the discrete case or critical points in the smooth case corresponds to the creation or destruction of a topological feature ($k$-cycle). \hyperref[defperspair]{Persistence} pairs critical simplexes $\sigma_a$|$\sigma_b$ (or critical points $P_a$|$P_b$) that create and destroy a given feature, their corresponding \hyperref[defpers]{persistence} being defined by the difference of their ``arrival time", $v\left(\sigma_a\right)-v\left(\sigma_b\right)$ (or $\rho\left(P_a\right)-\rho\left(P_b\right)$). It can also sometimes be useful to define a persistence ratio as the ratio of those values.  
\end{mydef}

The computation of \hyperref[defperspair]{persistence pairs} in a 2D \hyperref[deffiltration]{filtration} is illustrated in appendix \ref{sec_app_persistence} and intuitively, persistence describes how much a function would need to change to remove a topological feature.
\\
\begin{figure}
\centering
\includegraphics[width=\linewidth]{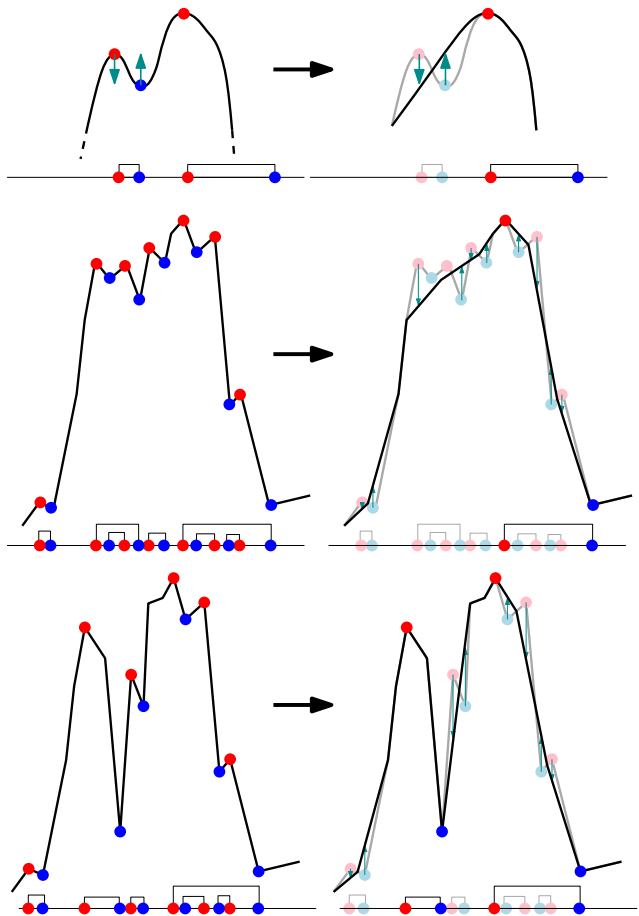}
\caption{Illustration of the topological simplification process applied to functions $A$, $A^\prime$ and $B^\prime$ defined on figure \ref{fig_per1D} (see top, central and bottom panel). The diagram under each function represents its Morse-Smale complex and \hyperref[defperspair]{persistence pairs}. \label{fig_per1Dcancel} }
\end{figure}

The main interest of being able to identify \hyperref[defperspair]{persistence pairs} of critical points (or simplexes) in a given function is that it yields an objective topological criterion to assess the significance of those critical points (or simplexes). Actually, one can go even further and show that it is actually always possible to {\em locally} modify the function to cancel non persistent pairs out and therefore remove topological noise. The process is illustrated on figure \ref{fig_per1Dcancel}. In 1D, a \hyperref[defperspair]{persistence pair} is always formed of a minimum and a maximum. If those two critical points are direct neighbors, one can in fact increase the value around the minimum and decrease the value around the maximum until the value at the maximum becomes smaller than that at the minimum. When this happens, both points are not critical anymore and none of the other critical points are affected. On the top panel for instance, the process is applied to the less persistent bump of function $A$. Note that the details of how the function  is modified are not important; what is is the fact that it is possible to cancel a non persistent pair and remove it from the Morse-complex (see the diagrams below the functions). For instance, if one considers that structures whose \hyperref[defpers]{persistence} is lower or equal to the \hyperref[defpers]{persistence} of the smaller bump of function $A$ are not significant (\ie generated by noise with high probability), then one can deduce cancel the corresponding topological features so that function $A$ becomes topologically equivalent to its simplified version (top right) with the corresponding Morse complex. Applying the same process to $A^\prime$, the noisy version of $A$, one actually obtains a function with identical topology and Morse complex (central panel). This means that even in the presence of a relatively important noise, it is still possible using \hyperref[defpers]{persistence} to recover the topology and Morse complex of the underlying function (see also the bottom panel to check how the topology of function $B$ on figure \ref{fig_per1D} can be recovered from its noisy counterpart, $B^\prime$). We detail in section \ref{sec_simplification} a generic algorithm that implements symbolic topological simplification in order to recover the structure of the Morse complex of matter distribution on large scale from a raw noisy version computed directly over a Delaunay tessellation.\\

\section{Discrete Morse complex }
\label{sec_implementation}
The basis of the necessary mathematical background being introduced in sections \ref{sec_morse}, \ref{sec_DMtheory} and \ref{sec_persistence}, we now start detailing the particular algorithm and implementation used in \progname. As previously mentioned, our purpose is to compute a {\em discrete} Morse complex and use its properties to identify and characterize the structure of the cosmic web. This approach has both advantages of being applicable to spaces with $3$ or more dimensions and having a solid mathematical framework see also \citet[\eg chap. 6,][]{phd}
(see also \citet[chap. 6][]{phd}). To summarize, a simplicial complex is computed from a discrete distribution (galaxy catalogue, N-body simulation, ...) using Delaunay tessellation and a density $\rho$ is set to each galaxy using DTFE (roughly speaking, the density at a vertex is proportional to the inverse volume of its dual Voronoi cell, see \citet{DTFE}). A \hyperref[defDMF]{discrete Morse function} is then defined by heuristically tagging a properly chosen value to each simplex in the complex (\ie the segments, facets and tetrahedron of the tessellation). From this discrete function, we then compute the \hyperref[defDG]{discrete gradient} and deduce the corresponding discrete \hyperref[defMSC]{Morse-Smale complex} (\hyperref[defDMC]{DMC} hereafter, see section \ref{sec_DMtheory}; \citet{forman}). The \hyperref[defDMC]{DMC} is then used as the link between the topological and geometrical properties of the density field. Its critical points together with their ascending and descending \hyperref[defmanifold]{manifold}s are identified to the peaks, filaments, walls and voids of the density field (see section \ref{sec_morse}). At this stage, the \hyperref[defDMC]{DMC} is mainly defined by Poisson sampling noise and measurement uncertainties, and we filter it using \hyperref[defpers]{persistence} theory (see section \ref{sec_persistence} and appendix \ref{sec_simp_homo} and \ref{sec_app_persistence}). For that purpose, we consider the \hyperref[deffiltration]{filtration} of the tessellation according to the values of the \hyperref[defDMF]{discrete Morse function} and use it to compute \hyperref[defperspair]{persistence pairs} of critical points. The \hyperref[defDMC]{DMC} is finally simplified by canceling the pairs that are likely to be generated by noise. This is achieved by computing the probability distribution function of the \hyperref[defpers]{persistence} ratio of all types of pairs in scale invariant Gaussian random fields and canceling the pairs with a \hyperref[defpers]{persistence} ratio whose probability is lower than a certain level.\\

\subsection{Discrete gradient}
\label{sec_disc_grad}
As stated in section \ref{sec_DMtheory}, a \hyperref[defDG]{discrete gradient} field is derived from a proper \hyperref[defDMF]{discrete Morse function}, which must satisfy definition \ref{def_DMfunction}. Although those conditions are restrictive enough to make the deduction of a valid \hyperref[defDMF]{discrete Morse function} difficult, they allow for a wide variety of such functions to exist; one has to keep in mind that the final \hyperref[defDG]{discrete gradient} field should be as similar as possible to its continuous counterpart $\nabla\rho$, the gradient of the density field $\rho$. We therefore note the \hyperref[defDMF]{discrete Morse function} $F_\rho$. An optimal method to define a \hyperref[defDG]{discrete gradient} has yet to be discovered, but \citet{lewiner_master} propose a nice review on the topic and relatively advanced solutions. Unfortunately, these solutions involve the computation of relatively complex hyper-graphs and are not easily applicable to large data sets. Instead, we implement here a modified version of the one presented in \citet{phd}, which present the advantage of not depending on an arbitrary labeling of the simplexes.\\

\begin{figure*}
\begin{centering}
\subfigure[Example of a smooth function and a simplicial tesselation of space]{\centering \includegraphics[height=0.15\textheight]{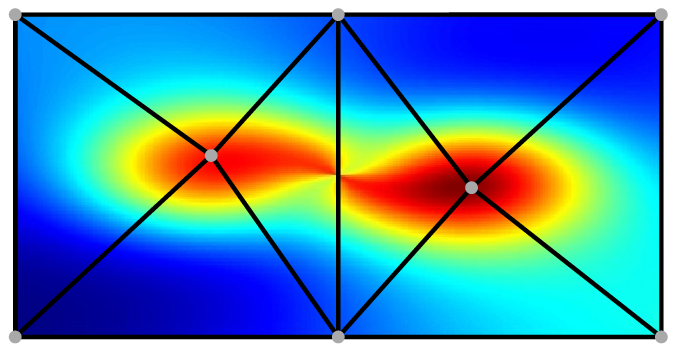}\label{fig_dg_algoA}}\\
\subfigure[The corresponding {discrete Morse function} defined over the simplicial complex (see equation \ref{eq_mf})]{\centering \includegraphics[height=0.18\textheight]{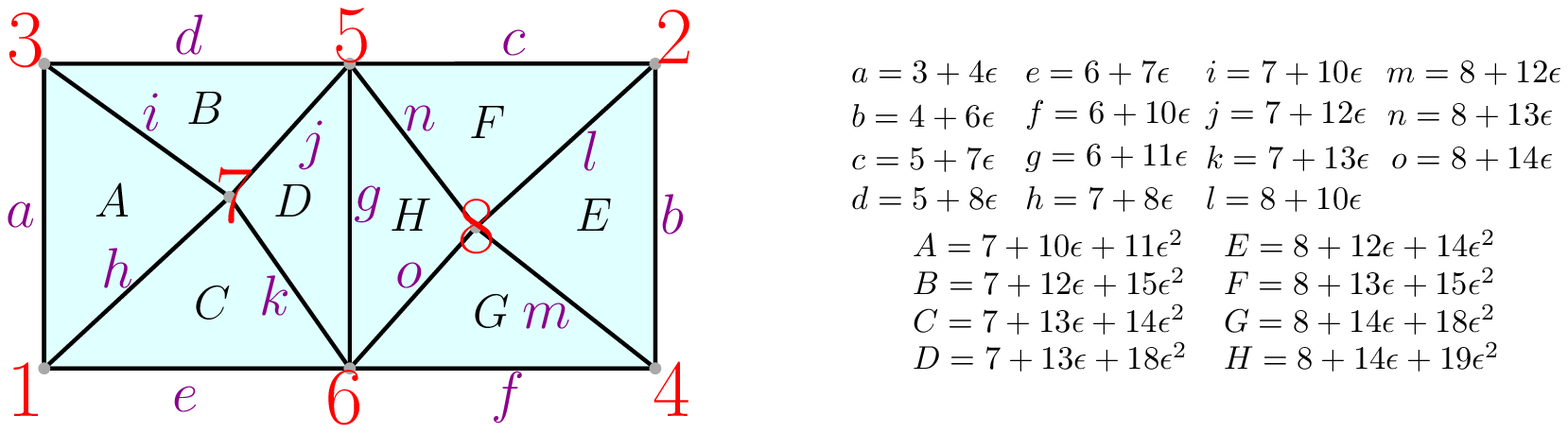}\label{fig_dg_algoB}}\\
\subfigure[Computation of the {discrete gradient}]{\centering \includegraphics[width=\linewidth]{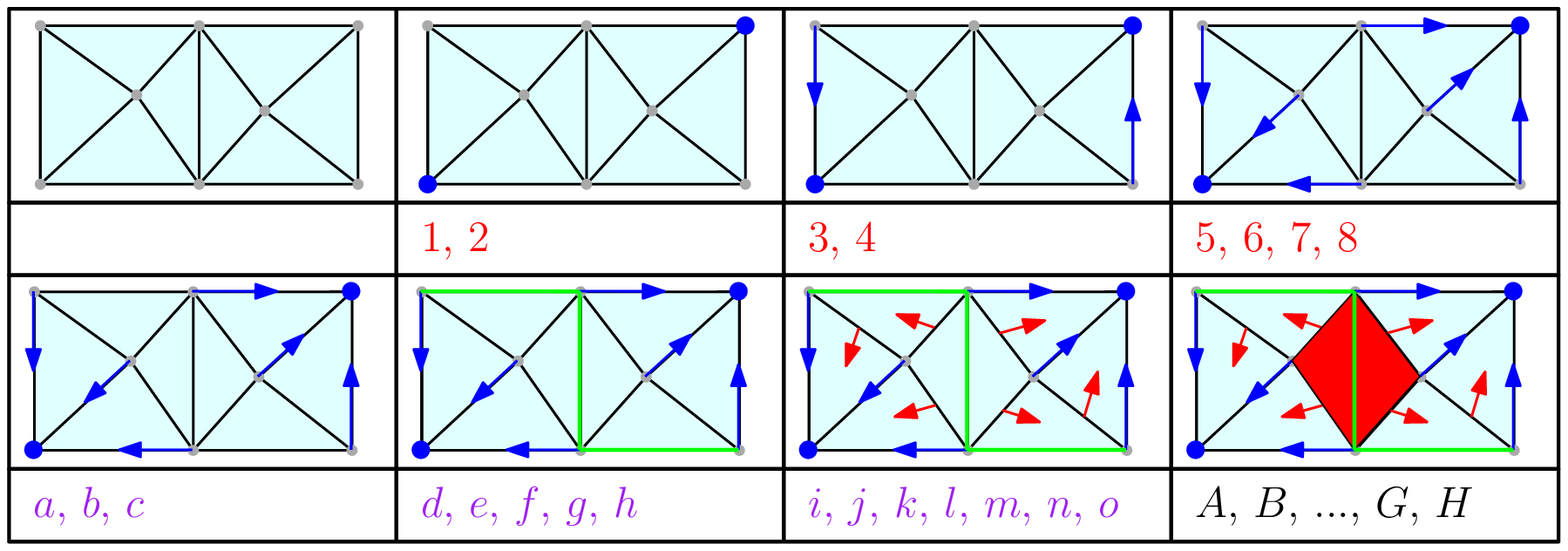}\label{fig_dg_algoC}}
\end{centering}
\caption{\label{fig_dg_algo} Illustration of the computation of a \hyperref[defDG]{discrete gradient} from a simple smooth function and a simplicial complex spanning its domain of definition, as shown on panel \ref{fig_dg_algoA}. The corresponding \hyperref[defDMF]{discrete Morse function} is represented on panel \ref{fig_dg_algoB}. Each vertex is labeled with the value of the corresponding smooth function, and the lower case and upper letters correspond to the labels of the segments and triangles respectively, for which the corresponding value of $F_\rho$ is shown on the right of the panel (see equation \ref{eq_mf}). Note that sorting segments or triangles labels according to alphabetical order also sorts them in increasing order of their value. Panel \ref{fig_dg_algoC} illustrates the computation of the \hyperref[defDG]{discrete gradient} according to the algorithm described in section \ref{sec_disc_grad}, which works by considering the vertexes, segments and triangles one after the other, in increasing order of their value (from left to right and top to bottom on the diagram). Starting with the first vertex, ${F_\rho}^{-1}\left(1\right)$ (lower left vertex), its \hyperref[defcoface]{cofaces} are the segments labeled $a$, $h$ and $e$ with value $3+4\epsilon$, $7+8\epsilon$ and $6+7\epsilon$ respectively. As none of those value differ from $1$ by a factor of $\epsilon$ only, no pair can be formed, and the vertex remains critical (\ie unpaired, represented by a blue disk on the diagram). The vertex with value $2$ presents the same configuration, and is therefore also critical, but the third one to enter, labeled $3$, has one available \hyperref[defcoface]{coface} labeled $a$ with value $3+4\epsilon$ that is only infinitesimally higher, which means the vertex and segment form a gradient pair (blue arrow between $3$ and $a$ on the diagram). The case of vertex $4$ is similar, and it is paired to segment $b$. The next vertex, labeled $5$ is problematic because it  presents two \hyperref[defcoface]{cofaces} with infinitesimally higher value, $c$ and $d$, but the conflict is easily solved by pairing with one with value closest from $5$, segment $c$. We then proceed until no vertex is available anymore, and start considering segments (leftmost box of the second row on the diagram). Segments $a$, $b$ and $c$ are skipped because they are already paired to vertex $3$, $4$ and $5$ respectively. Segment $d$ is free though but does not have an infinitesimally higher \hyperref[defcoface]{coface} (\ie triangle $B$), it is therefore a critical segment (\ie the equivalent of a saddle point, represent as in green). Segments $e$ and $h$ are paired while $f$ and $g$ are found to be critical. This leads to segments $i$ whose \hyperref[defcoface]{cofaces} are $A$ and $B$, whose value differ from that of $i$ by $11\epsilon^2$ and $15\epsilon^2$ respectively, $i$ is therefore paired to the closest triangle in value, $A$ (red arrow on the diagram). The remaining segments are processed the same way and one can then start reviewing the triangles. Only $D$ and $H$ are not paired, and as in 2D triangles have no \hyperref[defcoface]{cofaces}, they are critical (colored red on the diagram). The final \hyperref[defDG]{discrete gradient} is shown on the bottom right box of the figure \ref{fig_dg_algoC}. }
\end{figure*}

Let $F_\rho$ be the \hyperref[defDMF]{discrete Morse function} computed from a smooth function $\rho$ over a simplicial complex $K$, with $\sigma_k$ a \hyperref[defksimplex]{$k$-simplex} that belongs to $K$, ${\rm Facet}\left[\sigma_k\right] \in S$ the facets of $\sigma_k$ and ${\rm Vertex}\left[\sigma_k\right]\in S$ the faces of $\sigma_k$ with dimension $0$ (\ie its vertices). The value of the \hyperref[defDMF]{discrete Morse function} at $\sigma_k$ is defined as:

\begin{equation}
\label{eq_mf}
\left\{ \begin{array}{lrl}
\displaystyle k>0: & F_\rho\left(\sigma_k\right)&={\max}\left(F_\rho\left({\rm Facet}\left[\sigma_k\right]\right)\right) \\
 & & + \epsilon^k .\sum {F_\rho \left({\rm Vertex}\left[\sigma_k\right]\right)},   \\
 & & \\
k=0: & F_\rho\left(\sigma_0\right)&=\rho\left(\sigma_0\right),
\end{array}
\right.
\end{equation}

where ${\max}\left(\right)$ stands the maximal value of its arguments, $d$ is the number of dimensions, and $\epsilon$ is an infinitely small value. One can easily check that such a function does comply to the definition \ref{def_DMfunction} of a discrete Morse-function. In fact, the value of a simplex is always slightly higher than the value of its highest facet, and thanks to the factor of $\epsilon^k$, two simplexes sharing the same highest facet have different values if two vertice in $K$ cannot have the same density. In practice, this is always the case when computing densities using DTFE and in the following we will therefore assume that we are in such a situation.  For that reason, equation \ref{eq_mf} defines the value of $F_\rho$ uniquely from a given smooth function $\rho$, and independently of any arbitrary labeling of the simplexes. Note that to compute the \hyperref[defDMC]{DMC}, one only needs to be able to compare simplexes, and it is therefore not necessary to give a particular value to $\epsilon$, as only a comparison operator needs to be implemented. This definition of $F_\rho$ allows for a unique ordering over the simplexes of $K$.\\

As explained in section \ref{sec_DMtheory}, a \hyperref[defDG]{discrete gradient} can be defined over $K$ by grouping pairs of simplexes whose dimension differ only by $1$ (\ie a vertex and a segment, a segments and a triangle or a triangle and a tetrahedron) and such that conditions \ref{def_DGradient} are satisfied. A group of two paired simplexes form a gradient pair, and the remaining unpaired simplexes are critical (the equivalent of the critical point for a smooth density field\footnote{Note that a critical point of type $k$ from the smooth theory is equivalent to a critical \hyperref[defksimplex]{$k$-simplex} of the discrete theory. In 2D, minima are critical vertices, saddle-point critical segments and maxima are critical triangles.}). Looking at conditions \ref{def_DGradient}, one can see that for two simplexes to form a gradient pair, the simplex of lower dimension should always have a value higher than the other. But because $F_\rho$ has precisely been defined such that any simplex has a value higher than its facets, no pair may be formed, and all the simplexes in $K$ are therefore initially critical. As a consequence, the Morse complex of $F_\rho$ can be readily deduced: each \hyperref[defksimplex]{$k$-simplex} is a critical simplex of order $k$, and it is linked by and arc to each of its faces and \hyperref[defcoface]{cofaces}, which are also critical. Many of those arcs actually link critical simplexes whose \hyperref[defDMF]{discrete Morse function} $F_\rho$ only differ by an infinitesimal amount $\Delta F_\rho\propto \epsilon^p$ though, and we call such arcs $\epsilon$-persistent. Because along those arcs, the value of the function only changes infinitesimally, they can be canceled while only modifying the value of $F_\rho$ by an infinitely small amount. In fact, doing so one can basically exchange the values of $F_\rho$ given to each critical simplex at the extremity of the $\epsilon$-persistent arc and pair them within a gradient arrow. By repeating this process until no $\epsilon$-persistent arcs exist anymore, one can therefore deduce a correct \hyperref[defDG]{discrete gradient}.\\ 

In practice, we proceed by considering the sets of the \hyperref[defksimplex]{$k$-simplexes} of $K$ one by one, in ascending order of their dimension, and within each set, we iterate over the simplexes $\sigma_k$ in ascending order of their value $F_\rho\left(\sigma_k\right)$. For each of them, if it is not already in a gradient pair, we retrieve the lowest of its cofacets $\alpha_{k+1}\in\cofaceof{\sigma_k}$ that is not already in a gradient pair and which value is only infinitesimally higher than $F_\rho\left(\sigma_k\right)$. If it exists, we pair them and else, $\sigma_k$ remains unpaired. Note once again that the value of $F_\rho$ does not need to be explicitly modified in the actual implementation, as $\alpha_{k+1}$ and $\sigma_k$ may only differ infinitesimally if $\sigma_k$ is the highest facet of $\alpha_{k+1}$. The algorithm ends when all the simplexes have been checked once. We show on figure \ref{fig_dg_algo} a practical example of how the algorithm runs on a simple smooth function and a 2D simplicial complex spanning over its domain of definition.\\

\subsection{Discrete Morse complex computation}
\label{sec_dis_morse}
Once a proper \hyperref[defDG]{discrete gradient} has been defined over a simplicial complex, it becomes relatively straightforward to deduce its corresponding \hyperref[defDMC]{DMC}. According to definition \ref{def_manifolds}, the ascending (resp. descending) \hyperref[defmanifold]{manifold} of a critical point $P_k$ of order $k$ is the set of integral lines that end (resp. start) at $P_k$. The discrete analog of an integral line is a \hyperref[defVP]{V-path} (\ie a sequence of simplexes linked by the \hyperref[defDG]{discrete gradient}, see definition \ref{def_vpath}) and one can therefore identify ascending (resp. descending) \hyperref[defmanifold]{manifold}s by following the \hyperref[defVP]{V-pathes} that end (resp. start) at a critical simplex $C_k$. The core of the algorithm consists in a  simple  ``breadth first search" where sequences of \hyperref[defcoface]{cofaces} and gradient pairs are identified according to definition \ref{def_vpath}. Each \hyperref[defmanifold]{manifold} is stored in a separate set type data structure as one simplex may be reached by different \hyperref[defVP]{V-pathes} within one \hyperref[defmanifold]{manifold}. Let ${\cal A}\left(C_k\right)$ be the set that stores the ascending \hyperref[defmanifold]{manifold} of the critical \hyperref[defksimplex]{$k$-simplex} $C_k$. The recursive algorithm starts by considering the set of the {\em \hyperref[defcofacet]{cofacets}} of $C_k$, stored in an array $A_{cur}$ that will basically contain, at the $n^{\rm th}$ step of the algorithm, the set of \kp1-simplexes in the $n^{\rm th}$ gradient pair of any V-path starting at $C_k$. At each step, the content of $A_{cur}$ is scanned and for each \kp1-simplex, there exist four possibilities:
\begin{enumerate}
\item it is critical, in which case it is skipped as the V-path ends.
\item it is not critical, and is paired to a \hyperref[defksimplex]{$k$-simplex} in a gradient pair. In that case, the \hyperref[defksimplex]{$k$-simplex} is added to ${\cal A}\left(C_k\right)$ and stored in a temporary array $A_{tmp}$.
\item it is not critical, but is paired by discrete gradient to a \hyperref[defksimplex]{$k$-simplex} already in ${\cal A}\left(C_k\right)$. In that case, it is skipped.
\item it is not critical, and is paired to a $k+2$-simplex in a gradient pair. In that case, it is skipped.
\end{enumerate}
Once all simplexes in $A_{cur}$ have been treated, the content of $A_{cur}$ is replaced by the {\em \hyperref[defcofacet]{cofacets}} of the \hyperref[defksimplex]{$k$-simplexes} in $A_{tmp}$ and the process is iterated until $A_{cur}$ is empty at which stage all the simplexes in ${\cal A}\left(C_k\right)$ have been retrieved. The computation of the descending \hyperref[defmanifold]{manifold} $D\left(C_k\right)$ is achieved the exact same way, except that \hyperref[defcofacet]{cofacets} are replaced by \hyperref[deffacet]{facets}. A pseudo-code implementation is presented in Algorithm \ref{alg_manifold} (see the non tagged lines only). Note that in this implementation, only \hyperref[defksimplex]{$k$-simplexes} are stored to describe the \hyperref[defmanifold]{manifold} of a critical \hyperref[defksimplex]{$k$-simplex}, which reduces memory usage. It also implies that the algorithm does non compute the extended discrete \hyperref[defmanifold]{manifold}s of definition \ref{def_extdmanifolds} but rather those of definition \ref{def_dmanifolds}. This is indeed not a problem though as those \hyperref[defmanifold]{manifold}s can easily be extended at query time from the identified sets of \hyperref[defksimplex]{$k$-simplexes}. Practically, extending an ascending (resp. descending) $k$-\hyperref[defmanifold]{manifold}s consists in recursively adding the cofaces (resp. faces) of any simplex in the \hyperref[defmanifold]{manifold}, as well as the ascending (resp. descending) $p$-\hyperref[defmanifold]{manifold}s ($p>k$) of any of its critical $p$-simplexes.\\ 

\begin{figure*}
\begin{centering}
\subfigure[Reproduction of the example discrete Morse function and its {discrete gradient} (see figure \ref{fig_dg_algo})]{\centering \includegraphics[height=0.2\textheight]{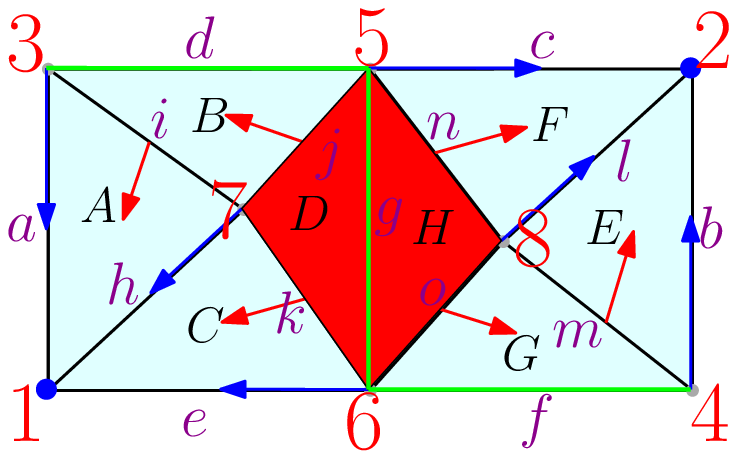}\label{fig_net_manifoldsA}}
\hfill\subfigure[Resulting discrete Morse-smale complex]{\centering \includegraphics[height=0.2\textheight]{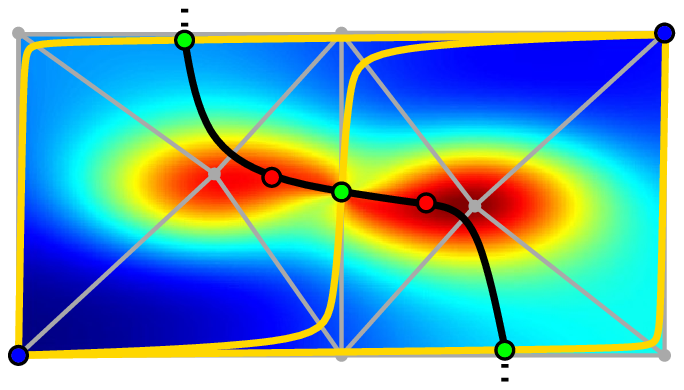}\label{fig_net_manifoldsB}}\\
\subfigure[Computation of the discrete extended ascending (left) and descending (right) $1$-{manifolds} (see definition \ref{def_extdmanifolds}) of the three critical $1$-simplexes (\ie saddle points, the green disks, critical simplexes being represented by disks for clarity)]{\centering \includegraphics[width=0.98\linewidth]{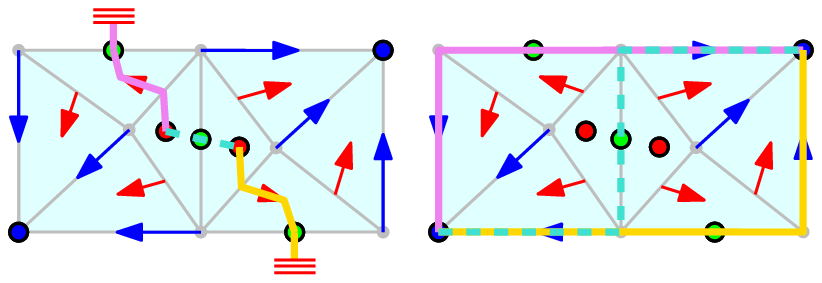}\label{fig_net_manifoldsC}}\\
\subfigure[Computation of the discrete extended ascending (left) and descending (right) $2$-{manifold}s (see definition \ref{def_extdmanifolds}) of the two critical $0$-simplexes (\ie minima, the blue disks, critical simplexes being represented by disks for clarity)]{\centering \includegraphics[width=0.98\linewidth]{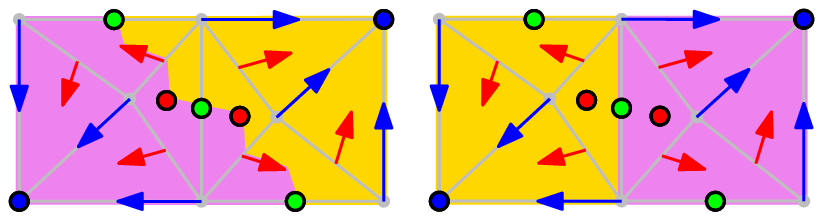}\label{fig_net_manifoldsD}}
\end{centering}
\caption{\label{fig_net_manifolds} Illustration of the computation of the discrete extended ascending and descending \hyperref[defmanifold]{manifold}s and corresponding Morse-Smale complex from a discrete gradient. The application of the algorithm described in section \ref{sec_dis_morse} over the simple discrete function and gradient shown on panel \ref{fig_net_manifoldsA} (see figure \ref{fig_dg_algo} for labels description) is illustrated on panels \ref{fig_net_manifoldsC} and \ref{fig_net_manifoldsD} for the ascending (left) and descending(right)  $1$-\hyperref[defmanifold]{manifold}s and $2$-\hyperref[defmanifold]{manifold}s respectively (critical simplexes are identified by colored disks in their center). On figure \ref{fig_net_manifoldsC}, the ascending (left diagram) and descending (right diagram) $1$-\hyperref[defmanifold]{manifold}s of the three critical $1$-simplexes (\ie equivalent of saddle points, represented by green disks) are represented as pink plain, cyan dashed and yellow plain broken lines respectively. The $1$-\hyperref[defmanifold]{manifold}s are represented as sets of segments joining the center of simplexes as according to definition \ref{def_vpath}, a \hyperref[defVP]{V-path} is an alternating sequences of $k$ and {\kp1}-simplexes linked by a face/coface relation or belonging to a gradient pair. Note on the right diagram how it is possible for two descending $1$-\hyperref[defmanifold]{manifold}s (blue dashed and plain yellow or plain pink) to share a portion of their path. On figure \ref{fig_net_manifoldsD}, the ascending (left diagram) and descending (right diagram) $2$-\hyperref[defmanifold]{manifold}s of the two critical $0$-simplexes (\ie equivalent of minima, represented by blue disks) are colored in pink and yellow respectively. The Morse-smale complex is the set of n-cell obtained by intersecting pairs ascending and descending \hyperref[defmanifold]{manifold}s (see definitions \ref{def_morse_ncell} and \ref{def_mscomplex}), and it is represented over the initial smooth function on panel \ref{fig_net_manifoldsB}. On this figure, the black and yellow curves represent the arcs (\ie 1-cells) linking maxima/saddle points and minima/saddle points respectively. It is very striking how the algorithm manages to correctly capture the essential features of the Morse Smale complex, even though it was only applied over a very crude simplicial tesselation of space: not only the critical points where correctly identified as critical simplexes, but the way they are connected by arcs is also correct (note that the arcs geometry was smoothed for clarity reasons). }
\end{figure*}

Figure \ref{fig_net_manifolds} illustrates the result of applying this algorithm over the simple \hyperref[defDG]{discrete gradient} of figure \ref{fig_dg_algo} (note that the corresponding discrete function and gradient has been reproduced on figure \ref{fig_net_manifoldsA}). The four diagrams displayed on figures \ref{fig_net_manifoldsC} and \ref{fig_net_manifoldsD} show the result obtained while computing the discrete extended ascending (left) and descending (right) $1$-\hyperref[defmanifold]{manifold}s of the three saddle points (pink, yellow, and blue dashed lines), and the discrete extended ascending (left) and descending (right) $2$-\hyperref[defmanifold]{manifold}s of the two minima and two maxima (pink and yellow shaded regions) respectively. 

As an example, let us detail first the process followed by our algorithm to measure the ascending $1$-\hyperref[defmanifold]{manifold} ${\cal A}\left(C_1\right)$ of $C_1$, the critical $1$-simplex (\ie saddle point) with label $d$ (see red path on left diagram of figure \ref{fig_net_manifoldsC}). We start by considering the \hyperref[defcofacet]{cofacets} of $C_1=d$ and as there is only one, labeled $B$, we initially set $A_{cur}=\left[ B\right]$. The $2$-simplex $B$ is linked to segment $j$ by a gradient arrow, $j$ is therefore added to ${\cal A}\left(C_1\right)$ and $A_{tmp}=j$. Segment $j$ has two \hyperref[defcofacet]{cofacets}, the triangles $B$ and $D$ and we therefore set $A_{cur}=\left[ B,D\right]$. We then proceed by considering all triangles in $A_{cur}$ one by one. The $2$-simplex $B$ is not critical but is paired to segment $j$ which already belongs to ${\cal A}\left(C_1\right)$, it is therefore skipped and we are left considering triangle $D$ which is critical and is therefore also skipped. Eventually, we obtain ${\cal A}\left(C_1\right)=\left[j\right]$. The pink path on the figure corresponds to the extended version of ${\cal A}\left(C_1\right)$, obtained by recursively also including the cofaces of the simplexes in ${\cal A}\left(C_1\right)$, namely the triangles $B$ and $D$. 

Similarly, the algorithm can be applied to the critical vertex $C_0$ with value $1$ to retrieve its ascending $2$-\hyperref[defmanifold]{manifold}s displayed in pink on the left frame of figure \ref{fig_net_manifoldsD}. The cofacets of vertex $1$ are segments $a$, $h$ and $e$ and, as none of them is critical, the algorithm starts with $A_{cur}=\left[a,h,e\right]$. The segments in $A_{cur}$ are paired with vertex $3$, $7$ and $6$ respectively, which are not critical vertexes and do not yet belong to ${\cal A}\left(C_0\right)$, they are therefore added to ${\cal A}\left(C_0\right)$ so that ${\cal A}\left(C_0\right)=A_{tmp}=\left[3,7,6\right]$. The content of $A_{cur}$ is then replaced by all the segments that are cofaces of at least one vertex in $A_{tmp}$, and we have $A_{cur}=\left[a,i,d,h,j,k,e,g,f\right]$. Considering the segments in $A_{cur}$ one by one, $a$, $h$ and $e$ are skipped because they are paired to vertex $3$, $7$ and $6$ respectively, which belong to ${\cal A}\left(C_0\right)$, $d$, $g$ and $f$ are skipped because they are critical and segments $i$, $j$ and $k$ are skipped because they are not paired to $1$-simplexes, but to the $2$-simplexes $A$, $B$ and $C$ respectively. This leaves $A_{tmp}$ empty, and as a consequence $A_{cur}$ becomes void which stops the algorithm with ${\cal A}\left(C_0\right)=A_{tmp}=3,7,6$. The pink shaded region on the figure corresponds to the extended version of ${\cal A}\left(C_0\right)$, obtained by also adding the \hyperref[defcoface]{cofaces} of vertex $3$, $7$ and $6$, which are segments $a$, $h$, $e$, $d$, $i$, $j$, $g$, $k$, $o$ and $f$ and triangles $A$, $B$, $C$, $D$, $G$ and $H$, as well as the extended ascending $1$-\hyperref[defmanifold]{manifold}s of critical $1$-simplexes $d$, $g$ and $f$.\\

The computation of the arcs in the \hyperref[defMSC]{Morse-Smale complex} is slightly more involved. A \hyperref[defMSC]{Morse-Smale complex} is formed by critical nodes and arcs linking them together. Those arcs are integral lines that start at a critical point of order $k+1$ and end at a critical point of order $k$, so they always have dimension $1$: they are represented by curves. Their discrete equivalents are \hyperref[defVP]{V-pathes} linking critical {\kp1}-simplexes and critical \hyperref[defksimplex]{$k$-simplexes}. In 2D, they are simply described by the ascending and descending $1$-\hyperref[defmanifold]{manifold}s (the dashed blue, pink and yellow lines on the upper part of figure \ref{fig_net_manifoldsB}), but this is not the case in higher dimensions where arcs are generally described by the one dimensional intersections of a descending and an ascending \hyperref[defmanifold]{manifold}. The bottom diagram of figure \ref{fig_net_manifoldsB} shows the discrete Morse Smale Complex computed from the simple density field $\rho$ represented by the background color. It was obtained thanks to a modification of the manifold algorithm: when computing an ascending (resp. descending) \hyperref[defmanifold]{manifold} of a critical \hyperref[defksimplex]{$k$-simplex}, we store the list of critical {\kp1}-simplexes (resp. {\km1}-simplexes) that are encountered and for each of them, trace the \hyperref[defVP]{V-path}es that led to them by storing in separated arrays all the simplexes in each path when the recursive procedure is returning. This way we obtain, for each pairs of critical \hyperref[defksimplex]{$k$-simplex} and {\kp1}-simplex that are linked by a \hyperref[defVP]{V-path}, the set of all simplexes in the \hyperref[defVP]{V-path} (\ie the arcs of the Morse-Complex). Note that on figure \ref{fig_net_manifoldsB}, each ascending (resp. descending) $1$-\hyperref[defmanifold]{manifold} is actually made of two arcs, each linking the same saddle point to a maximum (resp. minimum). Algorithm \ref{alg_manifold} presents the pseudo code for a function that computes the ascending or descending arcs and \hyperref[defmanifold]{manifold}s of a critical \hyperref[defksimplex]{$k$-simplex}, the \hyperref[defmanifold]{manifold} and arcs being returned in global simplex array $M$ and global list of simplex arrays $arcs$ respectively. In this code, the lines dedicated to identifying arcs are tagged to differentiate them from the simpler manifold identification algorithm. After this function is called on a critical simplex $C_k$, $M$ will contain the index of all the \hyperref[defksimplex]{$k$-simplexes} in the ascending (resp. descending) \hyperref[defmanifold]{manifold} of $C_k$ (not including $C_k$) and $Arcs$ will contain a list of arrays, each containing the \hyperref[defksimplex]{$k$-simplexes} in a \hyperref[defVP]{V-path} between $C_k$ and another critical simplex $C_{k+1}$ (resp. $C_{k-1}$), including $C_{k+1}$ (resp. $C_{k-1}$) and $C_k$.\\

\begin{algorithm}
\caption{Computes the ascending or descending \hyperref[defmanifold]{manifold} and arcs of a critical simplex $S_k$. Variables $arcs$ and $M$ store the retrieved Arcs and \hyperref[defmanifold]{manifold}. Triangular marks tag the lines dedicated to arcs identification only.}\label{alg_manifold}
\begin{algorithmic}[1]
\Function{GetManifold}{$\sigma_k$, $ascending$}
\Require $\sigma_k$ is a critical $k$-simplex
\Require $M$ is an empty list of simplexes
\Require $arcs$ is an empty list of arrays of simplexes
\If{$ascending$}
\State $A_{cur}\gets \Call{getCofaces}{\sigma_k}$
\Else
\State $A_{cur}\gets \Call{getFaces}{\sigma_k}$
\EndIf
\State $A_{tmp}\gets \lbrace\rbrace$
\State $curArcs\gets \lbrace\rbrace$\Comment{}

\ForAll{$c\gets A_{cur}$}
\State $p\gets \Call{getGradientPair}{c}$
\If{$\Call{getDimension}{p}==k$ and $p\notin M$} 
\State $M\rightarrow\Call{insert}{p}$
\If{not $\Call{isCritical}{p}$}
\State $A_{tmp}\rightarrow\Call{insert}{p}$
\Else\Comment{}
\State $newArc\gets \lbrace c\rbrace$\Comment{}
\State $arcs\rightarrow\Call{pushBack}{\& newArc}$\Comment{}
\State $curArcs\rightarrow\Call{insert}{\& newArc}$\Comment{}
\EndIf\Comment{}
\EndIf
\EndFor

\ForAll{$c\gets A_{tmp}$}
\State $newArcs\gets\Call{GetManifold}{c, ascending}$
\State $curArcs\rightarrow\Call{insert}{newArcs}$\Comment{}
\EndFor

\ForAll{$c\gets curArcs$}\Comment{}
\State $c\rightarrow\Call{pushBack}{\sigma_k}$\Comment{}
\EndFor\Comment{}
\State \textbf{return} $curArcs$\Comment{}
\EndFunction
\end{algorithmic}
\end{algorithm}

We end this subsection with a comment on our implementation. The ascending $\kmn[d]{k}$-\hyperref[defmanifold]{manifold} and descending $k$-\hyperref[defmanifold]{manifold} of a critical \hyperref[defksimplex]{$k$-simplex} are represented by lists of \hyperref[defksimplex]{$k$-simplexes}. This certainly makes sense for the descending $k$-\hyperref[defmanifold]{manifold}, as in 3D for instance, volumes will be represented by lists of tetrahedrons, surfaces by list of faces and lines by lists of segments. However, this is not the case for the ascending $\kmn[d]{k}$-\hyperref[defmanifold]{manifold}, where for instance, the ascending $3$-\hyperref[defmanifold]{manifold} of a minimum is represented by a list of vertice. To solve this issue one can choose to use extended \hyperref[defmanifold]{manifold}s instead of regular manifolds. This can be problematic though, for instance for visualization purpose, not only because it considerably increases the number of simplexes within each manifold, but also because in that case two neighboring $k$-\hyperref[defmanifold]{manifold}s will edge \hyperref[defksimplex]{$k$-simplexes}. For instance, on figure \ref{fig_net_manifoldsD}, the extended ascending $2$-\hyperref[defmanifold]{manifold}s should actually share $2$-simplexes $B$, $D$, $H$ and $G$ if our algorithm was followed. It is not the case on the figure though because we actually used the dual tessellation for the representation of the extended ascending \hyperref[defmanifold]{manifold}s (\ie the Voronoi tessellation in our case, where the complex is computed on a Delaunay tessellation). In fact, the dual tessellation associates a cell of dimension $\kmn[d]{k}$ to each \hyperref[defksimplex]{$k$-simplex}, and one only has to interpret the list of simplexes in an ascending \hyperref[defmanifold]{manifold}s in terms of its dual Voronoi cells, surfaces, lines or vertice, which does not necessitate any modification of the algorithm. Note that this point of view is also interesting as it enforces the fact that ascending and descending \hyperref[defmanifold]{manifold}s intersect transversely (\ie they cannot be tangent in any point), an essential property of a Morse-Smale function (see section \ref{sec_morse}). In practice, we always use the dual representation for visualization of the descending \hyperref[defmanifold]{manifold}s, $k$ dimensional regions being best represented by lists of \hyperref[defksimplex]{$k$-simplexes}, but only store $k$-\hyperref[defmanifold]{manifold}s as lists of \hyperref[defksimplex]{$k$-simplexes} as this is much more efficient.

\section{Dealing with noise: persistence and topological simplification}
\label{sec_toposimp}
Using the algorithms introduced in the previous two subsections, it is possible to compute efficiently the discrete \hyperref[defMSC]{Morse-Smale complex} (\hyperref[defDMC]{DMC}) of basically any function discretely sampled function via the delaunay tesselation of the sampling points. Applied directly to the Delaunay tessellation of a discrete galaxy catalogue or of a N-body dark matter simulation, the algorithm could therefore theoretically be used to identify the filaments, walls and void. However, because it cannot discriminate between the spurious Poisson noise induced detections and the actual cosmic web features, it is of no practical interest as is. As an example, we applied it to the output of a $50\Mpc$ large dark matter simulation down-sampled to $64^3$ particles. Running a simple friend-of-friend algorithm \citep{FOF} with a linking length equal to one twentieth of the inter particular distance and a minimal number of particle of $20$ leads to the identification of $~800$ bound structures (\ie potential dark matter haloes). Computing the Morse complex of the same distribution leads to the identification of $12771$ maxima (\ie potential haloes) and $32457$ type $1$ saddle points (\ie potential filaments). This suggests that only about $\sim 6\%$ of the detected structures are cosmologically significant and that most of the detected filaments actually link spurious noise induced maxima. In order to remedy this problem, we apply the concept of \hyperref[defpers]{persistence} \citep{edel00}, introduced in section \ref{sec_persistence}. Roughly speaking, \hyperref[defpers]{persistence} defines a mathematically rigorous framework to assess the significance of topological features while Morse theory, by the mean of the \hyperref[defMSC]{Morse-Smale complex}, establishes the link between local geometry and topology. We describe in the following how, using those theories together, the Morse complex can be simplified in order to get rid of its unwanted features.

\subsection{Pairing critical simplexes and persistence}
\label{sec_comp_pers}
\begin{figure}
\centering
\includegraphics[width=\linewidth]{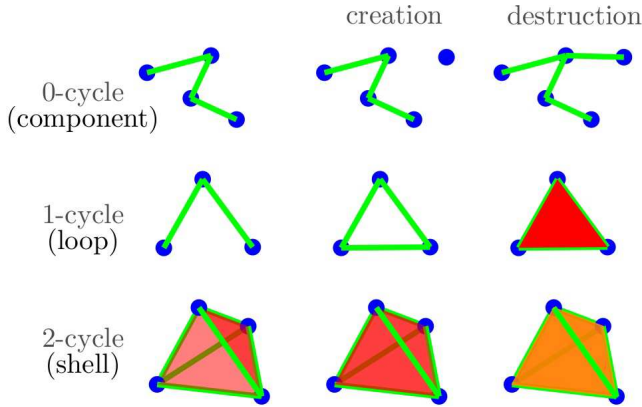}
\caption{\label{fig_simple_pair} Creation and destruction of $k$-cycles in a \hyperref[deffiltration]{filtration} according to $F_\rho$ (equation \ref{eq_mf}). An unlinked component creates a $0$-cycle , a loop around a hole creates a $1$-cycle and a shell around an empty volume creates a $2$-cycle.}
\end{figure}

Within the context of a smooth function, \hyperref[defpers]{persistence} can be understood as a measure of the life-time of a given topological feature (interpreted as the relative importance or significance of that feature) within the evolving sub-levels sets at levels varying from one extreme of the function possible values to the other. Within a discrete context, a very similar concept and interpretation can be defined for a \hyperref[deffiltration]{filtration} (see definition \ref{def_filtration}) of a simplicial complex $K$: new simplexes entering the filtration create or destroy topological features, defining their persistence in terms of how many new simplexes had to enter the filtration before a given topological feature was destroyed. In that case, one therefore measures the importance of the different topological features induced by the function that defines the order of entrance of each simplex in the filtration. As we are interested in the topology and geometry of the density function $\rho$, it is therefore natural to use its discrete counterpart $F_\rho$ (see equation \ref{eq_mf}) to define the time each simplex enters the filtration, as it associate a distinct value to each simplex. We therefore consider the \hyperref[deffiltration]{filtration} $F$ of $K$ according to the ascending values\footnote{note that the choice of ordering according to ascending or descending value is totally arbitrary and has no importance.} of $F_\rho$, similarly to what was done in section \ref{sec_dis_morse} to compute the Morse complex, and recast the persistence measure in terms of the difference of the value $F_\rho$ associated to the simplex that creates a feature, and the simplex that destroys it.\\

Because of the way $F_\rho$ was defined, any simplex enters $F$ before its \hyperref[defcoface]{cofaces}. In the 3D case for instance, this is illustrated on figure \ref{fig_simple_pair}. A vertex ($0$-simplex) is never linked to the rest of $F$ when it enters and therefore we say that it will always {\em create} a new component (\ie a $0$-cycle) in the \hyperref[deffiltration]{filtration}. Similarly, when a segment enters, its two faces already belong to $F$ while its \hyperref[defcoface]{cofaces} do not yet: it forms a bridge between two $0$-simplexes, and may therefore either {\em destroy} one component if those two $0$-simplexes belonged to distinct ``islands'' or {\em create} a new $1$-cycle (\ie a loop, a torus like structure) in the other case. The same way, a face could {\em destroy} a $1$-cycle by filling the hole in its center or {\em create} a $2$-cycle (\ie a shell), and a tetrahedron may only destroy a $2$-cycle (\ie fill a shell). A consequence of the fact that all simplexes create or destroy something is that all simplexes in the complex are initially critical, as was already noted in section \ref{sec_disc_grad}, and our goal is to establish which critical simplexes respectively create and destroy a given $k$-cycle of $F$ (\ie a component, a loop, a shell, ...).\\ 

Actually, that is exactly what the algorithm that computes the \hyperref[defDG]{discrete gradient} does for the so called $\epsilon$-persistent arcs (\ie arcs that link simplexes whose value $F_\rho$ only differs by an infinitesimal amount $\epsilon$, see section \ref{sec_disc_grad}). In fact, a simplex $\sigma_k$ and its face $\sigma_{k-1}$ may belong to a gradient pair if their value differ only by $\epsilon$ (\ie if they enter consecutively in the \hyperref[deffiltration]{filtration}). When this is the case, the value of $F_{\rho}$ is symbolically modified by an infinitesimal amount so that $\sigma_{k-1}$ actually enters just before $\sigma_k$ and none of them may create or destroy a cycle anymore. Whereas $\sigma_k$ created a new $k$-cycle destroyed by $\sigma_{k-1}$ in the initial \hyperref[deffiltration]{filtration}, this is not the case anymore after the modification, both simplexes are not critical anymore, and belong to gradient pair instead. We therefore only need to pair the critical simplexes that survive to the \hyperref[defDG]{discrete gradient} computation (\ie that belong to the discrete \hyperref[defMSC]{Morse-Smale complex}) into persistence pairs. \citet{edel00} first introduced and algorithm that does exactly that in 3D. Although, more general and efficient approaches have been developed since (\eg \citet{vineyard} or \citet{zomo}), we present here a variation of the original one, directly implemented over the morse complex. Note that given that only the critical simplexes identified in the \hyperref[defDMC]{DMC} of $K$ create or destroy cycles, one only needs to consider the Morse complex directly (\ie as opposed to considering each and every simplex in $K$) to identify \hyperref[defperspair]{persistence pairs}. From that point, it therefore does not matter anymore how the \hyperref[defMSC]{Morse-Smale complex} was computed, or whether it is discrete or not, as both have identical combinatorial properties anyway. Under the assumption that the \hyperref[defDMF]{discrete Morse function} was computed with enough care to correctly inherit the topology of the underlying density field, we can therefore indifferently talk about the critical points of the smooth density field $\rho$ or the critical \hyperref[defksimplex]{$k$-simplexes} $\sigma_k$ of the simplicial complex $K$ in the following. It is also equivalent to describe persistence in term of creation/destruction events in the level-sets of $\rho$ or in the filtration steps of the filtration induced by $F_\rho$ (by convention, we choose to order the entrance time by ascending values of $F_\rho$).\\

The algorithm starts by tagging each critical simplex $\sigma_k$ as {\em positive} or {\em negative} depending on whether it creates or destroys a cycle. As was noted before, in $3D$, the critical vertices and tetrahedrons (equivalent of minima and maxima) may only create a $0$-cycle and destroy a $2$-cycle respectively. The critical $0$-simplexes are therefore all tagged positive and the critical $3$-simplexes are negative. The sign of the rest of the critical simplexes is determined by following the growth and merging of each component in the \hyperref[deffiltration]{filtration} using a ``union-find'' type data structure\footnote{a Union-find data structure is particularly efficient at managing large number of sets of elements. It implements fast set merging (the ``union'' operation) and is able to recover efficiently to which set a given element belongs to (``find'' operation).}. Depending on whether a segment entering the filtration links two previously independant components (\ie destroys a $0$-cycle) or creates a new bridge within one unique component (\ie creates a $1$-cycle), it will be tagged negative or positive as it destroys a component or creates a $1$-cycle. Tracking the creation of $2$-cycles (\ie shells) or destruction of $1$-cycles by the critical $2$-simplexes in the filtration seems much more complex though, but it can actually be made easy by considering the filtration $F^\prime$ induced by $-F_\rho$, where simplexes enter in the opposite order to $F$. For symmetry reasons, a $2$-simplex creating a $2$-cycle in $F$ actually destroys a component in $F^\prime$, and is therefore positive,  while a simplex destroying a $1$-cycle in $F$ actually creates $1$-cycle in $F^\prime$ and is therefore negative. The exact same algorithm and ``union-find'' type data structure can therefore be used to track those events in $F^\prime$ and decide the sign of each critical $2$-simplex in $F$.\\

Practically, let us consider the \hyperref[deffiltration]{filtration} in the ascending order first. An entry is created in a ``union-find'' structure for each critical simplex in the \hyperref[defDMC]{DMC}, each of them is initially attributed a different group Id. Whenever a segment enters the \hyperref[deffiltration]{filtration}, the group Id of its two facets are retrieved and we check if they differ or are equal. In the first case, this means the segment created a bridge between two previously disjoint structures. It is therefore tagged negative and the groups of the two vertices and the segment are merged in the union find structure. In the second case, both vertice already belonged to the same structure, which means that the introduction of the segment created a new $1$-cycle (\ie a loop that passes through the newly created bridge). The segment is therefore tagged positive and its group is merged with that of its faces. The sign of the $2$-simplexes (triangles) is determined in the same way, but reversing the order of the \hyperref[deffiltration]{filtration}: a face is tagged positive whenever it creates a bridge between two previously unlinked tetrahedron and negative whenever it links two tetrahedron that where already linked.\\

\begin{algorithm}
\caption{Finds persistence cycles created by a negative critical simplex $\sigma_k$.}\label{alg_cycle}
\begin{algorithmic}[1]
\Function{cycleSearch}{$\sigma_k$}
\Require $\sigma_k$ is a negative critical $k$-simplex (parameter)
\Require $ppairs$ stores persistence pairs (global)
\Require $cycles$ store all previously computed cycles, each associated to a negative simplex (global)
\Require $CurSet$ is a $\mathbb{Z}_2$-Set of simplexes (see text), empty when first called (local)
\State $\alpha_{nei}\gets\Call{getMSCneighbors}{\sigma_k}$
\Comment{$\alpha$ contains the simplexes that share an arc with $\sigma_k$ in the DMC.}
\ForAll{$\beta \gets \alpha_{nei}$}
\If{$\Call{typeOf}{\beta}==\Call{typeOf}{\sigma_k}-1$ and not $\Call{signOf}{\beta}==\Call{signOf}{\sigma_k}$}
\State $CurSet\rightarrow\Call{insert}{\beta}$
\EndIf
\EndFor

\While{not $\Call{isEmpty}{CurSet}$}
\State$ \sigma^{cur}_{k-1}\gets\Call{getHighestOf}{CurSet}$
\If{$\Call{isEmpty}{cycles[\sigma^{cur}_{k-1}]}$}
\State $cycles[\sigma^{cur}_{k-1}]\gets CurSet$ 
\State $cycles[\sigma_k]\gets CurSet$
\State $ppairs\rightarrow\Call{insert}{\sigma^{cur}_{k-1},\sigma_k}$
\State \textbf{break}
\Else
\ForAll{$\beta \gets cycles[\sigma^{cur}_{k-1}]$}
\State $CurSet\rightarrow\Call{insert}{\beta}$ \Comment{note that adding the cycle of $\sigma^{cur}_{k-1}$ modulo 2 actually removes simplex $\sigma^{cur}_{k-1}$.}
\EndFor
\EndIf
\EndWhile

\EndFunction
\end{algorithmic}
\end{algorithm}

Now that each critical simplex $\sigma_k$ has been attributed a sign, we can reconsider the \hyperref[deffiltration]{filtration} $F$ of the critical simplexes in ascending order, and identify the \hyperref[defperspair]{persistence pairs} using algorithm \ref{alg_cycle}. Instead of detailing how this rather complex algorithm works, let us detail its application to a simple 2D example for the sake of clarity. Note that the method is very similar whatever the number of dimensions, as long as the sign of each critical simplex has been previously determined, and so deducing the 3D case from the 2D one should be straightforward. We first define a few variable names and types the algorithm uses. The purpose of the function $cycleSearch\left(\sigma_k\right)$ is to retrieve the $\left( k-1\right)$-cycle destroyed by the negative critical simplex $\sigma_k$. For each call, the result is stored in a variable $cycles$ that will in the end contain the description of all cycles, each associated to its creating and destroying critical $k$-simplex. Each cycle is stored as a list of critical \km1-simplexes that form a \km1-cycle within the Morse-Smale complex (for instance, a loop is stored as a list of critical segments). Another variable, labeled $ppairs$, stores pairs of critical simplexes that creates or destroys its corresponding a  given cycle $\left[\sigma_k,\sigma_{k-1}\right]$. Basically, the function $cycleSearch\left(\sigma_k\right)$ is called once every time a negative critical simplex $\sigma_k$ enters the filtration. Internally, the function uses a variable $CurSet$, of special type ``$\mathbb{Z}_2$-Set'',  to store a temporary list of critical \km1-simplexes considered at a given time. The type ``$\mathbb{Z}_2$-Set'' implements the $k$-chain group addition of definition \ref{def_kchain}, or in other words it behaves like regular ``Set type'' structure, that stores sets of elements, but contrary to normal ``Set'', adding an element already contained in the $\mathbb{Z}_2$-Set results in its actual removal\footnote{hence the name, as each element behaves as if it was counted modulo 2, with coefficients in $\mathbb{Z}_2$}.\\  

\begin{figure*}
\begin{minipage}[c][\textheight]{\linewidth}
\centering
\includegraphics[width=\linewidth]{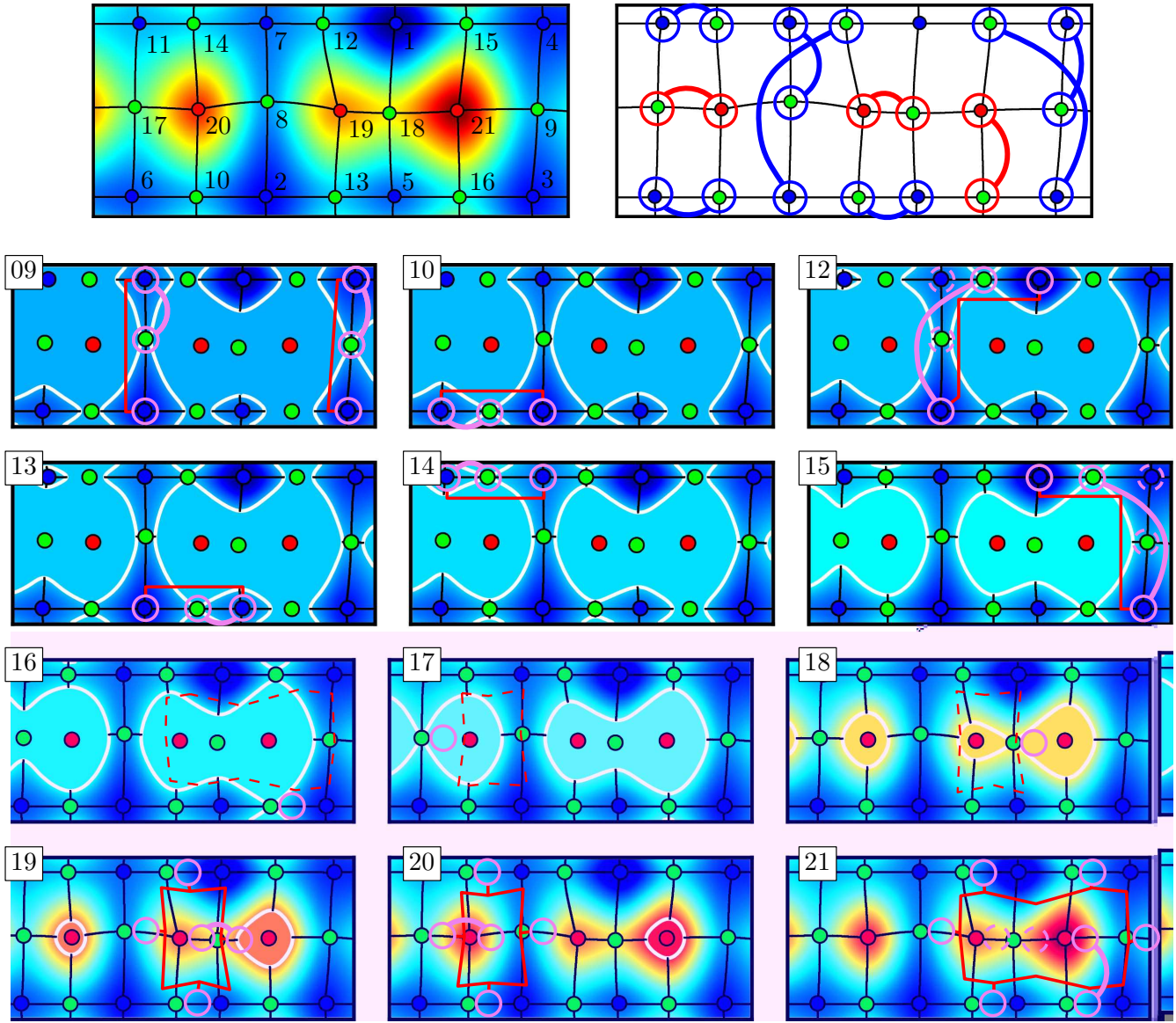}
\caption{\label{fig_cycle_search} Illustration of the computation of \hyperref[defperspair]{persistence pairs} using the algorithm \ref{alg_cycle} presented in section \ref{sec_comp_pers}. On the top left frame, the Morse-smale complex of the underlying smooth function $\rho$ is represented with blue,green and red disks, standing for minima, saddle points and maxima. Note that the Morse complex is actually a DMC computed from a discrete morse function $F_\rho$ (see equation \ref{eq_mf}) over a high resolution tessellation (not represented), so the blue, green and red disks, equally stand for critical vertexes, segments and triangles respectively (the two views are equivalent under the assumption that $F_\rho$ correctly identify the topology of $\rho$). The numbers $n$ beside the disk correspond to the corresponding values of the density $\rho$. On the $12$ panels in the bottom part, the evolution in the sub-level sets (\ie the set of points where density $\rho$ is smaller than a given threshold $\rho_n$) of the smooth density field is shown in the background, at levels $\rho_n=n$ corresponding to the value $n$ in the upper left corner of each panel. On each panel, the identification of a new \hyperref[defperspair]{persistence pair} in the DMC is represented by a pink arc, while the corresponding cycle is symbolized by the red line. Note that cycles and pairs are identified at the moment they are destroyed, not created, and red dashed lines on panels $16$, $17$ and $18$ roughly indicate the shape of the $1$-cycles (\ie 1D loop) at the moment of their creation, for information. The pink plain and dashed circles indicate all nodes of the DMC that are concerned by the creation or destruction of a cycle at a given step. Finally, all the identified \hyperref[defperspair]{persistence pairs} are represented on the top right frame, in blue or red depending on their type. A detailed description of the computation of the \hyperref[defperspair]{persistence pairs} and \hyperref[defkcycle]{$k$-cycles} as shown on this figure is given in the main text (see second half of section \ref{sec_comp_pers}).}
\end{minipage}
\end{figure*}

We show on figure \ref{fig_cycle_search} the aforementioned practical example of \hyperref[defperspair]{persistence pairs} and corresponding $k$-cycles computation over a simple \hyperref[defMSC]{Morse-Smale complex}. The upper left frame shows a \hyperref[defDMC]{DMC} computed from a high resolution triangulation of the underlying density field $\rho$ (note that only the smooth function is represented, not the simplicial complex). As mentioned earlier, only the structure of the Morse-Smale complex is necessary to identify the cycles, so on these figures, we represented in the background the sub-levels sets of the density field $\rho$ instead of steps in the filtration $F_\rho$ to show how cycles are created and destroyed. We could identically have shown subsets of a simplicial complex, and actually, at this stage, we could equally say that the colored disks represent minima/critical points/maxima of the smooth field $\rho$ or critical vertexes/segments/triangles (\ie $0$/$1$/$2$-simplexes) of the discrete Morse function $F_\rho$.\\

A selection of $12$ steps corresponding to the entrance in the ascending \hyperref[deffiltration]{filtration} of $12$ of the $21$ critical simplexes are represented on the frames in the bottom part. The entrance of the first $8$ critical simplexes (blue disks) is not represented and the first displayed step, step $9$, corresponds to the entrance of the rightmost critical $1$-simplex (green disk). Note however that before step $9$, the critical vertexes (\ie minima) from $1$ to $7$ already entered creating each one component in the filtration, and critical segment $8$ also entered, destroying the $0$-cycle created by critical vertex $7$ which was merged with that of vertex $2$ (this destruction is still represented at step $9$ by the pink and red lines though). Considering critical segment $\sigma_1=9$, we first retrieve its two neighboring critical $0$-simplexes, labeled $4$ and $3$, and we therefore have $CurSet=\lbrace3,4\rbrace$. We first consider the highest, $\sigma^{cur}_{0}=4$, and check if there is a cycle associated to it in $cycles\left[\sigma^{cur}_{0}\right]$. As this is not the case, it means that we have found the cycle of $\sigma_1$, and therefore set $cycles\left[\sigma_{1}\right]=cycles\left[\sigma^{cur}_{0}\right]=CurSet=\lbrace3,4\rbrace$ and insert pair $\left[\sigma_1,\sigma^{cur}_{0}\right]=[9,4]$. On panels $9$, all the critical simplexes involved in the cycle are circled in pink, the pink arc connects the critical simplexes in the identified pair and the red line represents the cycle. For instance, in that case, we identified that critical segment $9$ destroys the component ($0$-cycle) created by critical vertex $4$, which results in the components created by critical vertexes $3$ and $4$ merging into each other (see the sub-level sets in the background).\\

Step $10$ is very similar to step $9$, with critical segment $\sigma_1=10$ entering, and we therefore add persistence pair $[10,6]$ to $ppairs$ and set $cycles\left[\sigma_{1}\right]=cycles\left[\sigma^{cur}_{0}=6\right]=\lbrace6,2\rbrace$. Step $11$ is skipped as it corresponds to the entrance of a positive critical vertex (\ie the creation of a new component), but step $12$ is more interesting. Critical segment $12$ is negative, and we therefore start the algorithm as previously by setting  $CurSet=\lbrace 7,1\rbrace$, its neighbor critical vertex on the DMC. The highest critical vertex in $CurSet$ is $\sigma^{cur}_0=7$, which was already paired at step $8$ (represented on panel $9$). We therefore add the cycle associated to it, $cycles[\sigma^{cur}_0=7]=\lbrace2,7\rbrace$, to $CurSet$, which gives $CurSet=\lbrace 7,1\rbrace+\lbrace2,7\rbrace=\lbrace1,2\rbrace$, as the addition is performed modulo 2 (CurSet is of type $\mathbb{Z}_2$-Set). The new highest critical vertex in $CurSet$ is therefore $\sigma^{cur}_0=2$, which is not paired yet. We therefore add the new pair $\left[12,2\right]$ to $ppairs$, and set $cycles\left[\sigma_{1}=12\right]=cycles\left[\sigma^{cur}_{0}=2\right]=CurSet=\lbrace1,2\rbrace$, which basically means that when critical segment $8$ enters, the component created by vertex $2$ merges into that of vertex $1$.  Steps $13$ and $14$ correspond to simple pairings (similar to step $9$), and step $15$ is similar to step $12$, as critical vertex $4$ is already paired, resulting in variables $ppairs$ and $cycles$ being updated according to the diagrams of panel $13$, $14$, and $15$.\\

The critical segment entering at step $16$ is different though, as it creates a $1$-cycle (symbolized by red dashed lines on panel $16$). Indeed, its neighbors critical vertexes on the DMC are $3$ and $5$, which already belong to the same component at step $16$ (as can be seen on the underlying sub-level set or on the DMC, by observing that the path $\left[5,13,2,8,7,12,1,15,4,9,3\right]$ only has critical simplexes with values below $16$). As this critical segment is therefore positive, we skip it for now, but we will see later how its cycle will be identified when it gets destroyed by a critical $2$-simplex. The following steps $17$ and $18$ are similar, and corresponding critical segments are skipped.\\

The first negative critical $2$-simplex enters at step $19$. Following algorithm \ref{alg_cycle}, we start with $CurSet=\lbrace 8,12,18,13\rbrace$, the four critical segments neighbors of the critical triangle $19$ on the DMC. The highest valued critical vertex in $CurSet$ is $\sigma^{cur}_{1}=18$, which is not yet paired, and we therefore add pair $\left[19,18\right]$ to $ppairs$ and set $cycles\left[\sigma_{2}=19\right]=cycles\left[\sigma^{cur}_{1}=18\right]=\lbrace8,12,18,13\rbrace$, represented by the red loop on panel $19$ (see also red dashed loop of figure $18$, when the cycle was created). This means that critical segment $18$ created a new $1$-cycle that was destroyed by critical triangle $19$, and this cycle is represented by the closed path formed by critical segments $\lbrace8,12,18,13\rbrace$, which are linked to each other through their neighbor critical vertexes in the DMC, $1$, $7$, $2$ and $5$ (the $1$-cycle is given by sequence $\left[18,5,13,2,8,7,12,1,18\right]$, which can be easily retrieved at query time from the information in $ccyles$). Critical triangle $20$ also destroys a $1$-cycle. The process is similar to previous step and variables are updates accordingly. We finally proceed to step $21$, where the last critical simplex enters. It is also negative (as all critical triangles are anyway), and we start with $CurSet=\lbrace 9,15,16,18\rbrace$. The highest critical segment is $\sigma^{cur}_1=18$, which is already paired to critical triangle $19$, and its cycles is therefore added modulo $2$ to $CurSet$, giving $CurSet=\lbrace 9,15,16,18\rbrace+cycles[\sigma^{cur}_1=18]=\lbrace 9,15,16,18\rbrace+\lbrace8,12,18,13\rbrace=\lbrace 9,15,16,8,12,13\rbrace$. As critical segment $16$, the highest in $CurSet$, is free, this means we are done identifying the last cycle. We therefore update variables $ppairs$ and $cycles$ accordingly and the algorithm terminates.\\ 

The upper right frame shows all of the \hyperref[defperspair]{persistence pairs} over the \hyperref[defDMC]{DMC}, and one can convince himself of the correctness of the cycles retrieved at steps $19$, $20$ and $21$ by comparing them to what they actually looked like in the sub-level sets when they were created, at steps $16$, $17$ and $18$ (see the dashed red cycles and newly created closed loops in the white iso-contours in the background). Although we do not show examples here, the method is strictly similar in higher dimensions. 

\subsection{Simplification}
\label{sec_simplification}

The relative importance of topological features can be reliably assessed using \hyperref[defpers]{persistence} theory, and it was briefly shown in section \ref{sec_persistence} how it is possible in the 1D case to locally modify a smooth function in order to cancel a low \hyperref[defperspair]{persistence pair} of critical points without affecting other critical points. Although it would also seem a viable option to directly modify $\rho$ in order to cancel non persistent pairs in the higher dimensions, this may not be the best thing to do. From a purely technical point of view, for large data sets, the computational cost of actually modifying $\rho$ and recomputing the \hyperref[defMSC]{Morse-Smale complex} every time would be excessive. From a theoretical point of view, one would need to arbitrarily define a more or less natural way to smoothly transform $\rho$ so that the canceling pairs would disappear without affecting the remaining critical points. Fortunately, such a transformation does not need to be explicitly conducted and it is enough to know that it exists and how it affects the \hyperref[defMSC]{Morse-Smale complex}.\\

\begin{figure}
\centering
\includegraphics[width=8cm]{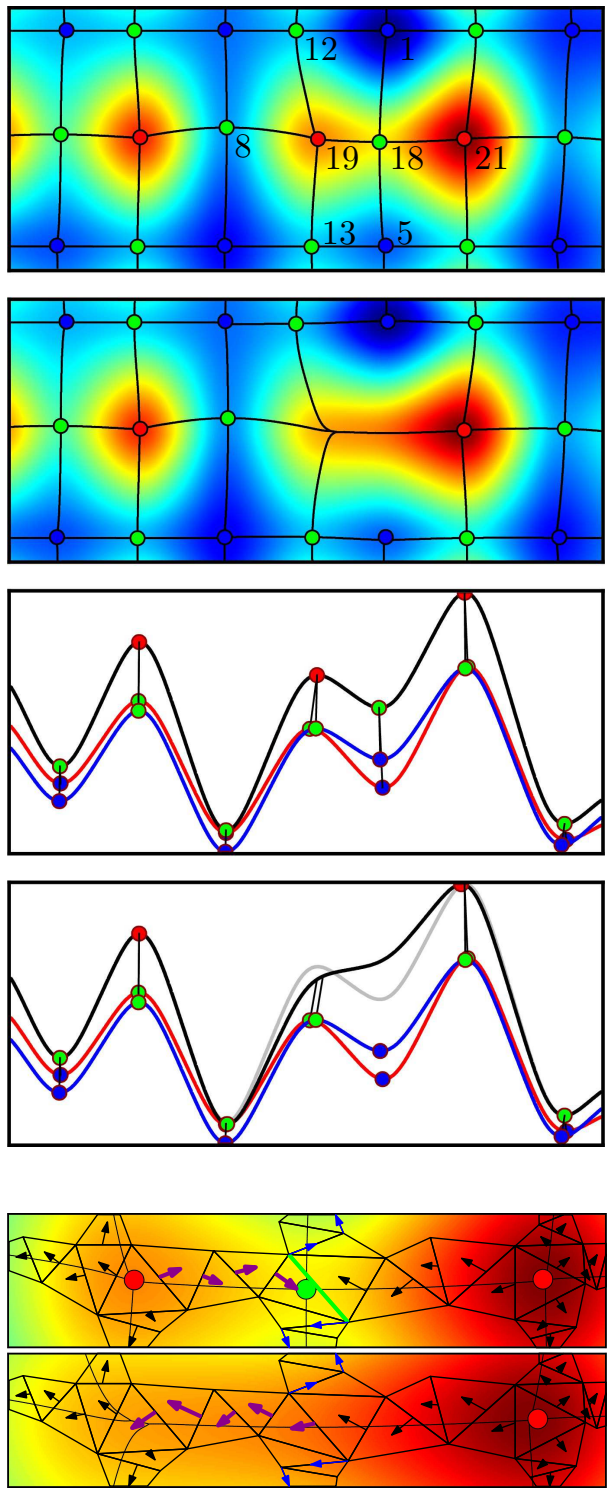}
\caption{ Topological simplification of a maximum and saddle point \hyperref[defperspair]{persistence pair} in the smooth 2D field $\rho$ of figure \ref{fig_cycle_search}. On the upper part, from top to bottom, the four frames display the Morse complex before and after simplification and the corresponding density profiles along the three horizontal axis of the Morse complex (red, black and blue for the upper, middle and lower axis respectively). The density profile before simplification is represented in gray. The lowest frame shows an equivalent cancellation of critical pairs in a discrete Morse complex by \hyperref[defDG]{discrete gradient} reversal. Note that non essential gradient pairs and simplexes have been omitted for clarity, as they are not affected by the path reversal. \label{fig_cycle_simp}}
\end{figure}

As a simple example, an arbitrary modification\footnote{Note that achieving the modification shown in this example was actually made easy by the fact that the function itself is analytically defined as a sum of Gaussian functions, but it would clearly have revealed much more challenging in the general case.} of the smooth density field of figure \ref{fig_cycle_search} that cancels the low \hyperref[defperspair]{persistence pair} $\ppair{18}{19}$ is presented on the top frames of figure \ref{fig_cycle_simp}. As expected, this modification of $\rho$ leads to the removal of the saddle-point and maximum, and a particular reorganization of the arcs in the Morse Complex. Before cancellation, the saddle point $18$ was linked to two minima ($1$ and $5$) and two maxima ($19$ and $21$). With its removal, the arcs emanating from the minima get rerouted to maximum $21$ (as maximum $19$ is also canceled in the operation), and they are therefore not critical anymore: they are removed from the Morse complex. The situation is different for saddle points $8$, $12$ and $13$ though, which were linked to maximum $19$. During the cancellation, the gradient of $\rho$ is reversed between the canceled points (see lower panels), and the arcs that led to maximum $19$ are therefore free to continue their ascension up to the former position of the canceled saddle point, and further along the arc $\left[18,21\right]$, to finally reach maximum $21$. Those field lines still link saddle-points to maxima, they are critical and therefore belong to the new modified Morse complex.\\

The field lines reorganization scheme during a cancellation can be intuitively understood in the general case by defining a similar minimalistic transformation of a \hyperref[defDMF]{discrete Morse function} and its \hyperref[defDG]{discrete gradient}. Basically, the essential feature lies in the reversal of the gradient path between the canceled critical points. Such an operation can easily be defined over a \hyperref[defDG]{discrete gradient} \citep{forman}. The corresponding modification is shown on the bottom frame of figure \ref{fig_cycle_simp}, in the case of a \hyperref[defDMF]{discrete Morse function} similar to $\rho$ and defined over a tessellation: by reversing the path of \hyperref[defDG]{discrete gradient} arrows between the critical points (purple shade), the two critical points are effectively removed while the rest of the \hyperref[defDG]{discrete gradient} is left unmodified, and it is easy to predict the consequences of this modification on the discrete Morse complex. Let us call $\sigma_k$ and $\sigma_{k+1}$ the critical $k$ and {\kp1}-simplexes to cancel, and $\alpha^i_{k+1}$ and $\beta^j_{k}$ the critical $k+1$ and \hyperref[defksimplex]{$k$-simplexes} respectively linked to $\sigma_k$ and $\sigma_{k+1}$ by an arc in the \hyperref[defDMC]{DMC}. By reversing the gradient path between $\sigma_k$ and $\sigma_{k+1}$, one also reroutes all the arcs and \hyperref[defmanifold]{manifold}s that previously reached one of those critical simplexes. After cancellation, an ascending arc emanating from $\beta^j_{k}$ still reaches the formerly critical simplex $\sigma_{k+1}$, and it can be extended through the reversed path and continue following any previously ascending arc emanating from $\sigma_k$, leading to a critical {\kp1}-simplex $\alpha^i_{k+1}$. Similarly, any descending {\kp1}-\hyperref[defmanifold]{manifold} of $\alpha^i_{k+1}$ now reaches $\sigma_{k+1}$ and can therefore be extended by the descending {\kp1}-\hyperref[defmanifold]{manifold} of $\sigma_{k+1}$. For the same reason, the ascending $\kmn[d]{k}$-\hyperref[defmanifold]{manifold}s of $\beta^j_{k}$ can be extended by the ascending $\kmn[d]{k}$ \hyperref[defmanifold]{manifold}s of $\sigma_{k+1}$. One therefore does not need to actually perform any gradient path reversal, and the cancellation of critical pair $\left[\sigma_k,\sigma_{k+1}\right]$  is achieved directly on the \hyperref[defDMC]{DMC} using the following procedure:
\begin{enumerate}
\item Let $\alpha^i_{k+1}$ and $\beta^j_{k}$ be the $N_\alpha$ and $N_\beta$ critical ${k+1}$ and $k$ critical simplexes sharing and arc in the \hyperref[defDMC]{DMC} with $\sigma_k$ and $\sigma_{k+1}$ respectively. 
\item Create a new arc between each of the $N_\alpha*N_\beta$ pair $\left[\alpha^i_{k+1},\beta^j_{k}\right]$ by joining arcs $\left[\alpha^i_{k+1},\sigma_k\right]$, $\left[\sigma_k,\sigma_{k+1}\right]$, and $\left[\sigma_{k+1},\beta^j_{k}\right]$. The path $\left[\sigma_k,\sigma_{k+1}\right]$ must be reversed during the operation.
\item Extend the descending  \hyperref[defmanifold]{manifold} of each $\alpha^i_{k+1}$ with the descending \hyperref[defmanifold]{manifold} of $\sigma_{k+1}$.
\item Extend the ascending \hyperref[defmanifold]{manifold} of each $\beta^j_{k}$ with the ascending \hyperref[defmanifold]{manifold} of $\sigma_{k}$.
\item Delete the critical simplexes $\sigma_{k}$ and $\sigma_{k+1}$, together with their $4$ ascending and descending \hyperref[defmanifold]{manifold}s and all of the arcs leading to or emanating from them.
\end{enumerate} 
It is important to remark that in general, the simplification of a pair may lead to an increase in the total number of arcs in the complex. This is particularly true when none of the canceling simplex is a $1$-saddle or a $D-1$ saddle, as in that case, the number of arcs is not bounded. Moreover, there exist two specific cases where a cancellation is impossible. The first is when critical simplexes do not share an arc in the \hyperref[defDMC]{DMC}. The second is when they share more than one arc, as in that case, gradient path reversal would lead to the creation of a looping path in the \hyperref[defDG]{discrete gradient}, which is forbidden (see section \ref{sec_DMtheory}). The detailed procedure to deal with this is explained in section \ref{sec_det_imp}.\\

\subsection{Filtering Poisson noise}
\label{sec_noise}
As mentioned previously, mainly because of Poisson noise, it is not possible to use the raw \hyperref[defDMC]{DMC} to identify structures in the cosmic web. In fact,  most of the critical points, arcs and \hyperref[defmanifold]{manifold}s are actually spurious artifacts created by sampling noise. This is especially true in the present case, where we wish to use DTFE density and a simplicial complex computed from the Delaunay tessellation of a discrete realization. As a matter of fact, the scale free nature of DTFE makes it locally very sensitive to Poisson noise, as information is always locally extracted at the sample resolution limit. Our approach to remedy this problem consists in computing a significance level for each \hyperref[defperspair]{persistence pair}, and canceling the \hyperref[defperspair]{persistence pairs} whose significance is below a given threshold.\\

\begin{figure}
\centering
\includegraphics[width=\linewidth]{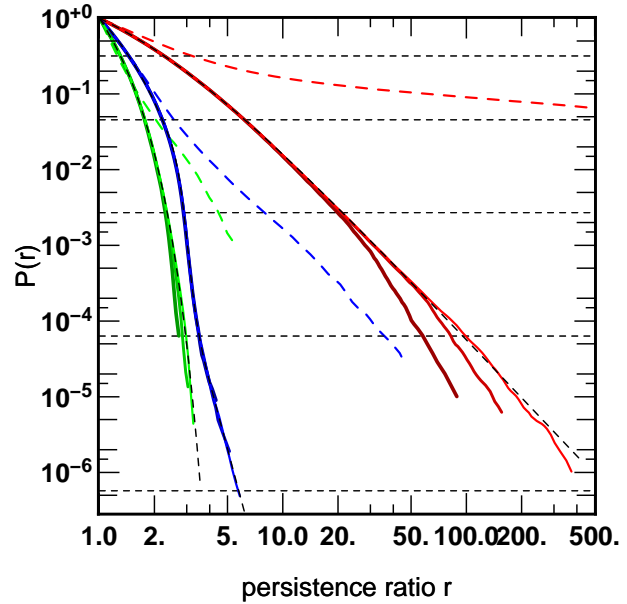}
\caption{The cumulative probability $P_k\left(r\right)$ that a \hyperref[defperspair]{persistence pair} of order $k$ with \hyperref[defpers]{persistence} ratio greater or equal to $r$ exists in a 3D scale free Gaussian random field (colored plain curves) and in a $50h^{-1}$ Mpc dark matter cosmological simulation (colored dashed curves). The red, blue and green curves correspond to maxima/$1$-saddle, $1$-saddle/$2$-saddle and $2$-saddle/minima pairs respectively. The different shades, from darker to lighter,  correspond $64^3$, $128^3$ and $192^3$ particles realizations respectively. The black dashed curves show fits to the Gaussian case, as presented in the main text, while the horizontal dashed lines correspond to different significance levels in units of ``sigma'', ranging from $S=\nsig{1}$ (top) to $S=\nsig{5}$ (bottom).\label{fig_ratio_proba}}
\end{figure}

Let $r$ be the \hyperref[defpers]{persistence} ratio of a \hyperref[defperspair]{persistence pair} $q_k=\ppair{\sigma_k}{\sigma_{k+1}}$, then 
\begin{equation}
r\left(q_k\right)=F_\rho\left(\sigma_{k+1}\right)/F_\rho\left(\sigma_{k}\right).
\end{equation} 
We note $P_k\left(r_0\right)$ the cumulative probability that a \hyperref[defperspair]{persistence pair} of critical simplexes of order $k$ and $k+1$ and with \hyperref[defpers]{persistence} ratio $r\geq r_0$ exists in the Delaunay tesselation of a random discrete Poisson distribution. It is then convenient to denote the relative importance of a given critical pair $q_k$ in terms of it significance, $S\left(q_k\right)$, expressed in units of ``sigmas'' with analogy to the Gaussian case:
\begin{equation}
 S\left(q_k\right) = S_k\left(r\left(q_k\right)\right)=\mbox{Erf}^{-1}\left(\frac{P_k\left( r\left(q_k\right)\right)+1}{2}\right),
\end{equation}
where $\mbox{Erf}$ is the error function. As a purely analytical derivation of $P_k\left(r\right)$ seems clearly out of reach, we use Monte-Carlo simulation to estimate it, measuring $P_k\left(r_0\right)$ as the fraction of \hyperref[defperspair]{persistence pairs} of order $k$ with \hyperref[defpers]{persistence} ratio $r\geq r_0$ in a Poisson sample. The results are shown on figure \ref{fig_ratio_proba}. On that figure, the values of  $P_k\left(r\right)$ is plotted as a function of $r$ in green, blue and red for $k=0$, $k=1$ and $k=2$ respectively and the horizontal dashed lines represent different significances level in units of ``sigma'', ranging from $S=\nsig{1}$ (top) to $S=\nsig{5}$ (bottom). From these results, the following fits can be extracted in the 3D case (represented as black dashed lines on figure \ref{fig_ratio_proba}):
\def\u {\left(r-1\right)}
\allowdisplaybreaks
\begin{eqnarray}
P_0\left(r\right) & = & \exp\left(-\alpha_0\u-\alpha_1\u^{\alpha_2}\right)\\
\nonumber & {\rm with }&\alpha\approx\left[3.694,0.441,2.538\right],\\
\eqvspace
P_1\left(r\right) & = & f_1\cdot (1-t) + f_2 \cdot t \\
\nonumber & {\rm with }&f_1=\exp\left(-\beta_0\u\right)\,,\quad
f_2=\beta_1r^{-\beta_2}\\
\nonumber &&t=\left(1+\beta_3/u^{\beta_4}\right)^{-1}\\
\nonumber &&\beta\approx\left[2.554,4.000,9.000,1.785,14.000\right],\\
\eqvspace
P_2\left(r\right) & = & \left(1+\gamma_0\u\right)^{-\gamma_1}\\
\nonumber & {\rm with }&\gamma\approx\left[0.449,2.563\right],
\end{eqnarray}
and in the 2D case, we obtain:
\allowdisplaybreaks
\begin{eqnarray}
S^{2D}_0\left(r\right) & = & \exp\left(-\alpha_0\cdot \u-\alpha_1\cdot\u^{\alpha_2}\right)\\
\nonumber & {\rm with }&\alpha\approx\left[2.00,0.01,3.50\right],\\
\eqvspace
S^{2D}_1\left(r\right) & = & \u^{-\beta_0\left(1+\beta_1\log\u\right)}\\
\nonumber & {\rm with }&\beta\approx\left[0.75,0.20\right].
\end{eqnarray}
A relatively similar approach was undertaken in a code named ZOBOV \citet{neyrinck} to measure the significance of cosmological voids. The approach developed in this article nevertheless differs from ours in that it is limited to voids and that they do not use \hyperref[defperspair]{persistence pairs}. Instead, they pair minima of the density field to the lowest $1$-saddle point on the surface of their ascending $3$-\hyperref[defmanifold]{manifold}s (\ie the voids themselves) that is not already paired to another minimum with higher density. This explains why our fit of $P_0\left(r\right)$ differs from theirs.\\

\begin{figure*}
\centering\includegraphics[width=\linewidth]{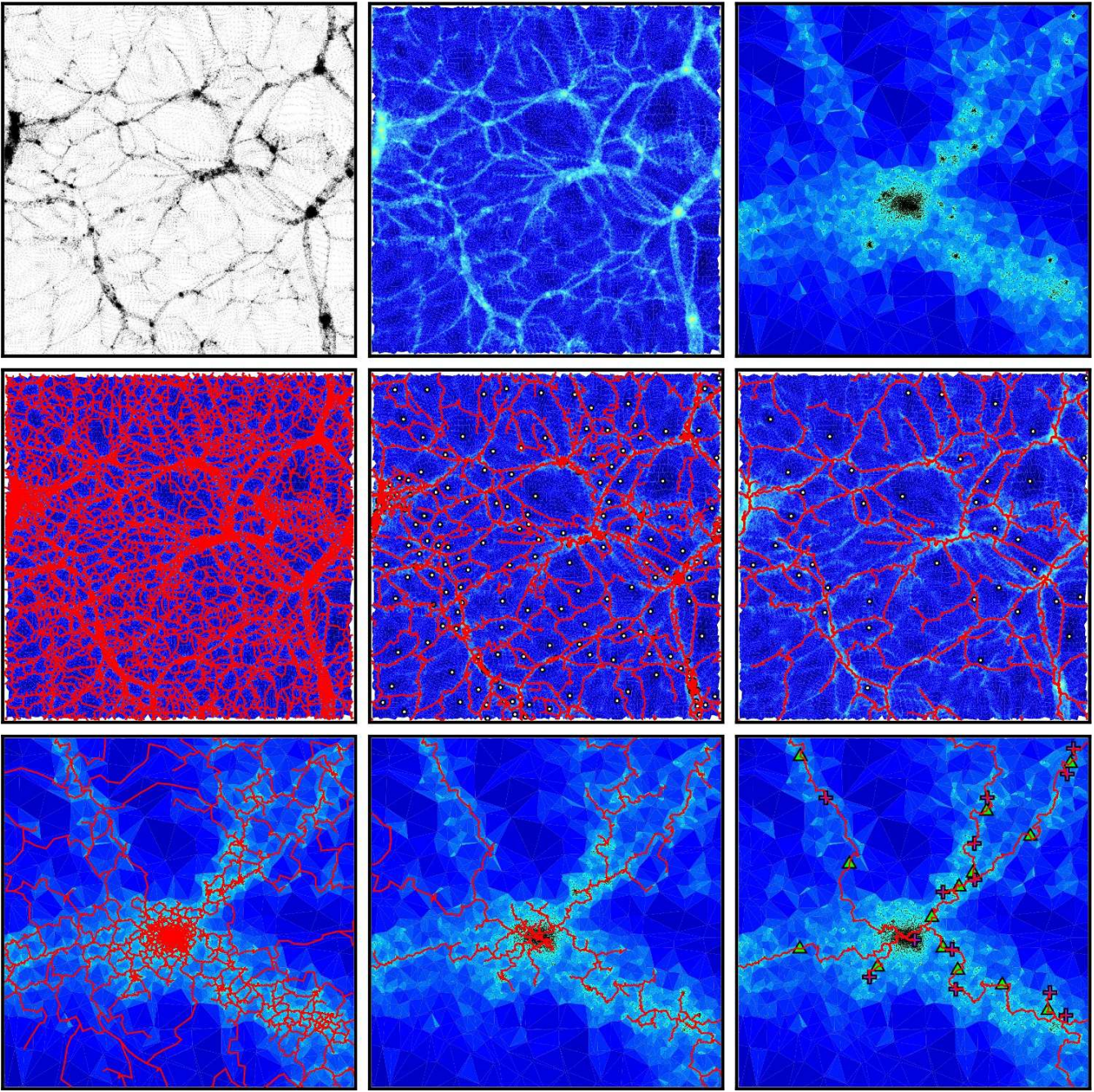}
\caption{The filaments measured in a 2D distribution obtained by projecting the particles from a slice of an N-Body cosmological simulation. The initial discrete distribution, its Delaunay tesselation and a zoom on a Halo (from the upper central part of the distribution) are displayed on the top raw, with colour corresponding to the DTFE density. The filamentary structure is traced in red on the middle raw, as the geometry of the arcs remaining after cancellation of \hyperref[defperspair]{persistence pairs} with significance less than $\nsig{0}$ (middle left), $\nsig{2}$ (center) and $\nsig{4}$ (middle right). A zoom around a projected Halo is shown on the bottom raw. The white disks, green triangles and purple crosses stand for the minima, saddle-points and maxima respectively (notes that they only represented on some panels for clarity). \label{fig_simu2D_persistence}}
\end{figure*}

The fact that the expression of the fits for $k=0$ and $k=2$ is relatively simple compared to the one for $k=1$ may seem intriguing at first sight. But if the fit for function $P_1\left(r\right)$ actually requires more coefficients, it is mainly because it undergoes some sort of transition around $r=4$, which roughly corresponds to a significance level of $\nsig{3.5}$. We believe that this only reflects the nature of DTFE itself, whose probability distribution function is clearly biased toward the high densities as the number of minima is limited by the comparatively larger volume of the Voronoi cells they occupy (see \cite{DTFE}). As the size of our Monte-Carlo sample is limited, just like this results in an increase of the number of $k=1$ type pairs when fewer and fewer comparatively lower density minima become available to form pairs. We also note that this tendency is present in the cumulative probabilities of \hyperref[defperspair]{persistence pairs} ratio measured in cosmological simulations as well (colored dashed curves). Nevertheless, those probabilities are significantly higher than that in the Poisson sample for any value of $k$ and it therefore seems that it should be reasonably easy to filter spurious \hyperref[defperspair]{persistence pairs} without affecting too much those storing precious information on the cosmic web topology.
\subsection{Illustration in 2D}
Figure \ref{fig_simu2D_persistence} shows the \hyperref[defDMC]{DMC} of a 2D discrete distribution of $\sim 350,000$ particles with periodic boundary conditions computed at different levels of significance. The discrete distribution (upper left) was obtained by projecting a sub-sampled $10\Mpc$ slice of a $50\Mpc$ large dark matter N-Body simulation at redshift $z=0$. The resulting delaunay tesselation, composed of $\sim 1,000,000$ $1$-faces and $670,000$ $2$-faces, and corresponding DTFE density field are shown on the upper central and upper right panels. Note that identifying the filamentary structure in such a distribution is particularly challenging because of its very high dynamic range and also because many filaments simply disappear into low density regions as they leave the slice in the original 3D distribution. The filamentary structure captured by the \hyperref[defDMC]{DMC} is depicted in red on the central and bottom rows through the representation of its ascending $1$-\hyperref[defmanifold]{manifold}s, after cancellation of the \hyperref[defperspair]{persistence pairs} at a significance level of $\nsig{0}$ (left, no simplification), $\nsig{2}$ (center) and $\nsig{4}$ (right). The central left panel nicely illustrates the strong influence of Poisson noise, as without simplification, filaments are basically detected almost everywhere in the distribution. This is particularly obvious when zooming on what was a dark matter halo in the former 3D distribution: whereas one can identify by eye a few obvious filaments connecting to the central clump, the algorithm (correctly) detects a swarm of local peaks and filaments locally created by random fluctuations in the distribution.\\

 It is quite striking though how much applying the above described \hyperref[defpers]{persistence} based simplification procedure succeeds at selecting what one would intuitively define as a filament. Already, at a $\nsig{2}$ level (central and middle bottom frame), it is clear that the large scale network of filaments is correctly identified as well as the valley resulting from the projected cosmic voids of the non projected distribution (the ascending $2$-\hyperref[defmanifold]{manifold}s associated to the minima, symbolized by the white disks). At a level of $\nsig{2}$, the probability that a topological feature such as an arc in the \hyperref[defDMC]{DMC} is the result of Poisson noise is $\sim 5\%$. At $\nsig{4}$, this probability goes down to $\sim 0.006\%$ and any arc in the \hyperref[defDMC]{DMC} can therefore safely be considered a feature of the underlying distribution. The lower right panel shows the arcs of the \hyperref[defDMC]{DMC} that link maxima (purple crosses) and saddle points (green triangles) at a significance level of $\nsig{4}$ around the projection of a large dark matter halo. At that level, the intricate initial network is reduced to a very neat set of filaments branching on a central clump. Note that while the network itself is simplified, the resolution is preserved which for instance allows for the correct identification of the merger of two relatively noisy filaments on the top right corner while preserving a very clean network on large scale (central right frame).\\

The application to 3D distribution and in particular galaxy catalogues and large scale N-body simulation is presented in the companion paper, \papertwo.

\section{Boundary conditions and technicalities}
\label{sec_det_imp}

\subsection{Boundary conditions}
\label{sec_boundary}
Whereas boundary conditions are not a concern in Morse theory, as it is defined over infinite or borderless spaces, things are clearly different when on tries applying it to real data sets. The easiest case corresponds to periodic boundary conditions data, such as those encountered in N-body simulation of matter distribution on large scales in the Universe. Because it is impossible to simulate the whole Universe and gravity has an infinite outreach, periodic boundary conditions are often used as a trick to obtain a smooth gravitational potential and emulate the isotropy of space within a restricted volume usually shaped as a box. Enforcing periodic boundary conditions over a cube basically amounts to assimilating opposite faces, any object leaving the cube through one face immediately entering the opposite one. Mathematically speaking such a space is called a torus $\torus^d$, where $d$ is the number of dimensions, and Morse theory readily applies to such spaces.  From a practical point of view, we use the periodic exact 3D periodic boundary conditions Delaunay tessellation \citep{CGAL_PB} implemented in CGAL\footnote{CGAL is the C++ Computational Geometry Algorithm Library, see {\em\tt http://www.cgal.org}} when the distribution is three dimensional. We also implemented our own periodic boundary conditions within CGAL for $d\neq 3$ case using a simpler, though less rigorous and optimized, technique. This method basically consists in building a larger distribution by replicating a fraction of the box to extend each boundary, computing the delaunay tesselation over this extended domain and then identifying the identical $k$-faces crossing opposite faces of the initial box (the Delaunay tesselation of figure \ref{fig_simu2D_persistence} was obtained using such method).\\

Of course, periodic boundary conditions only apply to periodic data sets, which is usually not the case of observational data, and one therefore needs to treat the boundaries of the distribution with special care. The simplest way to do so consists in transforming the definition domain of the data set into a boundaryless domain, a procedure called compactification. Usual compactification techniques consist in transforming the definition domain into a sphere by adding a point at infinity and attaching it to each boundary cell of the delaunay tesselation or transforming it into a torus, practically making it periodic by replicating a mirror image of the distribution through its boundaries. Both of these methods have pros and cons. Whereas sphere compactification is easy to build, whatever the geometry of the initial data set, it tends to pollute measurements around the boundaries by affecting the \hyperref[defDG]{discrete gradient} computation, therefore creating numerous fake manifolds and arcs that have to be ignored. This is not the case of the torus compactification, which creates relatively smooth conditions close to the boundaries, but it may only be easily implemented on cubic boxes and requires replicating the data set a large number of times (27 times in 3D), significantly increasing the necessary computational time and resources accordingly.\\

\begin{figure}
\centering
\includegraphics[width=\linewidth]{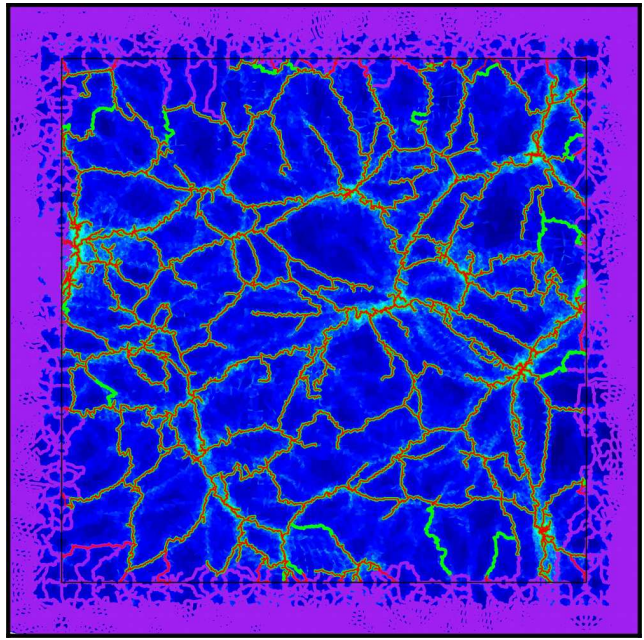}
\caption{Illustration of the computation of the \hyperref[defDMC]{DMC} with non periodic boundary condition. The boundary of the initial 2D distribution are delimited by the thin black square, and any particle in the distribution within a distance of $10\%$ the initial domain size is mirrored (see also figure \ref{fig_simu2D_persistence}). The thick green and purple network shows the $\nsig{3}$ filaments in measured in the non-periodic distribution, the purple part being discarded after topological simplification as belonging to the boundary. The filaments obtained in the periodic boundary situation are displayed in red for comparison.\label{fig_boundary_2D}}
\end{figure}

Our implementation of the boundary conditions is an hybrid between the sphere and torus compactification that tries to preserve the advantages of both while getting rid of their limitations. The idea is that the torus compactification is efficient because of the relatively natural extrapolation of the density field it allows, which as a result does not affect the computation of the \hyperref[defDG]{discrete gradient} at large distance from the boundary. We therefore allow the user to choose what fraction of the distribution should be mirrored on each boundary (a value around $10\sim 15\%$ of the initial distribution size seems to work fine), and then apply a sphere compactification on the new distribution by adding a point at infinity, with minus infinite density, that forms simplexes with the new boundary of the enlarged distribution. We then proceed by tagging as ``boundary'' any \hyperref[defksimplex]{$k$-simplex} of the Delaunay tesselation that contains a replicated vertex, the infinite vertex, or whose DTFE density may have been affected by the distribution outside the definition domain. This last condition in fact prevents the boundary simplexes, whose DTFE density may be wrong, to affect the resulting \hyperref[defDMC]{DMC} and we determine which simplexes may be affected by checking weather the circumsphere of each highest dimensional $d$-simplex intersects the boundary, in which case it is, with all its faces and vertice, tagged as ``boundary''.\\

 Note that one has to pay particular attention to the boundaries during the topological simplification process as the \hyperref[defpers]{persistence} of critical pairs formed with a boundary simplex have spurious \hyperref[defpers]{persistence} ratios. The point at infinity is special, it has a minus infinite density, and is allowed to form \hyperref[defperspair]{persistence pairs} with as many vertice as necessary, any of those \hyperref[defperspair]{persistence pair} having infinite \hyperref[defpers]{persistence}. The \hyperref[defperspair]{persistence pairs} formed between the non infinite boundary simplexes and those within the valid part of the distribution are treated normally during the simplification process simplification, but the topological feature they form are nevertheless spurious and any \hyperref[defperspair]{persistence pair} with at least one critical simplex on the boundary is therefore deleted\footnote{we really mean {\em deleted} in that case, and not canceled as a regular pair would be} after topological simplification. The whole process is illustrated on figure \ref{fig_boundary_2D} in the 2D case, using the same distribution as that of figure \ref{fig_simu2D_persistence}. On that picture, the filamentary structure detected at $\nsig{3}$ is represented for the same distribution when it is considered periodic (thin red network) and non-periodic (thick green network), and it is clear that both mostly agree. One can see nonetheless that, as should be expected, the small portions of the (red) periodic network crossing the boundaries cannot be detected in the non-periodic case and that a few portions of (green) filaments lying slightly farther away are detected only in the non-periodic case. This results from the fact that \hyperref[defperspair]{persistence pairs} of distant critical simplexes may be different in the non-periodic distribution because the $k$-cycles coupling them are not allowed to cross the boundaries. As a result, the \hyperref[defpers]{persistence} ratios of certain critical points may differ in both cases, and they may therefore still exist at $\nsig{3}$ level in the non-periodic case while they were canceled at $\nsig{2.5}$ level in the periodic one.

\subsection{Smoothing the manifolds}
\label{sec_smoothing}
Because the scale resolution of practical samples is always limited, so is the resolution of the ascending and descending \hyperref[defmanifold]{manifold}s of the \hyperref[defDMC]{DMC}. When identifying the filaments, voids or walls in cosmological distributions, their precise shape therefore becomes arbitrary at scales lower than the initial sampling resolution. Within our implementation, the \hyperref[defDMC]{DMC} features are sub-sets of the initial Delaunay tesselation or of its dual Voronoi tesselation and the identified structures therefore naturally tend to adapt to the measured sample much better than they would if one was using a regular sampling grid for instance. The \hyperref[defDMC]{DMC} is nevertheless always computed at the sampling resolution limit, and its geometry is mainly dictated by Poisson noise on that scale. It may therefore be desirable to have a way to enforce some continuity and differentiability even at the price of a loss of resolution (\eg for instance for representation purposes). The smoothing method that we use is pretty much the same as that presented in \citet{rsex} as it presents the advantages of being simple, robust and fast. The idea involves smoothing the filaments individually by fixing the critical points and averaging the position of each non-fixed segment's endpoint with the position of its closest neighbouring endpoints a given number of times. Within our implementation, a filament is defined as a sequence of $N$ linked vertexes, the vertexes corresponding to the centers of mass of simplexes in the Delaunay tesselation. Let $x^i_j$  be the $j^{th}$ coordinate of $i^{th}$ vertex. Then, after smoothing, its new coordinates $y^i_j$, are computed as:
\begin{equation}\label{eq_smoothing}
y^i_k = A^{ij} x^j_k,
\end{equation}
with
\begin{equation}
A^{ij}= \left\lbrace
\begin{array}{l}
3/4\quad{\rm if }\quad i=j=0\quad{\rm or }\quad i=j=N, \\
1/2\quad{\rm if }\quad i=j,\\
1/4\quad{\rm if }\quad i=j+1\quad{\rm or }\quad i=j-1, \\
0\quad{\rm elsewhere,}\\
\end{array}
\right.
\end{equation}
where equation (\ref{eq_smoothing}) is applied $s$ times in order to smooth over $s$ simplexes in the simplicial complex. The corresponding smoothing length is naturally adaptive and such a smoothing ensures continuity of the filaments location over $s$ Delaunay simplexes. Note that it is very easy to adapt this method to the ascending and descending \hyperref[defmanifold]{manifold}s of the \hyperref[defDMC]{DMC} (\ie the voids, walls, ...) as any of them is  defined as a simplicial complex within our implementation. The position of each vertex in a manifold can therefore similarly be averaged with that of its neighbors $s$ times to obtain sufficient smoothness.

\subsection{Essential implementation issues}
\label{sec_optim}
Finally, we close this section by presenting two technical issues that are essential to a practical implementation of the algorithm.

\subsubsection{Cancellation order}
When canceling \hyperref[defperspair]{persistence pairs}, the order in which the pairs are canceled has a crucial importance, both in terms of computational time and memory consumption. This is especially true in $3$D. In fact, whereas the number of arcs linking a given $1$-saddle (resp. $2$-saddle) and a maximum (resp. minimum) is always $2$, there is no bound on the number of arcs between two saddle points of different types. Following the arc redirection algorithm described in section \ref{sec_simplification}, the cancellation of two saddle-points of different type may therefore lead to a dramatic increase in the total number of arcs in the complex. Let $P$ and $Q$ be the $1$-saddle and $2$-saddle respectively. Then $P$ is linked to $P_\uparrow=2$ maxima and $P_\downarrow$ $2$-saddles, while $Q$ is linked to $P_\uparrow$ $1$-saddles and $P_\downarrow=2$ minima. The cancellation therefore creates $N_C =\left(P_\downarrow-1\right)\left(Q_\uparrow-1\right)$ arcs and destroys $N_d=2+2+P_\downarrow+Q_\uparrow-1$ arcs, and the number of additional arcs after cancellation is $N=N_c-N_d\propto P_\downarrow Q_\uparrow$ for large values of $P_\downarrow$ and $Q_\uparrow$. This means that the number of arcs in the complex may temporarily increase quadratically, and it is not uncommon to obtain saddle points with hundreds of thousands of arcs at a given moment\footnote{This does not means that hundreds of thousands of arcs will be present in the simplified complex though, as a single maximum/$1$-saddle or a minimum/$2$-saddle may later cancel all those arcs leading to a dramatic decrease in the total number.}. Within our implementation, we therefore always cancel the pair $\lbrace P,Q \rbrace$, with $P$ the critical point of highest type, that minimize the number $N$ of created arcs first, with $N=N_c-N_d=\left(P_\downarrow-1\right)\left(Q_\uparrow-1\right)-P_\downarrow-P_\uparrow-Q_\downarrow-Q_\uparrow+1$.\\

\subsubsection{Impossible cancellations}
There exist special configurations where two critical points are linked by two or more different arcs (think for instance of the circular crest around the crater of a volcano). Those particular configurations cannot be canceled, as applying a \hyperref[defDG]{discrete gradient} reversal (see  section \ref{sec_simplification}) would result in the formation a \hyperref[defVP]{V-path} (\ie discrete integral line) that loops onto itself; this is impossible as a \hyperref[defVP]{V-path} is a strictly decreasing alternating sequence of \hyperref[defksimplex]{$k$-simplexes} (see definition \ref{def_vpath}). This is not a problem for maximum/$1$-saddle and minimum/$2$-saddle cancellation though, as such \hyperref[defperspair]{persistence pairs} cannot be formed if the the critical points are linked by more than one arc (taking the example of the volcano, the highest point on its crest is a maximum, which are always positive (creating), and this is also the case of the lowest point on the crest which is a positive saddle point, as it creates the ring formed by the crest around the volcano). Yet the $3$D case of a $1$-saddle/$2$-saddle \hyperref[defperspair]{persistence pair} is different, as nothing prevents such configurations to occur. In practice, such configuration do not arise naturally and we noticed that using the order for canceling pairs  defined previously  drastically reduces the number of occurrence of such non-cancellable configurations (of order $\sim 10$ for a $128^3$ particles simulation cut at $\nsig{4}$). For those few remaining pairs, we offer the possibility in our implementation to skip them or force their removal after only keeping one of the arcs between the critical points within the \hyperref[defperspair]{persistence pair}. This last option is the preferred one, and although it seems difficult to justify from a theoretical point of view, the fact that the occurrence of non-cancellable pairs depends on the precise cancellation order suggests that it is acceptable to do so (note that the consequences on the resulting \hyperref[defMSC]{Morse-Smale complex} are quite minimal anyway).\\   

\section{Conclusion}
\label{sec_conclusion}

We presented a method that allows the scale-free and parameter-free coherent identification of all types of 3D astrophysical structure in potentially sparse discretely sampled density fields such as N-body simulations or observational galaxy catalogues. The method is based on Morse theory (section \ref{sec_morse}), {\em discrete} Morse theory (section \ref{sec_DMtheory}) and \hyperref[defpers]{persistence} theory (section \ref{sec_persistence}), and the  implementation of the corresponding algorithm was detailed in sections \ref{sec_implementation}, \ref{sec_toposimp} and \ref{sec_det_imp}. In particular, our specific algorithm was designed with astrophysical applications in mind, as it directly applies to the delaunay tesselation of point set samples\footnote{ in fact, the algorithm can also be used directly over structured regular meshes and we implemented a version that works directly on a regular grid.}, and we paid a particular attention to the computation of the \hyperref[defDMF]{discrete Morse function} so that it correctly represents the underlying DTFE density. From this \hyperref[defDMF]{discrete Morse function}, \progname basically computes the discrete Morse Smale complex of the density function and uses it to identify structures: the ascending $3$, $2$, $1$ and $0$ \hyperref[defmanifold]{manifold}s of the theory being identified to the voids, walls, filaments and clusters respectively. The implementation was designed so that each component of the cosmic web and its geometry can be easily identified and studied as individual objects or as group of objects and so that their relationship can be easily recovered: one can for instance identify the voids bordering a given wall or the clusters at the extremities of a given filament. Moreover, as the \hyperref[defpers]{persistence} criteria was re-casted in terms of confidence level with respect to noise, it make \progname very easy to use,  as it is the only parameter  required to identify structure at optimal resolution. It shows a great deal of potential for astrophysical applications, for the following reasons that distinguish it from traditional methods:
\begin{enumerate}
\item It applies directly to discrete data sets via their Delaunay tessellation, which makes it scale free and allows the identified structures to always be defined down to the resolution limit of the sample.
\item It is based on {\em discrete} Morse theory, which means that, as opposed to methods based on {\em smooth} Morse theory, the mathematical formalism does apply rigorously to the type of data sets one usually have to deal with in astrophysics. This implies that the well studied formalism of Morse theory readily applies to the numerically identified structures (which is not the case of watershed based methods for instance, see appendix \ref{sec_appwatershed}). 
\item All the different types of structures are defined  coherently: triangulated space can basically be divided into sets of volumes, surfaces, curves and points that correspond to voids, walls, filaments and clusters respectively. Each structure is identified {\em individually}, and the cosmic web can therefore for instance be rigorously divided into individual filaments, each corresponding to a given saddle-point.
\item It readily takes into account sampling and Poisson noise via \hyperref[defpers]{persistence} theory, allowing the user to define a detection confidence level in term of ``number of sigmas'' and provides the corresponding simplification of the DMC. As shown in \papertwo, this fact actually produces results obtained in highly sampled simulations and sparse galaxy catalogues which are qualitatively very similar, opening the way to a direct comparison of the properties of  the cosmic web in simulations and observational catalogues.  
\item Because the foundation of the method is based on topology and uses \hyperref[defpers]{persistence} theory, it also allows for a very robust computation of topological invariants such as Betti numbers or the Euler characteristic; this is possible even in the presence of an important shot noise, and without having to define any smoothing scale; it therefore takes into account the truly multi-scale nature of the cosmic web (see \papertwo\!).
\end{enumerate} 
Application to 3D cosmic simulated and observed data sets are presented in the companion paper, \papertwo\!. Let us emphasize however that even if there is a wide range of application in astrophysics already, the domain of application of \progname is undoubtedly wider than the cosmic large scale structures. 

\subsection*{Acknowledgements}
{ 

\sl The author gratefully acknowledges support from JSPS (Japan Society for the Promotion of Science) Postdoctoral Fellowhip for Foreign Researchers award P08324. 

\sl The author thanks C. Pichon for a careful reading and commenting of the manuscript, H.~Kawahara for his fruitful comments and Y.~Suto for his constant help and support. 

This work was made possible through an extensive usage of the Yorick programming language by D.~Munro (available at {\em\tt http://yorick.sourceforge.net/}) and also CGAL, the Computational Geometry Algorithms Library, ({\em\tt http://www.cgal.org}), which was used to compute the Delaunay tessellations.

}

\bibliographystyle{mn2e}
\bibliography{morse-theory}

\def\cprime{$'$}
\begin{thebibliography}{}

\bibitem[\protect\citeauthoryear{{Abazajian} et~al.,}{{Abazajian}
  et~al.}{2009}]{Abazajian09}
{Abazajian} K.~N.,  et~al., 2009, \apjs, 182, 543

\bibitem[\protect\citeauthoryear{{Aragon-Calvo}, {Platen}, {van de Weygaert} \&
  {Szalay}}{{Aragon-Calvo} et~al.}{2008}]{spine}
{Aragon-Calvo} M.~A.,  {Platen} E.,  {van de Weygaert} R.,    {Szalay} A.~S.,
  2008, ArXiv e-prints

\bibitem[\protect\citeauthoryear{{Aragon-Calvo}, {van de Weygaert},
  {Araya-Melo}, {Platen} \& {Szalay}}{{Aragon-Calvo}
  et~al.}{2010}]{calvo_voids}
{Aragon-Calvo} M.~A.,  {van de Weygaert} R.,  {Araya-Melo} P.~A.,  {Platen} E.,
     {Szalay} A.~S.,  2010, \mnras, 404, L89

\bibitem[\protect\citeauthoryear{{Arag{\'o}n-Calvo}, {van de Weygaert} \&
  {Jones}}{{Arag{\'o}n-Calvo} et~al.}{2010}]{calvo10}
{Arag{\'o}n-Calvo} M.~A.,  {van de Weygaert} R.,    {Jones} B.~J.~T.,  2010,
  \mnras, pp 1270--+

\bibitem[\protect\citeauthoryear{{Aubert}, {Pichon} \& {Colombi}}{{Aubert}
  et~al.}{2004}]{ADHOP}
{Aubert} D.,  {Pichon} C.,    {Colombi} S.,  2004, \mnras, 352, 376

\bibitem[\protect\citeauthoryear{{Bardeen}, {Bond}, {Kaiser} \&
  {Szalay}}{{Bardeen} et~al.}{1986}]{BBKS}
{Bardeen} J.~M.,  {Bond} J.~R.,  {Kaiser} N.,    {Szalay} A.~S.,  1986, \apj,
  304, 15

\bibitem[\protect\citeauthoryear{{Bertschinger}}{{Bertschinger}}{1985}]{bertsc%
hinger85}
{Bertschinger} E.,  1985, \apjs, 58, 1

\bibitem[\protect\citeauthoryear{Beucher \& Lantuéjoul}{Beucher \&
  Lantuéjoul}{1979}]{beucher}
Beucher S.,  Lantuéjoul C.,  1979, in Proceeding of International Workshop on
  Image Processing, Real-Time Edge and Motion Detection/Estimation Use of
  watersheds in contour detection.
pp 17--21

\bibitem[\protect\citeauthoryear{{Bond}, {Kofman} \& {Pogosyan}}{{Bond}
  et~al.}{1996}]{bond96}
{Bond} J.~R.,  {Kofman} L.,    {Pogosyan} D.,  1996, \nat, 380, 603

\bibitem[\protect\citeauthoryear{Caroli \& Teillaud}{Caroli \&
  Teillaud}{2010}]{CGAL_PB}
Caroli M.,  Teillaud M.,  2010, in , {CGAL} User and Reference Manual, {3.6}
  edn, {CGAL Editorial Board}

\bibitem[\protect\citeauthoryear{Cohen-Steiner, Edelsbrunner \&
  Morozov}{Cohen-Steiner et~al.}{2006}]{vineyard}
Cohen-Steiner D.,  Edelsbrunner H.,    Morozov D.,  2006, in , Computational
  geometry ({SCG}'06).
ACM, New York, pp 119--126

\bibitem[\protect\citeauthoryear{{Colberg}, {Pearce}, {Foster}, {Platen},
  {Brunino}, {Neyrinck}, {Basilakos}, {Fairall}, {Feldman}, {Gottl{\"o}ber},
  {Hahn} \& {Hoyle}}{{Colberg} et~al.}{2008}]{aspen}
{Colberg} J.~M.,  {Pearce} F.,  {Foster} C.,  {Platen} E.,  {Brunino} R.,
  {Neyrinck} M.,  {Basilakos} S.,  {Fairall} A.,  {Feldman} H.,
  {Gottl{\"o}ber} S.,  {Hahn} O.,    {Hoyle} F. e.~a.,  2008, \mnras, 387, 933

\bibitem[\protect\citeauthoryear{{Colless}, {Peterson}, {Jackson}, {Peacock},
  {Cole}, {Norberg}, {Baldry}, {Baugh}, {Bland-Hawthorn}, {Bridges} \&
  {Cannon}}{{Colless} et~al.}{2003}]{2Df}
{Colless} M.,  {Peterson} B.~A.,  {Jackson} C.,  {Peacock} J.~A.,  {Cole} S.,
  {Norberg} P.,  {Baldry} I.~K.,  {Baugh} C.~M.,  {Bland-Hawthorn} J.,
  {Bridges} T.,    {Cannon} R. e.~a.,  2003, ArXiv Astrophysics e-prints

\bibitem[\protect\citeauthoryear{{de Lapparent}, {Geller} \& {Huchra}}{{de
  Lapparent} et~al.}{1986}]{lapparent86}
{de Lapparent} V.,  {Geller} M.~J.,    {Huchra} J.~P.,  1986, \apjl, 302, L1

\bibitem[\protect\citeauthoryear{Delfinado \& Edelsbrunner}{Delfinado \&
  Edelsbrunner}{1995}]{delfinado}
Delfinado C. J.~A.,  Edelsbrunner H.,  1995, Comput. Aided Geom. Design, 12,
  771

\bibitem[\protect\citeauthoryear{Edelsbrunner, Harer, Natarajan \&
  Pascucci}{Edelsbrunner et~al.}{2003}]{edel3D}
Edelsbrunner H.,  Harer J.,  Natarajan V.,    Pascucci V.,  2003, in SCG '03:
  Proceedings of the nineteenth annual symposium on Computational geometry
  Morse-smale complexes for piecewise linear 3-manifolds.
ACM, New York, NY, USA, pp 361--370

\bibitem[\protect\citeauthoryear{Edelsbrunner, Harer \&
  Zomorodian}{Edelsbrunner et~al.}{2003}]{edel2D}
Edelsbrunner H.,  Harer J.,    Zomorodian A.,  2003, Discrete Comput. Geom.,
  30, 87

\bibitem[\protect\citeauthoryear{Edelsbrunner, Letscher \&
  Zomorodian}{Edelsbrunner et~al.}{2000}]{edel00}
Edelsbrunner H.,  Letscher D.,    Zomorodian A.,  2000, in , 41st {A}nnual
  {S}ymposium on {F}oundations of {C}omputer {S}cience ({R}edondo {B}each,
  {CA}, 2000).
IEEE Comput. Soc. Press, Los Alamitos, CA, pp 454--463

\bibitem[\protect\citeauthoryear{Edelsbrunner, Letscher \&
  Zomorodian}{Edelsbrunner et~al.}{2002}]{edel}
Edelsbrunner H.,  Letscher D.,    Zomorodian A.,  2002, Discrete Comput. Geom.,
  28, 511

\bibitem[\protect\citeauthoryear{{Forero-Romero}, {Hoffman}, {Gottl{\"o}ber},
  {Klypin} \& {Yepes}}{{Forero-Romero} et~al.}{2009}]{jaime}
{Forero-Romero} J.~E.,  {Hoffman} Y.,  {Gottl{\"o}ber} S.,  {Klypin} A.,
  {Yepes} G.,  2009, \mnras, 396, 1815

\bibitem[\protect\citeauthoryear{Forman}{Forman}{1998a}]{formanvfield}
Forman R.,  1998a, Math. Z., 228, 629

\bibitem[\protect\citeauthoryear{Forman}{Forman}{1998b}]{forman98}
Forman R.,  1998b, Adv. Math., 134, 90

\bibitem[\protect\citeauthoryear{Forman}{Forman}{2002}]{forman}
Forman R.,  2002, S\'em. Lothar. Combin., 48, Art.\ B48c, 35 pp. (electronic)

\bibitem[\protect\citeauthoryear{{Gay}, {Pichon}, {Le Borgne}, {Teyssier},
  {Sousbie} \& {Devriendt}}{{Gay} et~al.}{2010}]{gay10}
{Gay} C.,  {Pichon} C.,  {Le Borgne} D.,  {Teyssier} R.,  {Sousbie} T.,
  {Devriendt} J.,  2010, \mnras, 404, 1801

\bibitem[\protect\citeauthoryear{{Gottloeber}}{{Gottloeber}}{1998}]{gottlober9%
8}
{Gottloeber} S.,  1998, in {V.~Mueller, S.~Gottloeber, J.~P.~Muecket, \&
  J.~Wambsganss} ed., Large Scale Structure: Tracks and Traces {Galaxy Tracers
  in Cosmological N-Body Simulations}.
pp 43--46

\bibitem[\protect\citeauthoryear{Gyulassy}{Gyulassy}{2008}]{phd}
Gyulassy A.,  2008, PhD thesis, Univ. California Berkeley

\bibitem[\protect\citeauthoryear{{Hahn}, {Porciani}, {Carollo} \&
  {Dekel}}{{Hahn} et~al.}{2007}]{hahn}
{Hahn} O.,  {Porciani} C.,  {Carollo} C.~M.,    {Dekel} A.,  2007, \mnras, 375,
  489

\bibitem[\protect\citeauthoryear{Hatcher}{Hatcher}{2002}]{hatcher}
Hatcher A.,  2002, Algebraic topology.
Cambridge University Press, Cambridge

\bibitem[\protect\citeauthoryear{{Hoffman} \& {Shaham}}{{Hoffman} \&
  {Shaham}}{1982}]{hoffman82}
{Hoffman} Y.,  {Shaham} J.,  1982, \apjl, 262, L23

\bibitem[\protect\citeauthoryear{{Huchra} \& {Geller}}{{Huchra} \&
  {Geller}}{1982}]{FOF}
{Huchra} J.~P.,  {Geller} M.~J.,  1982, \apj, 257, 423

\bibitem[\protect\citeauthoryear{{Icke}}{{Icke}}{1984}]{icke84}
{Icke} V.,  1984, \mnras, 206, 1P

\bibitem[\protect\citeauthoryear{Jost}{Jost}{2008}]{jost}
Jost J.,  2008, Riemannian geometry and geometric analysis, fifth edn.
Universitext, Springer-Verlag, Berlin

\bibitem[\protect\citeauthoryear{{Kirshner}, {Oemler} Jr., {Schechter} \&
  {Shectman}}{{Kirshner} et~al.}{1981}]{kirshner81}
{Kirshner} R.~P.,  {Oemler} Jr. A.,  {Schechter} P.~L.,    {Shectman} S.~A.,
  1981, \apjl, 248, L57

\bibitem[\protect\citeauthoryear{Lewiner}{Lewiner}{2002}]{lewiner_master}
Lewiner T.,  2002, Master's thesis, Department of Mathematics, PUC-Rio

\bibitem[\protect\citeauthoryear{Milnor}{Milnor}{1963}]{milnor}
Milnor J.,  1963, Morse theory.
Based on lecture notes by M. Spivak and R. Wells. Annals of Mathematics
  Studies, No. 51, Princeton University Press, Princeton, N.J.

\bibitem[\protect\citeauthoryear{{Neyrinck}}{{Neyrinck}}{2008}]{neyrinck}
{Neyrinck} M.~C.,  2008, \mnras, 386, 2101

\bibitem[\protect\citeauthoryear{{Neyrinck}, {Gnedin} \& {Hamilton}}{{Neyrinck}
  et~al.}{2005}]{neyrinck05}
{Neyrinck} M.~C.,  {Gnedin} N.~Y.,    {Hamilton} A.~J.~S.,  2005, \mnras, 356,
  1222

\bibitem[\protect\citeauthoryear{{Novikov}, {Colombi} \& {Dor{\'e}}}{{Novikov}
  et~al.}{2006}]{skel2D}
{Novikov} D.,  {Colombi} S.,    {Dor{\'e}} O.,  2006, \mnras, 366, 1201

\bibitem[\protect\citeauthoryear{{Okabe}}{{Okabe}}{2000}]{okabe00}
{Okabe} A.,  ed. 2000, {Spatial tessellations : concepts and applications of
  voronoi diagrams}

\bibitem[\protect\citeauthoryear{{Platen}, {van de Weygaert} \&
  {Jones}}{{Platen} et~al.}{2007}]{platen}
{Platen} E.,  {van de Weygaert} R.,    {Jones} B.~J.~T.,  2007, \mnras, 380,
  551

\bibitem[\protect\citeauthoryear{{Platen}, {van de Weygaert} \&
  {Jones}}{{Platen} et~al.}{2008}]{platen2}
{Platen} E.,  {van de Weygaert} R.,    {Jones} B.~J.~T.,  2008, \mnras, 387,
  128

\bibitem[\protect\citeauthoryear{{Pogosyan}, {Bond}, {Kofman} \&
  {Wadsley}}{{Pogosyan} et~al.}{1996}]{pogo96}
{Pogosyan} D.,  {Bond} J.~R.,  {Kofman} L.,    {Wadsley} J.,  1996, in Bulletin
  of the American Astronomical Society Vol.~28 of Bulletin of the American
  Astronomical Society, {The Cosmic Web and Filaments in Cluster Patches}.
pp 1289--+

\bibitem[\protect\citeauthoryear{{Pogosyan}, {Pichon}, {Gay}, {Prunet},
  {Cardoso}, {Sousbie} \& {Colombi}}{{Pogosyan} et~al.}{2009}]{pogo}
{Pogosyan} D.,  {Pichon} C.,  {Gay} C.,  {Prunet} S.,  {Cardoso} J.~F.,
  {Sousbie} T.,    {Colombi} S.,  2009, \mnras, 396, 635

\bibitem[\protect\citeauthoryear{Robins}{Robins}{2000}]{robins}
Robins V.,  2000, PhD thesis, Department of Applied Mathematics, University of
  Colorado

\bibitem[\protect\citeauthoryear{Roerdink \& Meijster}{Roerdink \&
  Meijster}{2000}]{watershed}
Roerdink J. B. T.~M.,  Meijster A.,  2000, Fund. Inform., 41, 187

\bibitem[\protect\citeauthoryear{{Schaap} \& {van de Weygaert}}{{Schaap} \&
  {van de Weygaert}}{2000}]{DTFE}
{Schaap} W.~E.,  {van de Weygaert} R.,  2000, \aap, 363, L29

\bibitem[\protect\citeauthoryear{{Sousbie}, {Colombi} \& {Pichon}}{{Sousbie}
  et~al.}{2009}]{rsex}
{Sousbie} T.,  {Colombi} S.,    {Pichon} C.,  2009, \mnras, 393, 457

\bibitem[\protect\citeauthoryear{{Sousbie}, {Pichon}, {Colombi}, {Novikov} \&
  {Pogosyan}}{{Sousbie} et~al.}{2008}]{skel}
{Sousbie} T.,  {Pichon} C.,  {Colombi} S.,  {Novikov} D.,    {Pogosyan} D.,
  2008, \mnras, 383, 1655

\bibitem[\protect\citeauthoryear{{Sousbie}, {Pichon}, {Courtois}, {Colombi} \&
  {Novikov}}{{Sousbie} et~al.}{2008}]{SDSSskel}
{Sousbie} T.,  {Pichon} C.,  {Courtois} H.,  {Colombi} S.,    {Novikov} D.,
  2008, \apjl, 672, L1

\bibitem[\protect\citeauthoryear{{Springel}, {White}, {Tormen} \&
  {Kauffmann}}{{Springel} et~al.}{2001}]{springel01}
{Springel} V.,  {White} S.~D.~M.,  {Tormen} G.,    {Kauffmann} G.,  2001,
  \mnras, 328, 726

\bibitem[\protect\citeauthoryear{{Stoica}, {Mart{\'{\i}}nez}, {Mateu} \&
  {Saar}}{{Stoica} et~al.}{2005}]{stoica05}
{Stoica} R.~S.,  {Mart{\'{\i}}nez} V.~J.,  {Mateu} J.,    {Saar} E.,  2005,
  \aap, 434, 423

\bibitem[\protect\citeauthoryear{{Stoica}, {Mart{\'{\i}}nez} \&
  {Saar}}{{Stoica} et~al.}{2010}]{stoica10}
{Stoica} R.~S.,  {Mart{\'{\i}}nez} V.~J.,    {Saar} E.,  2010, \aap, 510, A38+

\bibitem[\protect\citeauthoryear{{Tweed}, {Devriendt}, {Blaizot}, {Colombi} \&
  {Slyz}}{{Tweed} et~al.}{2009}]{ADHOP2}
{Tweed} D.,  {Devriendt} J.,  {Blaizot} J.,  {Colombi} S.,    {Slyz} A.,  2009,
  \aap, 506, 647

\bibitem[\protect\citeauthoryear{Zomorodian}{Zomorodian}{2009}]{zomo}
Zomorodian A.~J.,  2009, Topology for computing.
Vol.~16 of Cambridge Monographs on Applied and Computational Mathematics,
  Cambridge University Press, Cambridge

\end{thebibliography}


\appendix
%
\section{Applicability of Morse theory to practical data-sets}
\label{sec_appwatershed}

There exist a large number of methods to reconstruct a smooth density field from the discrete sample of galaxies in a catalogue or a dark matter particles distribution in a cosmological simulation. Whether one uses a simple constant resolution uniform grid to sample the original distribution or a more sophisticated scale free method such as DTFE \citep{DTFE}, that is able to reconstruct the unbiased density field over the full dynamic range of the sample, the initial sampling always defines some lower scale resolution below which one is free to infer the behavior of the distribution. As the constraints undergone by a \hyperref[defMF]{Morse function} (definition \ref{def_morse_function}) are essentially local (continuity, differentiability and non degeneracy of the critical points), one could imagine designing some sophisticated interpolation scheme that would enforce Morse properties on the distribution. In practice, designing such an interpolation scheme seems extremely difficult though and to our knowledge, this kind of solution have never been implemented. Another solution consists in relaxing Morse conditions by computing the manifolds and Morse complex of a non-\hyperref[defMF]{Morse function}, and later correct for this ommission by enforcing the correct combinatorial properties on the pseudo Morse complex (see definition \ref{prop_mscomplex}). The approach has been successfully developed by \citet{edel2D} and \citet{edel3D} in the 2D and 3D case respectively, but at the price of a very high algorithmic complexity. In fact, whereas the method for the 2D case has been implemented and tested, there exist no implementation to date in the case of a 3D function, although the method has been mathematically proved to be correct. Another more radical approach simply consists in abandoning the idea of rigorously computing the Morse-complex and rather rely directly on a pseudo-Morse complex. A pseudo-Morse complex is an approximation of a Morse complex and its combinatorial properties are not guaranteed by Morse theory anymore. This is mainly the result of a fundamental property of the paths defined by following the gradient arrows, the so-called integral lines, being violated: they are not guaranteed not to cross anymore, as they should with a \hyperref[defMF]{Morse function} (see definitions \ref{def_int_line} and \ref{prop_int_line}). The second approach recently became relatively popular in astrophysics as a way to identify cosmologically significant structures, mainly using the Watershed transform. The Watershed technique (see \citet{beucher,watershed}) was first applied to this kind of problem by \citet{platen}  as a mean of identifying voids in large scale structures (see also \citet{platen2}, \citet{aspen} or \citet{calvo_voids}), it was latter extended to the identification of walls and filament through a pseudo Morse complex by \citet{rsex} and it is also used by \citet{spine}. But although promising, these techniques seem to be doomed by the lack of a consistent theory and therefore of a good understanding of the properties of the pseudo Morse complex, as illustrated in the following.\\ 

\begin{figure*}
\centering\centering  \subfigure[2D case]{\includegraphics[width=0.49\linewidth]{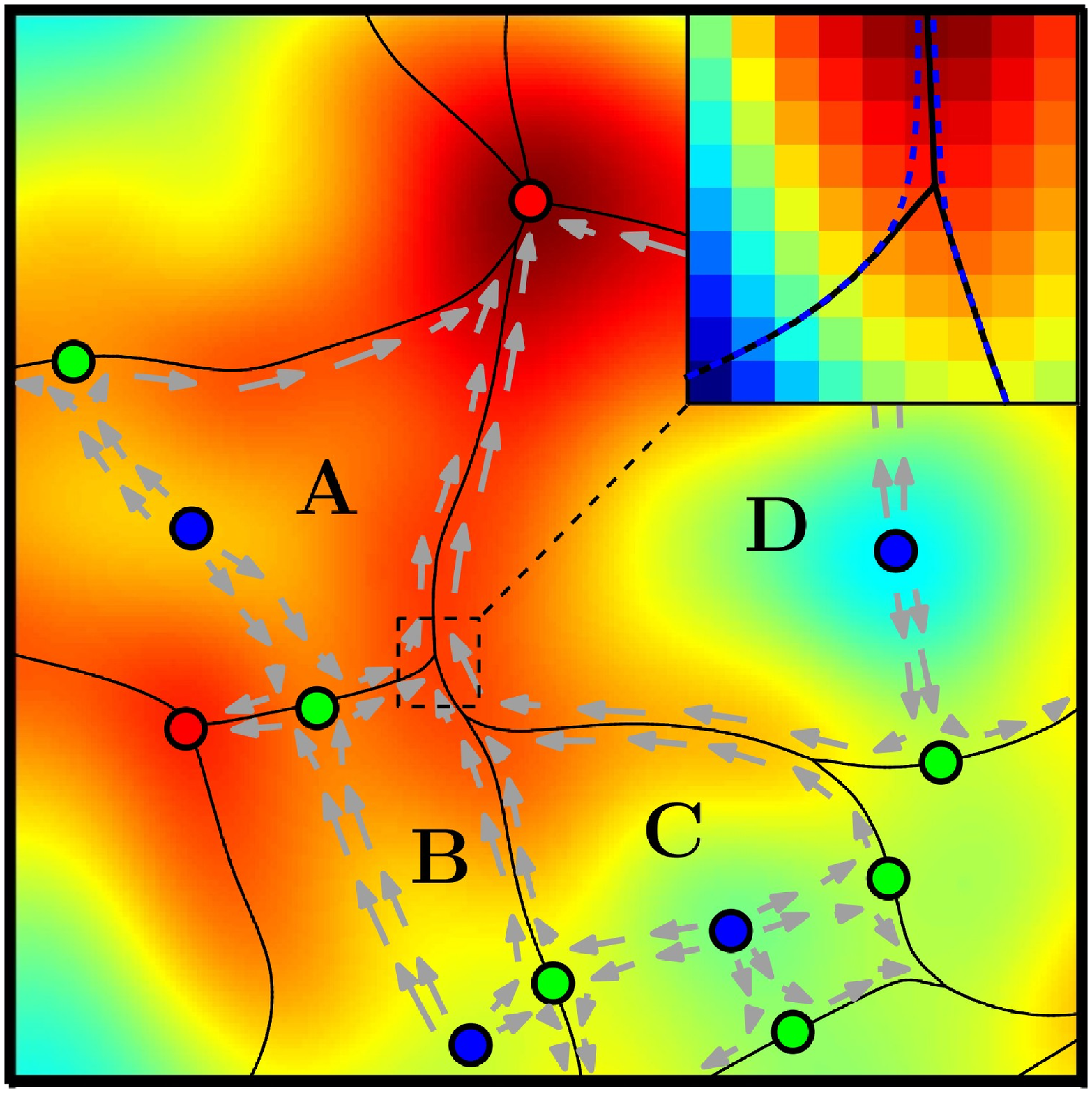}\label{fig_watershed_problem_2D} }
\hfill  \subfigure[3D case]{\includegraphics[width=0.49\linewidth]{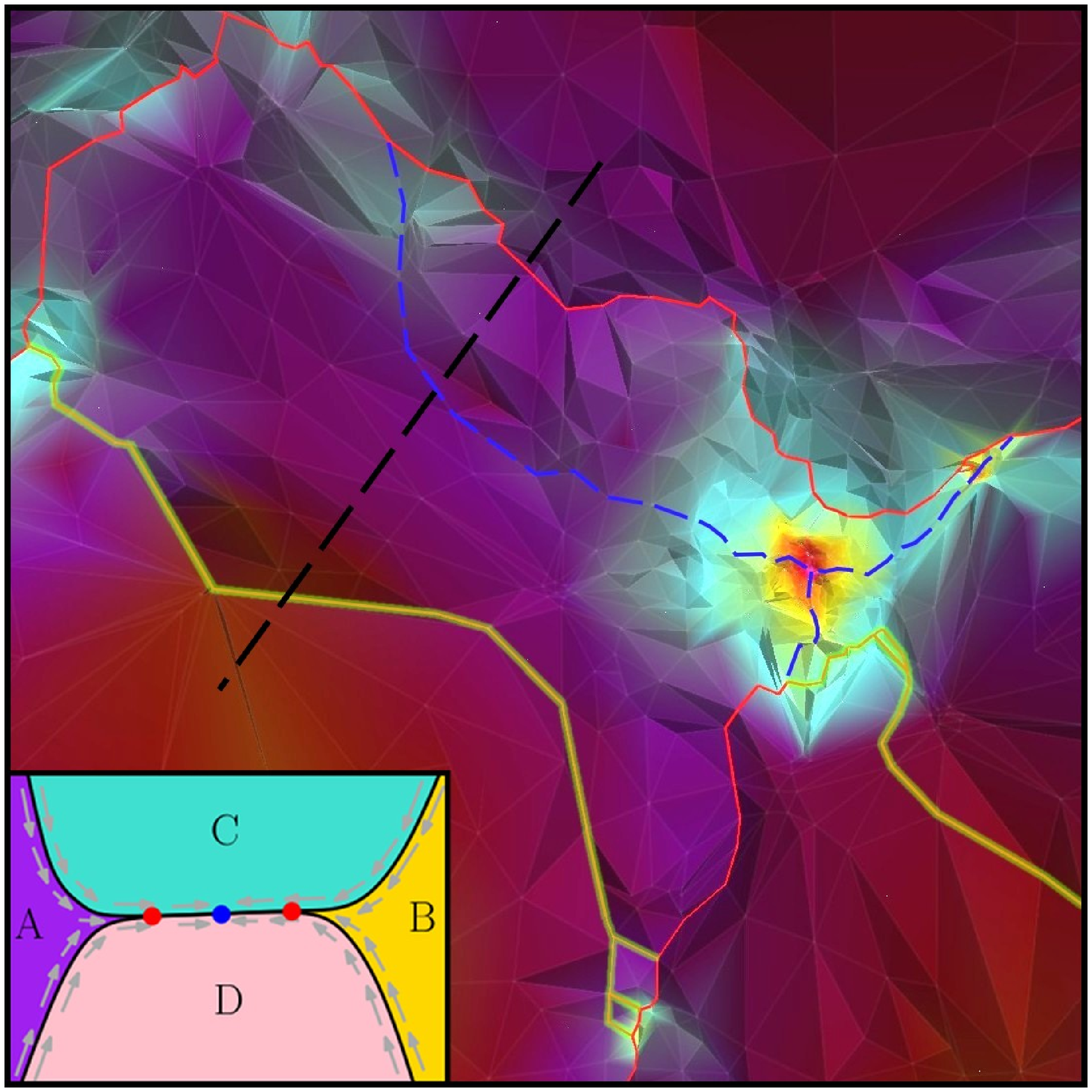}\label{fig_watershed_problem_3D}}
\caption{\label{fig_watershed_problem} Illustration of a problem in the identification of the filaments when using the watershed technique to recover the Morse complex directly from a non Morse function. Figure \ref{fig_watershed_problem_2D}: filaments (black curves) of a 2D field sampled at discrete locations over a high resolution grid, with gray arrows showing the gradient direction along those filaments. The maxima/saddle points/minima are represented as red/green/blue disks respectively and the letters designate regions delimited by filaments. Figure \ref{fig_watershed_problem_3D}: filaments identified on a 3D delaunay tessellation of $50h^{-1}$ Mpc large dark matter simulation with density computed using DTFE. The surface represents the boundary of a void, shaded according to the logarithm of the density and viewed from its corresponding minimum. The red and yellow curves show the filaments detected by a multi-scale watershed method. See main text for more explanations.}
\end{figure*}

The watershed transform segments a field into isolated regions called basins, the analogs of the ascending \hyperref[defmanifold]{manifold}s of the minima (or equivalently $0$-\hyperref[defmanifold]{manifold}s, see the top left frame of figure \ref{fig_watershed_problem}). The boundary of those basins delineate the walls (see bottom left frame) and the regions at the boundary of three basins describe the filaments as an approximation of the  ascending \hyperref[defmanifold]{manifold}s of the first kind saddle points. We show on figure \ref{fig_watershed_problem} how the fact that only a pseudo Morse complex is computed can lead to subtle but significant errors in the identification of the filaments in galaxy distribution. Figure \ref{fig_watershed_problem_2D} illustrates the problem in 2D, using a similar implementation as the one presented in \citet{rsex}. A density field is sampled on a high resolution Cartesian grid and a watershed transform is applied, generating basins (labeled by letters). The filaments are therefore identified as the basins boundary (black curves) and form a pseudo Morse complex: a network that links the critical points together (the red, green and blue disks, standing for the maxima, saddle points, and minima respectively). One can see that the filaments seem correctly identified but according to Morse theory, if the Watershed transform yielded a correct Morse complex, field lines would only cross at critical points and the bifurcation points, located at the intersection of at least three basins (for instance $A$, $B$ and $D$ or $B$, $C$ and $D$), would therefore be maxima. This is not the case on figure \ref{fig_watershed_problem_2D} because the function is not a \hyperref[defMF]{Morse function}, and its gradient lines may therefore intersect where the filaments seem to bifurcate (the gradient direction along critical lines is represented by the gray arrows). If the function complied to Morse criteria, these bifurcation points would actually look like the blue dashed line in the framed zoom in the upper right corner of the picture. This is not a significant problem for the identification of filaments in 2D, as it could theoretically be corrected for through some post-treatment, but as shown on figure \ref{fig_watershed_problem_3D}, the consequences are more dramatic in the 3D case.\\

In order to assess the extent of this problem, a 3D multi-scale version of the probabilistic watershed transform presented in \citet{rsex} was implemented directly over a Delaunay tessellation computed from a discrete point sample. Each vertex of the tessellation is attributed a density using the DTFE method \citep{DTFE}, and the probabilistic watershed transform is applied, using the natural neighborhood defined by the dual Voronoi cells to propagate the probabilities. Basically, the minima and maxima are identified as those vertice with only higher or lower density neighbors respectively, and the probability that each vertex belongs to the integral line of a given extremum is computed according to \citet{rsex}. This defines the watershed basins attached to minima and maxima (\ie the void patches and peak patches according to the terminology of \citet{rsex}, or the pseudo - ascending and descending 3-\hyperref[defmanifold]{manifold}s, according to Morse theory terminology). Figure \ref{fig_watershed_problem_3D} shows the triangulated interface between void patches (\ie the boundary of the cosmological voids), computed over the delaunay tessellation of a sub-sampled $512^3$ particles dark-matter cosmological simulation in a $50h^{-1}$ Mpc box. This surface represents the density ``walls'' of the cosmic web, shaded according to the locally interpolated density. The surface is seen from the point of view of the minimum inside the void patch and one can identify a dark matter halo on the central-right part of the image. Following \citet{rsex} (see also \citet{spine}) the filaments are identified as those segments located at the one dimensional interface of at least three different void patches, and are represented by the non-dashed red and yellow lines. It is clear on this picture that the yellow shaded lines are spurious as they do not correspond to any filament visible in the overdensity field projected onto the surface. One can also remark that the network does not pass through the local maximum located at the center of the halo, which should obviously be the case for a cosmological filament. Actually, a more reasonable network could be obtained by displacing the red lines to match the blue dashed ones and removing the yellow shaded spurious identifications. The cause of those errors is actually similar to the one described in the previous paragraph for the 2D case: the density function does not comply to Morse criterion and its field lines may therefore cross. The sketch in the lower left illustrates what happens along the dashed black line, in the plane perpendicular to the surface: the void patches A and B are sandwiched between C and D, resulting in the identification of critical lines at the spurious intersection of ADC and BCD, symbolized by two red dots on the sketch, and the intersection of the dashed black line and the red and yellow critical lines on the 3D image. Actually, the only real critical line is at the true intersection of the four patches, symbolized by the blue dot on the sketch and blue dashed line on the picture (\ie where the field lines really end, as represented by the gray arrows).\\

This tendency of the void and peak patches to get sandwiched between each other is perfectly natural and understood in Morse theory, and it is not a simple consequence of the particular selected sampling method, but rather of the fact that sampling is used at all. Moreover, it seems to be particularly the case in the large scale cosmological dark matter density fields, probably as a consequence of the nature of the initial Gaussian random field from which tiny perturbations evolve to form the cosmic web (see the discussion on bifurcations points in \cite{pogo}). In short, this shows that the simple approach that consists in requiring filaments to be at the intersection of walls which are at the intersection of voids is a bit naive as in practice, when the field is sampled and/or noisy, these boundaries do not have the right properties and do not trace the cosmic network correctly. These problems, among others, severely limit the domain of application of watershed based method (for instance it renders practically impossible their usage to count the number of filaments attached to a given halo, or the measurement of the physical properties of individual filaments) and demonstrate the necessity to adopt a different, mathematically more consistent approach.\\

\section{Simplicial homology}
\label{sec_simp_homo}

Homology theory studies the topological properties of a spaces (intuitively, its number of component, how they are connected or if holes exist ...). Roughly speaking, it does so by studying the properties of deformable chains and loops over these spaces and giving a method to relate them to sequences of Abelian groups, the so-called Homology groups. The goal of this section is only to give the reader enough intuitive understanding of its restriction to simplicial complexes - the weaker simplicial homology - to grasp the concept of topological \hyperref[defpers]{persistence} as introduced by \citet{edel}. For that reason, although we give a few necessary mathematical definitions, we always try to explain them in a less formal and more intuitive manner. One could always refer to \cite[chap. 4]{zomo} for a very interesting and somewhat more rigorous introduction or \cite{hatcher} for a thorough reference.\\

In order to understand simplicial homology, one should first define the $k$-chain group over a simplicial complex $K$ that contains $p$ simplexes.
\begin{mydef}[$k$-chain group]
\label{def_kchain}
  Let $k\in\lbrace0,..,d\rbrace$ the dimension of the $k$-chain, then $\lbrace\sigma_1,..,\sigma_p\rbrace$ is the set of all the $k$-simplexes in $K$. Any $k$-chain $c_k$ can be written:
 \begin{displaymath} 
   c_k=\sum\limits_{i=1}^{p} n_i \sigma_i,\;n_i\in\mathbb{Z}/2\mathbb{Z}=\lbrace0,1\rbrace.
\end{displaymath} 
The $k$-chain group, $C_k\left(K\right)$, is the group with element $c_k$ and addition defined as
 \begin{displaymath} 
   c_k+{c^\prime}_k = \sum\limits_{i=1}^{p} \left(n_i+{n^\prime}_i\right) \sigma_i.
\end{displaymath} 
\end{mydef}
In other words, a $k$-chain is a subset of the simplexes in $K$ with dimension $k$. For a 3D simplical complex such as the delaunay tessellation of a galaxy catalogue, it would be a set of vertice, segments, facets or tetrahedrons. Note that in this definition, although the more general case could be considered, the coefficients $n_i$ are chosen to be positive integers modulo $2$ which, as we will see, is sufficient to capture interesting topological properties. This means that a given simplex can only be absent or present once in a $k$-chain. Adding a simplex to a $k$-chain of $C_k\left(K\right)$ that already contains it therefore results in its actual removal (the addition being performed modulo 2). This definition alone only relates simplexes of identical dimensions, but for different values of $k$, the $C_k\left(K\right)$ are independent. The notion of topology (\ie the connectivity of the simplexes in $K$) can be introduced through the definition of a boundary operator. Intuitively, the boundary of a simplex is the set of its faces:
\begin{mydef}[boundary operator]
Let $v_i$ be $k+1$ vertice of $K$, and $\sigma=\left[v_0,v_1,..,v_k\right]\in C_k\left(K\right)$ a \hyperref[defksimplex]{$k$-simplex}, then the boundary of $\sigma$ is
\begin{displaymath} 
\partial_k\left(\sigma\right)=\sum\limits_{i=0}^{k}\left[v_0,..,\hat{v}_i,..,v_k\right],
\end{displaymath} 
where $\hat{v}_i$ means that vertex $v_i$ is removed from the list. By extension, the boundary of a $k$-chain is defined as:
\begin{displaymath}
\begin{tabular}{rcl}
$\partial: C_k\left( K\right)$&$\mapsto$&$ C_{k-1}\left( K\right)$ \\
$c$&$\mapsto $&$\partial c = \sum_{\sigma\in C_k\left( K\right)} \partial\sigma$\\
\end{tabular}
\end{displaymath}
\end{mydef}

\begin{figure*}
\centering
\includegraphics[width=11cm]{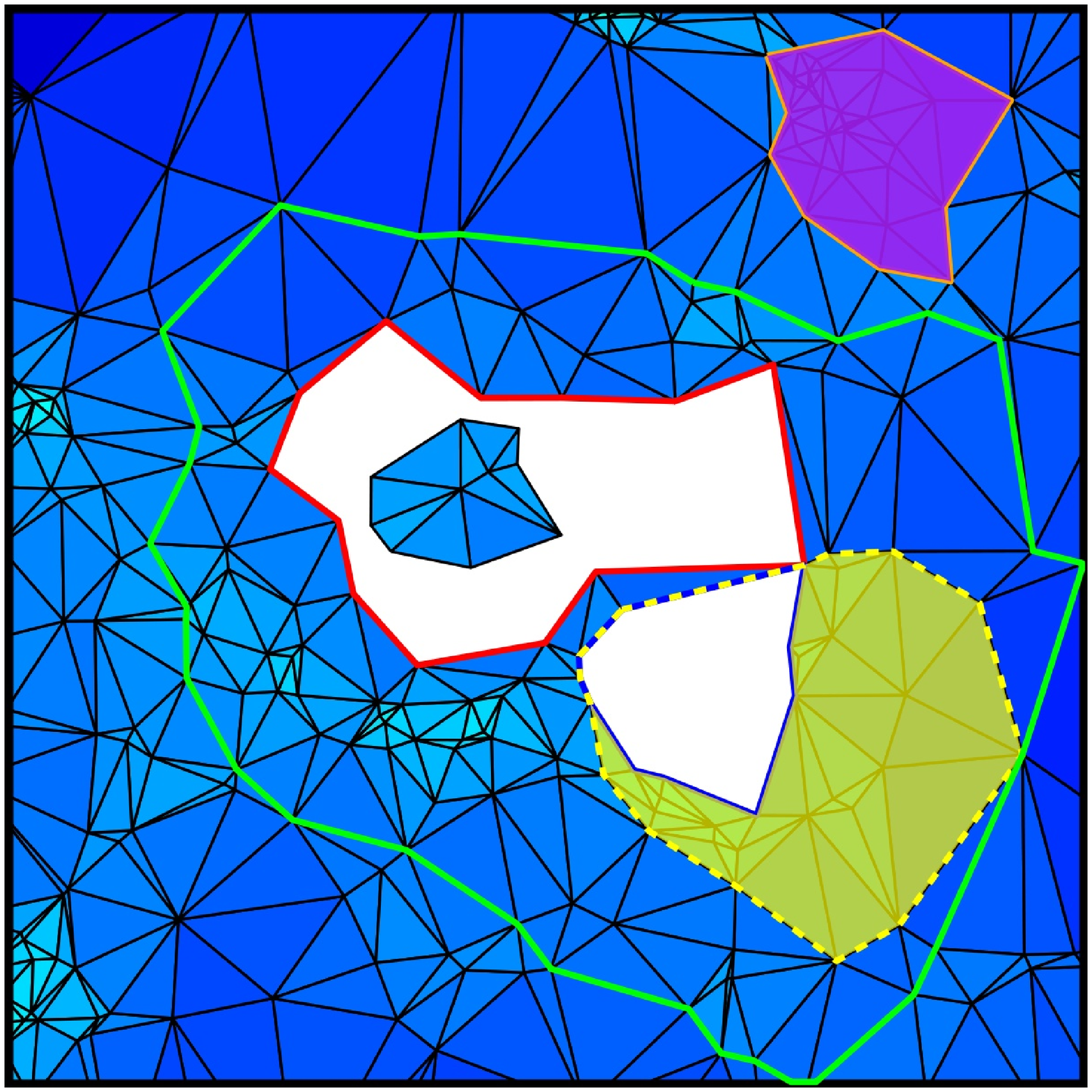}
\caption{Illustartion of $1$-boudaries and $1$-cycles of a 2D simplicial complex extracted from a \hyperref[deffiltration]{filtration} of a Delaunay tessellation. The facets present in the filtration are colored with different shades of blue, depending on the local density, and there are two holes at this stage (white parts). See main text for explanations.\label{fig_net_cycles}. }
\end{figure*}

Following this definition, the boundary of a $k$-chain only contains the {\km1}-simplexes that are faces of exactly $1$ \hyperref[defksimplex]{$k$-simplex} in the chain. On figure \ref{fig_net_cycles} for instance, the segments in the orange contour (see upper right corner of the figure) are the boundary of the facets within the purple shaded area; all other purple shaded segments being faces of two facets, they cancel each other because of the addition modulo 2 in definition \ref{def_kchain}. A very important property of the boundary operator is that $\partial_{k-1}\partial_k=0$: the boundary of a boundary is void. This is intuitively easy to understand as a boundary is a cycle an cycles do not have boundaries. The orange boundary of figure \ref{fig_net_cycles} for instance forms a chain $c_1$ that does not have boundary, as its segments all share the vertice at their extremity with exactly one other segment in $c_k$, and therefore appear twice when applying $\partial_1$ to $c_1$. The subgroup of $C_k\left(K\right)$ formed by the chains which are the boundary of a chain in $C_{k+1}\left(K\right)$ is called the image of $\partial_{k+1}$.
\begin{mydef}[$k^{\rm{th}}$ boundary group $B_k$]
Let $B_k=\rm{im}\,\partial_{k+1}$ be the image of $C_{k+1}\left(K\right)$ under the boundary operator. Then $B_k$ is a subgroup of $C_{k+1}\left(K\right)$ called the $k^{\rm{th}}$ boundary group. Its elements form cycles called bounding cycles, and therefore do not have boundary.\vspace{1mm} 
\end{mydef}
On figure \ref{fig_net_cycles}, a $1$-chain of segments belongs to $B_1$ if it is the boundary of a $2$-chain of 2D facets, which is the case of the orange contour (boundary of the purple shaded facets) or the boundary of the yellow shaded area. This is nevertheless not the case of the green, red, blue and yellow dashed contours as no set of facets can have these contours as boundary due to the presence of the two holes. At best, the boundary of a 2-chain formed by a ring around a hole could include them, but it would necessarily contain additional cycles (the boundary of the hole). These contours are nevertheless cycles and therefore neither do they have boundaries. They all belong to the wider $k^{\rm{th}}$ cycle group:
 \begin{mydef}[$k^{\rm{th}}$ cycle group $Z_k$]
Let $Z_k=\rm{ker}\,\partial_{k}$ be the subset of $C_{k}\left(K\right)$ whose image under $\partial_k$ is the null {\km1}-chain. Then $Z_k$ forms a subgroup of $C_k$ called the $k^{\rm{th}}$ cycle group, and the $k^{\rm{th}}$ boundary group $B_k$ is included in $Z_k$.\vspace{1mm}     
\end{mydef} 
The elements of $Z_k$ are any chain that form a cycle (or equivalently that have no boundary), and the green, red, blue and yellow dashed contours of figure \ref{fig_net_cycles} do belong to $Z_1$.\\

These elements are enough to get an idea of how simplicial homology works. It involves trying to count how many different types of cycles it is possible to define for each dimension. To achieve this, one first needs to define what one means by ``different types of cycles'', and to do so, homology define an equivalence relation over the $k$-chains:
\begin{mydef}[simplicial homology]
Two cycles $c$ and $c^\prime$ in the $k^{\rm{th}}$ cycle group $Z_k$ are said to be homologous if there exist a bounding cycle $b\in B_k$ such that:
\begin{displaymath}
c+b=c^\prime.
\end{displaymath}
This equivalence relation can be used to define the class of equivalence of $z\in Z_k$, $[z]$, which contains all the elements of $Z_k$ that are homologous to $z$ (\ie all $z^ \prime\in Z_k$ that can be written $z+b=z^\prime$ with $b\in B_k$).\vspace{1mm} 
\label{def_homology}
\end{mydef} 
In a nutshell, definition \ref{def_homology} formalizes, for  simplicial complex, the intuitive idea that two cycles are equivalent if they can be continuously deformed into each other. This definition is at the core of regular Homology theory. For instance, the $1$-chains represented by the blue and yellow dashed contours of figure \ref{fig_net_cycles} are homologous, as one can obtain the yellow one by adding the boundary of the yellow shaded $2$-chain to the blue $1$-chain. At the contrary, the red and yellow dashed $1$-chains are clearly not homologous as it is impossible to find a chain that is both a boundary of a $2$-chain and transform one into the other through addition. This impossibility clearly comes from the fact that there exist holes in the simplicial complex, and homology shows that the presence of these hole directly affects the maximum number of non homologous cycles one can create. This link can be established through the so called $k^{\rm{th}}$ Homology group, which elements are the sets of homologous $k$-chains:
\begin{mydef}[$k^{\rm{th}}$ Homology group]
The $k^{\rm{th}}$ Homology group is the group which elements are the sets of homologous $k$-chains. It is defined as the quotient group of the $k^{\rm{th}}$ cycle group $Z_k$ by the $k^{\rm{th}}$ boundary group $B_k$:
\begin{displaymath}
H_k=Z_k/B_k=\rm{ker}\,\partial_{k} / \rm{im}\,\partial_{k+1}.
\end{displaymath}
An element $h$ of $H_k$ is represented by the class of equivalence [z] of all chains homologous to $z\in Z_k$.\vspace{1mm} 
\end{mydef}
In other words, on figure \ref{fig_net_cycles}, an element of $H_1$ could be represented by the blue $1$-chains around the smaller hole, as well as chains homologous to it such as the yellow dashed one. Another element is the red $1$-chain and its homologous chains, and yet another one is the class of equivalence of the green contour. But there is something different with the green $1$-chains: it may not be homologous to the blue and red ones, but it could be obtained by adding to cycles homologous to the red and blue ones respectively. This leads us to the definition of the Betti numbers, the mean by which homology describe the topology of a space:
\begin{mydef}[$k^{\rm{th}}$ Betti number]
the $k^{\rm{th}}$ Betti number $\beta_k$ is the rank of the free\footnote{The term ``free'' in the definition actually excludes some specific cycles that may exist when the space has torsion (think about a m\"{o}bius strip for instance)} part of $H_k$:
\begin{displaymath}
\beta_k=\rm{rank}{H_k}=\rm{rank}Z_k - \rm{rank}{B_k},
\end{displaymath}
\end{mydef}
To put it simply, the $k^{\rm{th}}$ Betti number really is the minimal number of $k$-cycles equivalence classes (\ie sets of homologous $k$-cycles) that one needs to generate any possible cycle through homology. Betti numbers are interesting because they are characteristic of the topological properties of a given space, and in that sense allow quantifying and comparing the topologies of different spaces.\\ 

 

\section{Persistence and Betti numbers in a simplicial complex}
\label{sec_app_persistence}

In order to explain the computation of \hyperref[defperspair]{persistence pair} over a simplicial complex we use figure \ref{fig_persistence}, a figure inspired from \citet[figure 3][]{edel}. Although the reader can always refer to page \pageref{sec_terminology} for an explanation of the terminology, it is advisable to read appendix \ref{sec_simp_homo} for a quick introduction to simplicial homology. The initial discovery of \hyperref[defpers]{persistence} was triggered by the design of a simple algorithm to compute the Betti numbers over a \hyperref[deffiltration]{filtration} of a simplicial complex, first presented in \citet{delfinado}. A \hyperref[deffiltration]{filtration} of a simplicial complex (definition \ref{def_filtration}) is a concept related to the one of sub-level set (definition \ref{def_sublevel}).
Basically, it consists in a set of sub-complexes which are given a particular order. Figure \ref{fig_persistence} shows the sub-complexes $K^i$ in a \hyperref[deffiltration]{filtration} $F$ of a simplicial complex $K$, the index $i$ being represented in the bottom left part of each box. It is the counterpart of a sub-level set in the sense that the arrival order of each simplex in the \hyperref[deffiltration]{filtration} can be defined by a function that affects a value to each simplex, in which case each sub-complex $K^i$ in the \hyperref[deffiltration]{filtration} can be defined as the set of simplexes with value higher or lower than a given threshold $v_i$. Note that the complex $K$ is always the last to enter the \hyperref[deffiltration]{filtration}, and is therefore represented in box number $17$. In this particular \hyperref[deffiltration]{filtration}, the simplexes of $K$ enter one at a time (we skipped a few steps for the sake of conciseness, as symbolized by the gray hatched box). This does not have to be the case in general though, but because each sub-complex in the \hyperref[deffiltration]{filtration} is a simplicial complex, a particular simplex may never enter a \hyperref[deffiltration]{filtration} before any of its faces. In each frame, the newly entering simplex is colored in red or blue, and the two numbers following the index are the Betti numbers ${\bf \beta}^i=\left(\beta^i_0,\beta^i_1\right)$ of $K^i$. As detailed in appendix \ref{sec_simp_homo}, $\beta_0$ represents the number of components in a complex (\ie how many separated ``islands'' exist) while $\beta_1$ is the number of holes or, equivalently, the number of independent non-homologous $1$-cycles one can create in $K^i$ in the more sophisticated language of homology.\\

\begin{figure*}
\centering
\includegraphics[width=\linewidth]{img/persistence}
\caption{\label{fig_persistence} Illustration of the filtration of a 2D simplicial complex. Each box represent one step of the filtration, with index $i$ (lower left corner) and Betti numbers $\left(\beta_0,\beta_1\right)$. The gray hatched box symbolizes the fact that a few steps are not represented.}
\end{figure*}

 Let us see how Betti numbers can be computed using this particular algorithm. $K^0$ is always the empty set, and so the algorithm starts with ${\bf \beta}^0=\left(0,0\right)$. A vertex first enters $F$ to form $K^1$, this adds one new component in the \hyperref[deffiltration]{filtration}, but no one cycle can still be created, so ${\bf \beta}^1=\left(1,0\right)$. As the entering vertex {\em created} a new component, it is represented in red and is labeled ``positive''. Step $2$ is essentially the same, and therefore ${\bf \beta}^2=\left(2,0\right)$. In $K^3$ though, the first segment enters $F$. Although we had two distinct components in $K^2$, the segment creates a link between them, and only one component remains. As one component was {\em destroyed}, the entering segment is represented in blue and labeled ``negative'', and $\beta_0$ decreases, leading to ${\bf \beta}^3=\left(1,0\right)$ again. Nothing special happens up to $K^{8}$, every entering vertex creating a new component therefore increasing $\beta_1$ while each new segment destroys a components, therefore decreasing the value of $\beta_0$, leading to ${\bf \beta}^7=\left(1,0\right)$. The segment entering $K^{8}$ is different though, as it does not destroy any component: all the simplex in $K^7$ where already linked and the new segment only links two vertice that already belonged to the same component. Actually, it {\em creates} a new class of $1$-cycles (black rounded arrow) as it is now possible to draw a segments path that starts and ends at the same segment while passing through each other segment in the path only once (equivalently, it creates a hole within the cycle). The value of $\beta_1$ is therefore increased and  ${\bf \beta}^8=\left(1,1\right)$. The entering segment is labeled ``positive'' and represented in red. The new segment in $K^9$ is of the same kind: it creates a second hole, or equivalently a second class of cycles (black circular arrows) that is not homologous to the previous one . In fact, one cannot transform one into the other by adding the boundary of a set of facets, as there is no facet in the complex yet anyway. The entrance of a facet in $K^{10}$ changes this fact, as this facet does fill one of the previously created hole: by adding the edges of this facet to the cycle created in $K^8$ one obtains a cycle created in $K^9$, the two classes therefore becoming homologous (remember that by adding a simplex to a complex already containing it, one actually removes it). This leads to $\beta_1$ being decreased and therefore ${\bf \beta}^{10}=\left(1,1\right)$. The \hyperref[deffiltration]{filtration} then goes on until all simplexes in $K$ have entered, and ${\bf \beta}^{19}=\left(1,0\right)$.\\

 Although we only presented a 2D example here, the procedure works for any number of dimensions and one can in general think of a $k$-cycle as the shell of a deformed {\kp1}-dimensional sphere triangulated with \hyperref[defksimplex]{$k$-simplexes}, the simplest $k$-cycle being the faces of a {\kp1}-simplex. The algorithm therefore consists in labeling each \hyperref[defksimplex]{$k$-simplex} of $K$ as ``positive'' if it creates a $k$-cycle and ``negative'' if it destroys one when entering the \hyperref[deffiltration]{filtration}.\footnote{The question of how to decide weather a newly entered simplex actually belongs to a cycle or not is addressed in \citet{edel}, but we do not present the method here as it is not essential to understand the concept of \hyperref[defpers]{persistence}. The implementation of such an algorithm is detailed in section \ref{sec_simplification}.} Going a bit farther, one can see that actually any cycle destroyed by an entering simplex were created earlier in the \hyperref[deffiltration]{filtration}. For instance, the segment that enters in $K^3$ destroys the component created by the entering vertex in $K^1$ or $K^2$. By convention, we will say that it destroys the most recently created, the vertex entering $K^2$. Identically, the new segment in $K^6$ destroys the new component created in $K^4$, and the loop created in $K^9$ is destroyed by the facet entering $K^{10}$ while the facet entering $K^{11}$ destroys the cycle created by the segment entering $K^8$. This defines pairs of negative and positive simplexes that create and destroy cycles, the partner of a positive (resp. negative) \hyperref[defksimplex]{$k$-simplex} being a negative {\km1}-simplex (resp. positive {\kp1}-simplex). All the cycles can therefore be attributed some sort of ``lifetime'' in the \hyperref[deffiltration]{filtration}, equal to the index difference of their creating and destroying simplexes. This lifetime is called their \hyperref[defpers]{persistence}. In the case of figure \ref{fig_persistence} for example, the most persistent topological feature of $K$ would be that $K$ has two main components, joined by a central bridge: the segment entering $K^{19}$ destroys the component created by the vertex entering $K^{12}$, the \hyperref[defpers]{persistence} of this topological feature therefore is $19-12=7$, which is larger than any other in the \hyperref[deffiltration]{filtration}. Of course, for a given complex, the \hyperref[defpers]{persistence} of each cycle (and actually the cycles themselves) depends on their precise order of arrival and what \hyperref[defpers]{persistence} really assesses is the topological properties of a function defined on the simplicial complex (\ie the function that defines the order of arrival of the simplexes in the \hyperref[deffiltration]{filtration}).



\cleardoublepage
\onecolumn

\section*{Terminology}
\label{sec_terminology}
\description
\item[{\bf Arc}]\label{defarc} An arc is a $1$-cell: an integral line (or a V-path in the discrete theory) whose origin and destinations are critical points. The arcs of \hyperref[defMSC]{Morse-Smale complex} comply to conditions \ref{prop_mscomplex}, in particular, an arc always connects two critical points of order difference $1$ (\ie in 2D, a minimum and a saddle-point or a maximum and a saddle-point).   
\item[{\bf $n$-cell}]\label{defcell} A $n$-cell is a region of space of dimension $n$ such that all the integral lines in the $n$-cell have a common origin and destination. The $n$-cells basically partition space into regions of uniform gradient flow (see definition \ref{def_morse_ncell}) 
\item[{\bf Coface}] \label{defcoface} A coface of a $k$-simplex $\alpha_k$ is any $p$-simplex $\beta_p$, with $p\geq q$, such that $\alpha_k$ is a face of $\beta_p$. In 3D, the cofaces of a segment (\ie a $1$-simplex) are any triangle or tetrahedron (\ie $2$ or $3$-simplex) whose set of summits (\ie vertexes) contains the two vertexes at the extremities of the segment, as well as the segment itself. (see definition \ref{def_face})
\item[{\bf Cofacet}]\label{defcofacet}  A cofacet of a $k$-simplex $\alpha_k$ is a coface $\beta_{k+1}$ of $\alpha_k$ with dimension $k+1$. Equivalently, $\alpha_k$ is a facet of $\beta_{k+1}$. (see definition \ref{def_face})
\item[{\bf Critical point of order $k$}] For a smooth function $f$, a critical point of order $k$ is a point such that the gradient of $f$ is null and the Hessian (matrix of second derivatives) has exactly $k$ negative eigenvalues. in $2$D, a minimum, saddle point and maximum are critical points of order $0$, $1$ and $3$ respectively. (see definition \ref{def_crit})
\item[{\bf Critical $k$-simplex}] A critical $k$-simplex is the equivalent in discrete Morse theory of the critical point of order $k$ in its smooth counterpart. Note that in $2$D, the equivalent of a minimum is a critical vertex ($0$-simplex), a saddle-point is a critical segment ($1$-simplex) and a maximum is a critical triangle ($2$-simplex). (see definition \ref{def_critical_simplex})
\item[{\bf Crystal}]\label{defcristal} A crystal is a $3$-cell: a 3D region delimited by $6$ quads and $12$ arcs, within which all the integral lines (or V-pathes in the discrete case) have identical origin and destinations.
\item[{\bf $k$-cycle}]\label{defkcycle}  A $k$-cycle in a simplicial complex corresponds to a $k$ dimensionnal topological feature. in $3D$, $0$-cycles correspond to independant components, $1$-cycles to loops and $2$-cycles to shells (see definition \ref{def_kcycle} and appendix \ref{sec_simp_homo})
\item[{\bf Discrete Gradient}]\label{defDG} A discrete gradient of a discrete Morse-Smale function $f$ defined over a simplicial complex $K$ pairs simplexes of $K$ according to the rules of definition \ref{def_DGradient}. Within a gradient pair, the simplex with lower value is called the tail and the other the head, and any unpaired simplex is critical (see definition \ref{def_DGradient}).
\item[{\bf Discrete Morse-Smale complex (DMC)}]\label{defDMC} The discrete Morse-Smale complex (DMC for short) is the equivalent of the Morse-Smale complex applied to simplicial complexes (see discrete Morse theory as introduced in section \ref{sec_DMtheory}) (see definition \ref{def_morse_complex}).
\item[{\bf Discrete Morse-Smale function}]\label{defDMF} A discrete Morse-Smale function $f$ defined over a simplicial complex $K$ associates a real value $f\left(\sigma_k\right)$ to each simplex $\sigma_k\in K$ and that obey the condition described in definition \ref{def_DMfunction}.
\item[{\bf Excursion set}] see sub-level set.
\item[{\bf Face}]\label{defface} A face of a $k$-simplex $\alpha_k$ is any $p$-simplex $\beta_p$ with $p\leq q$, such that all vertexes of $\beta_p$ are also vertexes of $\alpha_k$. In 3D, the faces of a $3$-simplex (\ie a tetrahedron) are the tetrahedron itself, the $4$ triangles that form its boundaries, the $6$ segments that form its edges, and its $4$ summits (\ie vertexes). (see definition \ref{def_face})
\item[{\bf Facet}]\label{deffacet} A facet of a $k$-simplex $\alpha_k$ is a face $\beta_{k-1}$ of $\alpha_k$ with dimension $k-1$. The facets of a $3$-simplex (\ie a tetrahedron) are the $4$ triangles (\ie $2$-simplexes) that form its boundaries (see definition \ref{def_face})
\item[{\bf Filtration}]\label{deffiltration} A filtration of a simplicial complex $K$ is a {\em growing} sequence of sub-complexes $K_i$ of $K$, such that each $K_i$ is also a simplicial complex. If the different $K_i$ are defined by a discrete function $F_\rho$ as the set of simplexes of $K$ with value $F_\rho\left(\sigma\right)$ less or equal to a given threshold, a filtration can be though of as the discrete equivalent of a sequence of growing sub-level sets of a smooth function. (see definition \ref{def_filtration})
\item[{\bf Gradient pair / arrow}]\label{defGP} A Gradient pair or arrow is a set of two simplex, one being the facet of the other, and such that they are paired within a discrete gradient. Within a gradient pair, the simplex with lower value is called the tail and the other the head.
\item[{\bf Integral line}] An integral line of a scalar function $\rho\left({\mathbf x}\right)$ is a curve whose tangent vector agrees with the gradient of $\rho\left({\mathbf x}\right)$. An integral line obeys properties \ref{def_int_line} (see definition \ref{def_int_line})
\item[{\bf Level set / Sub-level set)}] A level set, also called iso-contour, of a function $\rho\left({\mathbf x}\right)$ at level $\rho_0$ is the set of points such that $\rho\left({\mathbf x}\right)=\rho_0$. The corresponding Sub-level set is the set of points such that $\rho\left({\mathbf x}\right)\geq\rho_0$ (see definition \ref{def_sublevel})
\item[{\bf Ascending/Descending $p$-manifold}]\label{defmanifold} Within a space of dimension $d$, an ascending $p$-manifold is the set of points from which, following minus the gradient, one reaches a given critical point of order $d-p$. A descending $p$-manifold is the set of points from which, following the gradient, one reaches a given critical point of order $p$. For istance, ascending $1$-manifolds in 3D can be associated to the filaments, and ascending $3$-manifolds describe the voids (see definition \ref{def_manifolds})
\item[{\bf Morse function}]\label{defMF} A Morse function is a continuous, twice differentiable smooth function whose critical points are non degenerate. In particular the eigenvalues of the Hessian matrix (\ie the matrix of the second derivatives) must be non-null (see definition \ref{def_morse_function})
\item[{\bf Morse complex}] The Morse complex of a Morse function is the set of its its ascending (or descending) manifolds (see definition \ref{def_morse_complex})
\item[{\bf Morse-Smale function}] A Morse-Smale function is a Morse function whose ascending and descending manifolds intersect {\em transversely}. This means that there exist no point where an ascending and a descending manifold may be tangent (see definition \ref{def_morseSmale_function} or \ref{def_dmanifolds} for the discrete case)
\item[{\bf Morse-Smale complex}]\label{defMSC} The Morse-Smale complex is the intersection of the ascending and descending manifolds of a Morse-Smale function. One can think of the Morse-Smale complex as a network of critical points connected by $n$-cells, defining a notion of hierarchy and neighborhood among them. In particular, the geometry of the arcs (\ie $1$-cells) is determined by the critical integral lines (\ie integral lines that join critical points) and the order of two critical points connected by an arc may only differ by $1$.  
\item[{\bf Peak/Void patch}] In 3D, a peak patch is a descending $3$-manifold (\ie the region of space from which, following the gradient, one reaches a given maximum), and a void patch an ascending $3$-manifold (\ie the region of space from which, following minus the gradient, one reaches a given minimum). 
\item[{\bf Persistence}] \label{defpers} The persistence of a persistence pair (or equivalently of the corresponding $k$-cycle it creates and destroys) is defined as the difference between the value of the two critical points (or critical simplexes in the discrete case) in the pair. It basically represents its life time within the evolving sub-level sets, or filtration in the discrete case. (see section \ref{sec_persistence} and definition \ref{def_persistence})
\item[{\bf Persistence pair}]\label{defperspair} In the smooth context of a function $\rho$, persistence pairs critical points $P_a$ and $P_b$ of $\rho$ that respectively create and destroy a topological feature (or $k$-cycle) in the sub-level sets of $\rho$, at levels $\rho\left(P_a\right)$ and $\rho\left(P_b\right)$. In the discrete case of a simplicial complex $K$, a persistence pair is a pair of critical simplexes $\sigma_a$ and $\sigma_b$ of a given discrete function $F_\rho\left(\sigma\right)$, such that $\sigma_a$ creates a $k$-cycle (\ie topological feature) when it enters the filtration of $K$ according to $F_\rho$ and $\sigma_b$ destroys it when it enters. (see section \ref{sec_persistence} or appendix \ref{sec_app_persistence} for more details)
\item[{\bf Persistence ratio}] \label{defpersR} The persistence ratio of a persistence pair (or equivalently of the corresponding $k$-cycle it creates and destroys) is the ratio of the value of the two critical points (or critical simplexes in the discrete case) in the pair. Persistence ratio is preferred to regular persistence in the case of strictly positive functions such as the density field of matter on large scales in the universe. (see also the definition of persistence)   
\item[{\bf Quad}] \label{defquad}A quad is a $2$-cell : a 2D region delimited by four arcs within which all the integral lines (or V-pathes in the discrete case) have identical origin and destinations.
\item[{\bf $k$-simplex}] \label{defksimplex}A $k$-simplex is basically the $k$ dimensional analog of a triangle: the simplest geometrical object with $k+1$ summits, called vertex. It is the building block of simplicial complexes (see definition \ref{def_simplex})
\item[{\bf Simplicial complex}] A simplicial complex $K$ is a set of simplexes such that if a $k$-simplex $\alpha_k$ belongs to $K$, then all its faces also belong to $K$. Moreover, the intersection of two simplexes in $K$ must be a simplex that also belongs to $K$ (see definition \ref{def_simp_complex})
\item[{\bf Vertex}] A vertex is a $0$-simplex or simply a point.
\item[{\bf V-path}] \label{defVP}A V-path is the discrete equivalent of an integral line: it is a set of simplexes linked by discrete gradient arrows and face/coface relation. Tracing a V-path basically consists in intuitively following the direction of the gradient pairs of a discrete gradient from a critical simplex to another. (see definition \ref{def_vpath})
\enddescription

\label{lastpage}

\end{document}